\documentclass[twocolumn,amsmath,amssymb,superscriptaddress]{paper}

\newcommand{\av}[1]{\langle #1 \rangle}

\usepackage[english]{babel}
\usepackage{graphicx}                       
\usepackage{dcolumn}                        
\usepackage[colorlinks=false]{hyperref}     
\usepackage{bm}                             
\usepackage{subfigure}
\usepackage{longtable}
\usepackage{mathtools}
\usepackage[hang,small]{caption2}

\usepackage{amssymb}
\usepackage{amsmath}
\usepackage[affil-it]{authblk}

\graphicspath{{figures/}}

\begin{document}
\title{Asymptotic theory for the dynamic of networks with heterogenous social capital allocation}

\author[1,2]{\large Enrico Ubaldi}
\author[3,4]{\large Nicola Perra}
\author[5]{\large M{\'a}rton Karsai}
\author[1,6]{\large Alessandro Vezzani}
\author[1,2]{\large Raffaella Burioni}
\author[4]{\large Alessandro Vespignani}

\affil[1]{\small Dipartimento di Fisica e Scienza della Terra, Universit\`a di Parma, Parco Area delle Scienze 7/A, 43124 Parma, Italy}
\affil[2]{\small INFN, Gruppo Collegato di Parma, Parco Area delle Scienze 7/A, 43124 Parma, Italy}
\affil[3]{\small Centre for Business Network Analysis, University of Greenwich, Park Row, London SE10 9LS, United Kingdom}
\affil[4]{\small Laboratory for the Modeling of Biological and Socio-technical Systems, Northeastern University, Boston MA 02115 USA}
\affil[5]{\small Laboratoire de l'Informatique du Parall\'elisme, INRIA-UMR 5668, IXXI, ENS de Lyon, 69364 Lyon, France}
\affil[6]{\small Centro S3, CNR-Istituto di Nanoscienze, Via Campi 213A, 41125 Modena Italy}

\date{August 2015}

\maketitle{}

\begin{abstract}
The structure and dynamic of social network are largely determined by the heterogeneous interaction activity 
and social capital allocation of individuals. These features interplay in a non-trivial way in the formation of network and challenge a rigorous dynamical system theory of network evolution. Here we study seven real networks describing temporal human
interactions in three different settings: scientific collaborations, Twitter
mentions, and mobile phone calls. We find that the node's activity and social capital allocation can be described by two general functional forms that can be used to define a simple stochastic model for social network dynamic. This model allows the explicit asymptotic solution of the Master Equation describing the system dynamic, and provides the scaling laws characterizing the time evolution of the social network degree distribution and individual node's ego network. The analytical predictions reproduce with accuracy the empirical observations validating the
theoretical approach. Our results provide a rigorous dynamical system framework that can be extended to include other features of networks' formation and to generate data driven predictions for the asymptotic behavior of large-scale social networks.
\end{abstract}

The formation of social networks requires investments in time and energy by each individual actor with the anticipation that collective benefits can arise for individuals and groups. Individuals however invest in developing social interactions heterogeneously and according to very diverse strategies. In the first place not all individuals are equally active in a given social network. Furthermore, individuals may allocate their social capital in very diverse way, for instance by favoring the strengthening of a limited number of strong ties (bonding capital) as opposed to favor the exploration of weak ties opening access to new information and communities (bridging capital)~\cite{granovetter1973strength,friedkin1980test,lin1981social,granovetter1983,Brown:1987kx,nelson1989strength,levin2004strength,Demeo2014_Facebook}.   The origins of such heterogeneities are rooted in the trade off between competing factors such as the need for close relationships~\cite{holt2010social}, the efforts required to keep social ties~\cite{dunbar1998social}, temporal and cognitive constraints~\cite{miritello2013time,stiller2007perspective,powell2012orbital}, and  have long been acknowledged as key elements in the description of social networks' properties~\cite{Onnela01052007,PhysRevE.83.025102,Karsai:2014aa}, dynamical features ~\cite{Holme:2012lr,holme2013temporal,cattuto2010dynamics,Isella:2011,PhysRevE.81.035101,iribarren2009impact,Perra:2012uq,Saramaki21012014,PhysRevE.83.025102,clauset07}, and the  the behavior of processes unfolding in social systems~\cite{Onnela01052007,PhysRevE.83.025102,Karsai:2014aa,Holme:2012lr,morris93-1,rocha2010information,Perra:2012fk,ribeiro13-1,pfitzner13-1,starnini_rw_temp_nets,bakshy2012role}.  However, it is still lacking a general dynamical system framework able to relate the emerging connectivity pattern of social networks to the combined action of social actors activity and their heterogeneity in distributing resources in social capital allocation.

Here we analyze seven time-resolved datasets describing three different types of social interactions: scientific collaborations, Twitter mentions, and mobile phone calls. For all network datasets we define two functions statistically encoding the instantaneous  activity of nodes and the allocation of social capital, respectively. The latter function is regulated by two parameters -  system dependent- that define a simple reinforcement mechanism. In particular we observe in all datasets that the larger the number of social ties already activated by each node and the smaller is the probability of creating a new tie. We provide a thorough statistical characterization of the activity and reinforcement dynamics at play in each network and identify the basic parameters defining the dynamic of ties evolution. 

Prompted by this statistical analysis, we propose a dynamic network model that includes the heterogenous activity of nodes and the the tie formation mechanisms. This model allows the definition of a formal Master Equation (ME) describing the evolution of the network connectivity structure that can be solved in the asymptotic regime (large network size and long time evolution). The solution of the ME provides the asymptotic form the degree distribution and the scaling relations relating  degree, activity and  and the functions characterizing the social capital allocation. The analytical solutions are capturing very well the empirical behavior measured in the analyzed datasets, connecting explicitly the evolution of social networks to the parameter regulating the emergence of heterogenous social ties. The proposed analytical framework is remarkably general and it can be solved for statistically different activity patterns. The presented results have the potential to open the path to a general asymptotic theory of the dynamic of social networks by progressively integrating further social capital allocation strategies for the formation of social ties.

\section{Results} 

We analyze seven datasets containing time-stamped information about three different type of social
interactions: scientific collaborations, Twitter mentions, and
mobile phone calls. While we refer the reader to the Material and Methods section
for the details of each data set, we represent all datasets as time-varying networks. Each node
describes an individual. Each time-resolved link describes a social act. The
nature of connections is different according to the specific dataset. Links
might represent a collaboration resulting in a publication in a scientific journal, a Twitter mention, or a
mobile phone call. We considered five scientific collaborations networks
obtained from five different journals ($PRA$, $PRB$, $PRD$, $PRE$, and $PRL$) of
the American Physical Society (APS), one Twitter mentions network ($TMN$), and
one mobile phone network ($MPN$).

In order to characterize the time-varying properties of such networks we first
measure the activity $a_i$. Formally,  $a_i$ is defined as the fraction
of interactions  in which node $i$ is engaged per unit
of time with respect to all the interactions per unit time occurring in the network. 
This quantity describes the propensity of nodes $i$ to be involved in social
interactions.  Empirical measurements in a wide set of social networks show
broad distributions of activity \cite{Perra:2012uq,
Perra:2012fk,ribeiro13-1,Karsai:2014aa,tomasello14-1}. As shown in
Figure~\ref{fig:rhoa} [A-D], we confirm these observations in our datasets. In
particular we find that  in the APS and MPN datasets the activity is well fitted
by a truncated power law, while in the TMN we find a Log-Normal distribution
(see Material and Methods and Supplementary Online Materials for details).

\subsection{Social capital allocation}
The activity $a_i$ sets the clock for the activation of each node, however it
does not provide any information on how each node invests its social capital in
exploring new ties or reinforcing already established
ties~\cite{miritello2013limited}.
In order to measure the formation of new ties, we group nodes in classes with
similar activity $a$ and final degree $k$, so that each class $b$ contains
actors with statistically equivalent characteristics (see SI for details).
We then measure the probability $p_b(k)$ that the next social act
for the nodes in the class $b$  that have already contacted $k$ nodes will result in the
establishment of a new, $k + 1$-th, tie. 
As shown in Figure~\ref{fig:rhoa} [E-H] $p_b(k)$ is in general a decreasing function of
$k$. This observation resonates with previous research and empirical findings suggesting
that our social interactions are bounded by cognitive and temporal
constrains~\cite{dunbar1998social,miritello2013time,stiller2007perspective,powell2012orbital}.
Indeed, the larger the number of alters in our social circle, the smaller the probability
that the next social act will be towards a new tie.

The above empirical findings  suggest that the mechanism governing the allocation of social capital follow a general form that in its 
simplest analytical form can be written as:
\begin{equation}
    p_b(k) = \left( 1 + \frac{k}{c_b} \right)^{-\beta_b}.
    \label{eq:pn}
\end{equation}
In this expression, $\beta_b$ modulate the tendency to explore new connections, while  $c_b$ define the intrinsic characteristic limit of the individual to maintain multiple ties. Although one could imagine more complicate analytical forms, we use this parsimonious approach to characterize the different data sets. Interestingly, we find that in the five co-authorship networks and Twitter, the exponent $\beta$ is the same regardless of the class $b$. Furthermore, the values of $c_b$ are typically peaked around a
well defined value (see SI for details). More in detail, we can rescale the
proposed functional form in each class $b$ by defining the variable $x_b=k/c_b$, yielding
 \begin{equation}
p_b(x_b)^{\frac{1}{\beta}}=(1+x_b)^{-1}.
\end{equation}
In the presence of a single exponent $\beta$ characterizing the system, as shown
in Figure~\ref{fig:rhoa} [I-K], all empirical curves do collapse on the
reference function $(1+x)^{-1}$. The data collapse however is not occurring in
the case of the MPN dataset. In the latter we find a more heterogeneous scenario in which different nodes' classes  are characterized by different values of $\beta_b$ and $c_b$, see
Figure~\ref{fig:rhoa} L. In the Supplementary Online Material we provide further evidence for the evidence of a single or distirbuted value of $\beta$ in different datasets.

\subsection {Stochastic model for the network dynamic}
By leveraging on the empirical evidence gathered here, it is possible to define a basic generative model of network formation based on two stochastic processes. Defined the network $\mathcal{G}$
containing $N$ nodes, at each time step a  node $i$ is active according to a
probability $a_i$ drawn from distribution $F(a)$.~\cite{Perra:2012uq,
Perra:2012fk,ribeiro13-1,Karsai:2014aa,tomasello14-1}. Once active, the node $i$ that has already contacted $k$ different agents will contact
a new, randomly chosen node with probability $p_i(k) = (1 + k/c_i)^{-\beta_i}$. Otherwise, with probability $1 - p_i(k)$, it will interact with an already contacted node chosen at random. Interactions are considered to last one single time step.
For this model it is possible to write explicitly the master equation (ME) describing the evolution of the probability distribution $P_i(k,t)$ that a node $i$ has degree $k$ at time $t$:
{
\small
\begin{align}
    \label{eq:ME_mm}
        &P_i(k, t + 1) = &\nonumber \\
        &P_i(k - 1, t)  \bigg[ {a_i p_i(k-1)} + \sum_{j\nsim
        i}{ a_j \sum_{k_j}{p_j(k_j)\over (N - j)}  P_j(k_j, t) } \bigg] + &\nonumber \\
        &P_i(k_, t)  \bigg[ {a_i [1 - p_i(k)]} + \sum_{j\nsim
        i}{ a_j \sum_{k_j}{\Big( 1 - {p_j(k_j)\over (N - j)}\Big) P_j(k_j, t)}
        } \bigg] + &\nonumber \\
        &P_i(k, t)  \bigg[ 1 - \sum_{j}{a_j} \bigg].&
\end{align}
}
In the above equation ${\small j\sim i}$ and ${\small j\nsim i}$ are the sum over the nodes
already contacted and not yet contacted by ${\small i}$, respectively. Within
these sums, we use ${\small k_j}$ as the degree of the node ${\small j}$.
The first two terms on the right hand side of Eq.~(\ref{eq:ME_mm}) account for
the creation of nodes of degree $k$ which occurs when a node of degree $k-1$
gets active and contacts a new node, or when it gets in contact with a new node of
previous degree $k_j$ that activates and attaches to node $i$.
The third and fourth terms of the r.h.s. of the equation account for the
conservation of nodes of degree $k$, i.e. nodes that either get active and
contact one of their neighbors with probability $a(1-p(k))$ or get contacted by
one of their neighbors. The last line of Eq.~(\ref{eq:ME_mm}) takes into account for the case in which no node
gets active in the current evolution time step, thus conserving the $P_i(k,t)$.

\subsection{Asymptotic theory for network with $\beta_b=\beta$}

In the case of networks characterized by a single exponent $\beta$ it is possible to consider for the ME
the large time and large ${\small k}$ limit, so that $k$ can be approximated by a continuous variable.
By neglecting the subleading terms of order $1 \backslash t$ we can thus write the continuous asymptotic
version of  Eq.~(\ref{eq:ME_mm}) as
{
    \small
    \begin{align}
        \label{eq:ME_cont}
        \frac{\partial P}{\partial t}  = &
        - \frac{a c^\beta}{k^\beta} \frac{\partial P}{\partial k} +
        \frac{a c^\beta}{2 k^\beta} \frac{\partial^2 P}{\partial k^2} +
        \frac{a \beta c^\beta}{k^{\beta+1}} P(a,k,t) +&\\ \nonumber
        &\left( \frac{1}{2} \frac{\partial^2 P}{\partial k^2} -
        \frac{\partial P}{\partial k} \right)
        \int da  \rho(a) a \int dh  \frac{c^\beta }{h^\beta}P(a,h,t).&
        %
    \end{align}
}
This equation can be solved explicitly (see SI for details), yielding the asymptotic form:
\begin{equation}
    P_i(k, t) = A \exp{\left[ -\frac{\left( k - B(a_i,c_i) t^{\frac{1}{1 + \beta}}
    \right)^2}{C t^{\frac{1}{1 + \beta}}} \right]},
    \label{eq:Pakt}
\end{equation}
where $A$ is a normalization constant, $C$ a constant and $B(a_i,c_i)$ a
multiplicative factor of the $t^{1/(1+\beta)}$ term that depends on the activity
$a_i$ and $c_i$ of the considered agent. Its implicit expression is given in the
SI.

A first general result concerns the evolution in time of the average degree
$\av{k(a,t)}$ of nodes belonging to a given activity class that follows the scaling laws
\begin{align}
    \av{k(a, t)} \propto \left( at \right)^{1\over 1 + \beta}.
    \label{evol}
\end{align}
The growth of the system is thus modulated by the parameter $\beta$ that sets the strength of the reinforcement process in the process ruling the establishment of new social ties. In the limit case $\beta=0$ the growth would be linear. Indeed, the reinforcement of previously activated ties would be zero and nodes would keep connecting randomly to other vertices, thus increasing indefinitely their social circle. 
In the opposite limit $\beta \rightarrow \infty$ each node would invest is
social capital on just one single connection, i.e. the first established. In the
six datasets described by a single $\beta$ value, we observe the range $0.13 \le
\beta \le 0.47$  that indicates a sub-linear growth of the social system. In
Figure~\ref{fig:kat} we find a very good agreement between the analytical
prediction of Eq.~(\ref{evol}) and the empirical $\av{k(a,t)}$ curves, obtaining the first empirical validation of the modeling framework proposed and its ability at capturing the network formation dynamic. 

Furthermore, Eq.~(\ref{evol})  connects, at a given time $t$, the degree $k$ and the
activity $a$ of a given node, as $k \propto a^{1\over 1+\beta}$.
Thus, given any specific activity distribution $F(a)$,
we can infer the functional form of the degree distribution $\rho(k)$ by
substituting $a\to k^{1\over 1+\beta}$, finding:
\begin{align}
    \rho(k) dk \propto   F(k^{(1+\beta)}) k^{\beta} dk.
    \label{rhok}
\end{align}
It is important stressing that the analytical framework is not limited to a
specific functional form of the activity. Indeed, with an arbitrary functional
form of $F(a)$, Eq.~(\ref{evol}) gives us the possibility to predict
the behavior and parameters of the corresponding degree distribution. In
Table~\ref{tab:rhos} we report the degree distribution predicted by
Eq.~(\ref{evol}) for activities following a common set of heavy-tailed
distributions, i.e. power-laws, truncated power-law, stretched exponentials, and
log-normal, that are usually find in empirical data. In
Figure~\ref{fig:kat}[E-G] we compare the degree distributions $\rho(k)$
predicted by Eq.~(\ref{rhok}) with real data. Interestingly, also in this case the functional form obtained from the analytical solution of the model fit remarkably well the empirical evidence. It is important to notice that $\rho(k)$ is also function of the parameter $\beta$. In other words, the connectivity patterns emerging from social interactions can be inferred knowing the propensity of individuals to be involved in social acts, the activity, and the strength of the reinforcement towards previously establish ties, $\beta$.
Finally it is worth remarking that Eqs.~(\ref{evol}, \ref{rhok})  are not affected
by the distribution of $c_i$. This is an important result as it reduces the number of
relevant parameters necessary to define the temporal evolution of the system. 

\subsection{Asymptotic theory for networks with distributed $\beta$}
As we already mentioned, in the $MPN$ dataset we find the evolution of social
ties described by a distribution of $\beta$ rather than a single
value of it.  This observation points to a more heterogeneous distribution of
social attitudes with respect to the other six analyzed datasets. Arguably, such
tendency might be driven by the different functions phone calls serve enabling us
to communicate with relatives, friends or rather to companies, clients etc..
The need to introduce different values of $\beta$ in the system complicates the
model beyond analytical tractability (see SI for details).
Nevertheless, we find that the leading term of the evolving average degree can be described by introducing a simplified model, in which
the nodes of the system feature different values of $\beta$ and undergo a
simplified dynamics (see SI for further information) that neglects, for every
node, the effects of links established by others. In these settings we can solve the ME and show that the minimum value
of $\beta$, $\beta_{\rm min}$, rules the leading term of the evolving average
degree. In other words, we find that even in this case $\langle k(a,t) \rangle$ evolves as in
Eq.~(\ref{evol}) but with $\beta$ substituted by $\beta_{\rm min}$. As shown in Figure~\ref{fig:kat}-D the analytical predictions coming from the simplified model find good agreement with the empirical evidence. 
It is interesting to notice that the nodes characterized by $\beta_{\rm min}$ are those with the weak tendency to reinforce already established social ties. They are social explorers~\cite{miritello2013limited}. 
Notably, our results, indicate that they lead the growth of average connectivity of the network.


\section{Discussion} 
The empirical finding presented here shows clearly that the ``cost" associated to the establishment of a new social tie is not constant but is function of the number of already activated ties, thus supporting the idea that social capabilities are limited by cognitive, temporal or other forms of constraints~\cite{miritello2013time,stiller2007perspective,powell2012orbital}. Framing this empirical finding in a simple stochastic model of network formation, we can derive a general asymptotic theory of the network dynamic and derive the general scaling laws for the behavior in time of the node's degree and degree distribution. 

The model comes with some shortcomings. Indeed, it does not capture the modular
structure or, more in general, correlations beyond the nearest neighborhood that
are typical of many social networks~\cite{fortunato2010community}. In fact,
individuals tend to organize their social circles in tight, often hierarchical,
communities. The model does not capture the burstiness typical of social
acts~\cite{barabasi2005origin,karsai2012universal}. We consider a simplified
Poissonian scheme of nodes activation. A recent extension of the activity driven
framework, without the reinforcement mechanism acting on social ties, has been
proposed to account for non Poissonian node
dynamics~\cite{moinet2015burstiness}. This is the natural starting point to
generalize our model to bursty activities. Furthermore, the model does not
consider the turnover of social ties~\cite{miritello2013limited}. Indeed, in our
framework once a social connection has been established it cannot be eliminated
in favor of others. Clearly, this feature is of particular importance when
considering social systems evolving on longer time scales, as the scientific
journals we studied here, and might influence the measurement of the parameters
describing evolution of the ego-networks.

Notwithstanding these limitations, the modeling framework we propose pave the
way to a deeper understanding of the emergence and evolution of social ties. The
agreement between the analytical predictions and observed behaviors in seven
real datasets, describing different types of social interactions, are
encouraging steps in this direction. Finally, our results are a starting point for the development of predictive tools able to
forecast the growth and evolution of social systems based not just of regression
models or simplified toy models but on a more rigorous analysis of ego-network dynamics.

\section{Materials}

\subsection{Datasets} 
\label{sub:Datasets}
We analyzed seven large-scale and time resolved networks describing three different types of social interactions.
\begin{itemize}
\item Five networks from the APS datasets takes into account the co-authorship networks found
in the Journals of the American Physical Society.
Specifically, the PRA dataset covers the period from Jan. 1970 to Dec. 2006 and
contains 36,880 papers written by 34,093 authors and
connected by 100,683 edges.
The PRB dataset refers to the Jan. 1970 to Dec. 2007 period and contains
104,047 papers published by 84,367 authors which are
connected by 416,048 links.
The PRD datasets covers the same period as the PRB one and it is composed by
33,376 papers, 21,202 authors and 60,033 edges.
The PRE dataset refers to the Jan. 1993 to Dec. 2006 period with
24,204 papers published by 28,188 authors connected by 
68,029 edges.
Finally, the PRL dataset contains all the 66,422 papers published
between Jan. 1960 to Dec. 2006 and written by 78,763 authors forming
 299,017 edges.
\item One network dataset describing Twitter mentions (TMN), exchanged by users from January to September 2008. The network has 536,210 nodes performing about 160M events and connected by
2.6M edges.
\item One Network dataset describing the mobile phone calls network (MPN) of
6,779,063 users of a single operator with about 20\% market share in an
undisclosed European country from January to July 2008.  The datasets contains
all the phone calls to and from company users thus including the calls towards
or from 33,160,589 users in the country connected by 92,784,825 edges.
\end{itemize}

\subsection{Asymptotic solution of the ME for distributed $\beta_i$ values}
The solution of Eq.~(\ref{eq:ME_cont}) found in
Eq.~(\ref{eq:Pakt}) holds if the system feature a single value of
$\beta$.
As already discussed in the MPN dataset we find multiple values of ${\small
\beta}$ ranging from a minimum value, {\small $\beta_{\rm min}$} to a maximum one
${\small \beta_{\rm max}}$.
To find a prediction of the long time behavior of such a system, let us propose
a simplified model in which we focus on a single agent whose parameters are
${\small a_i}$, ${\small \beta_i}$ and ${\small c_i}$. In this simplified
version the agent can only call other nodes in the network, i.e. we neglect the
contribution coming from the incoming calls).
In this approximation we have to solve a modified version of
Eq.~(\ref{eq:ME_mm}), obtained by discarding all the terms containing
the activity $a_j$ of the nodes ${\small j\neq i}$.
By repeating the same procedure above, we get to the continuum limit that reads:
{
    \small
    \begin{align}
        \label{eq:ME_multi_b}
        {\partial P_i(k,t) \over \partial t}= -a_i\left({c_i\over
        k}\right)^{\beta_i}
        \left[ {\partial P_i(k,t)\over\partial k} -
            {1\over2}{\partial^2 P(k,t)\over \partial k^2}\right],
    \end{align}
}
whose solution is similar to Eq.~(\ref{eq:Pakt}), the only differences
being the value of ${\small \beta=\beta_i}$ and the behavior of the $B(a_i,c_i)$
constant (see Materials and Methods and the SI for details).
Interestingly, even in this case we find an average degree {\small$\av{k(a,t)}$} growing accordingly
to the exponent {$\small\beta_i$}, i.e. {\small$\av{k(a,t)}\propto (at)^{1\over
1+\beta_i}$}.
Now, let us create a reservoir of ${\small N}$ distinct nodes of equal activity
$a$ and assign to each of them a different value of {\small$\beta_i$} drawn from an arbitrary
distribution {\small$P(\beta_i)$}. Let us also group these nodes in {\small$B$}
classes, defined so that each class $i$ contains all the nodes featuring a similar value of
{\small$\beta\sim\beta_i$}.
If we now let these ${\small N}$ nodes evolve following the simplified model
above, the average degree of each class {\small$i$} will grow as ${\small
\av{k_i(a,t)}\propto t^{1\over1+\beta_i}}$.
Then, in the long time limit, the minimum value of ${\small\beta_i}$, i.e. ${\small \beta_{\rm
min}}$, will lead the growth of the ensemble's average degree (see SI for
further details), i.e.
{
    \small
    \begin{align}
        \av{k(t)}\propto t^{1\over 1+\beta_{\rm min}}.
        \label{eq:ltl_multib}
    \end{align}
}

\subsection{$F(a)$ and $\rho(k)$ distributions from real data}
\begin{center}
\begin{figure*}
    \centering
    \includegraphics[width=1.75\columnwidth,angle=0]{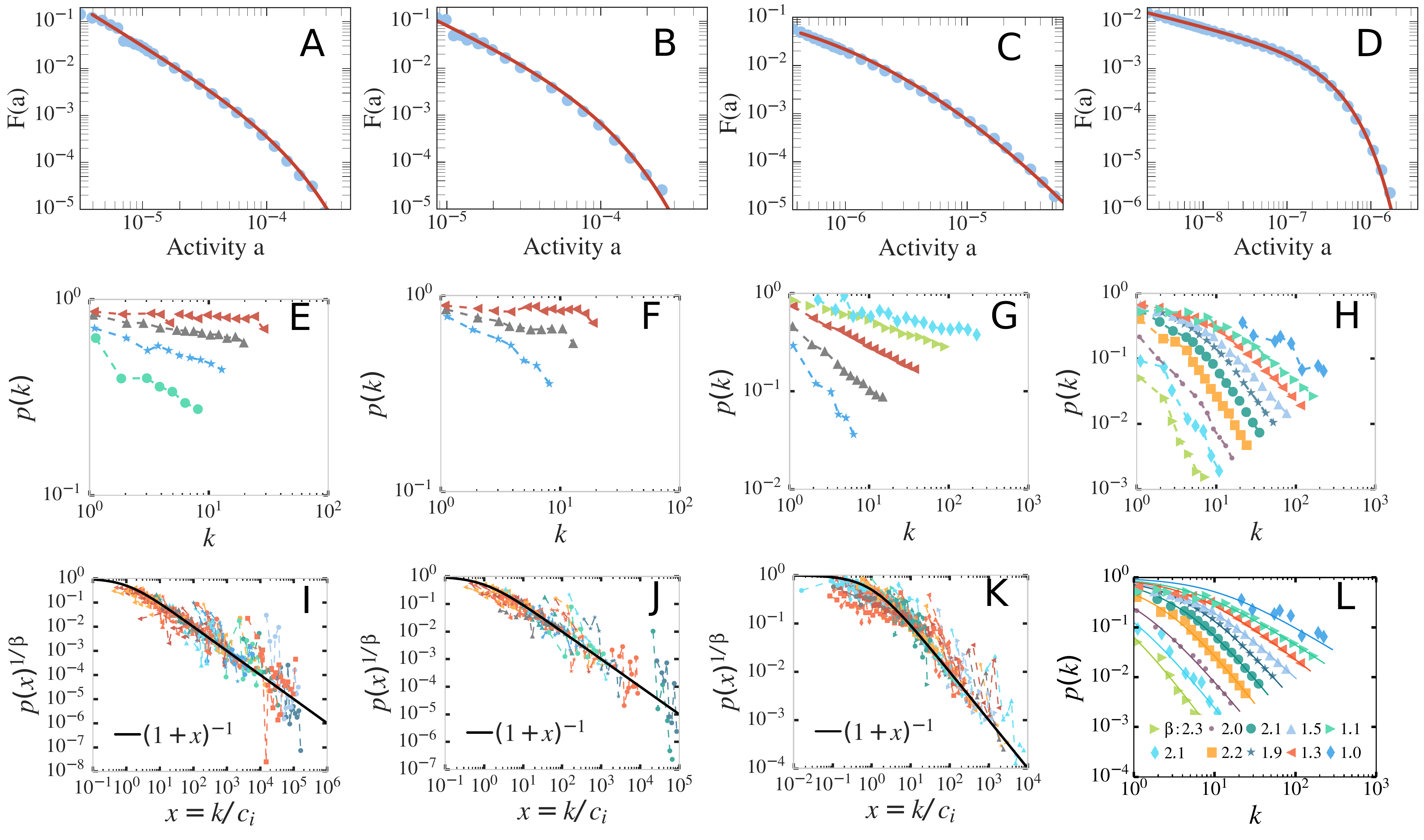}
    \caption{
        \label{fig:rhoa}
        ($A-D$) The activity distribution $F(a)$ for PRB ($A$), PRL
        ($B4$, TMN ($C$) and the MPN ($D$) dataset.
        The solid lines represent the fit $F(a)$ with the best functional
        form for each dataset. The latter are a truncated power law for the PRB,
        PRL and MPN case, while we find a lognormal for the TWT case (see SI and
        supplementary materials for details).
        In these plots we show the experimental data ranging from the lower
        bound of the fit to the $99.9\%$ of the total amount, thus excluding the
        higer $0.1\%$ of the measured activity values from the visible area (see
        materials and SI for details).
        ($E-H$) The measured $p_b(k)$ curves for selected nodes classes
        belonging to the PRB ($E$), PRL ($F$), TMN ($G$) and MPC ($H$) datasets.
        Each data sequence (different colors and markers) corresponds to a
        selected nodes class of the system.
        As one can see different nodes classes feature a differently behaving
        attachment rate function $p_b(k)$: for some nodes the probability to
        attach to a new node quickly drops to $0$ at degree $\lesssim 10$ while
        for some others the attachment probability is still $\gtrsim 0.1$ even
        at very large degree ($k\sim 10^2$).
        ($I-K$) We rescale the attachment rate curves of all the nodes classes
        of the PRB ($I$), PRL ($J$) and TMN ($K$) datasets by sending  $k \to
        x_b = k/c_b$ and then plotting the $p_b(x_b)^{1/\beta}$, where $\beta$ has
        the same value for every curve.
        For the MPC dataset ($L$) we show the  original $p_b(k)$ curves
        belonging to a single nodes class with their fit. The resulting values
        of $\beta_b$ are shown in the legend. The latter are found to fall in
        the $1.0 \lesssim \beta_i \lesssim 2.5$ range.
    }
\end{figure*}
\end{center}

We implement the method found in \cite{clauset2009power} to determine
the most likely functional form of both the activity and degree distributions.
The fitting procedure is as follows: for each functional form of the
distribution considered (power law, log-normal, truncated power law and stretched
exponential) we first determine the $x_{\rm min}$ value, i.e. the lower
bound to the functional form behavior. The $x_{\rm min}$ value is defined as the
value that minimizes the Kolmgorov-Smirnov (KS) distance between the analytical
complementary cumulative distribution (CDF) and the CDF of the data. The latter
are found for each value of $x_{\rm min}$ by computing the optimal parameters of the distribution using
the maximum-likelihood estimator (MLE).
Then, comparing the CDF$(x\ge x_{\rm min})$ of the data $S(x)$ with the
analytical one $S(x)$, we compute the KS-distance as the maximum distance
between the two CDF, i.e. ${\rm KS}_d(x_{\rm min}) = \max_{x_i\ge x_{\rm min}} |S(x_i)-P(x_i)|$.
Once all the distances are computed we determine $x_{\rm min}$ as the values at
which the minimum distance is recorded, i.e. $x_{\rm min} = \min_{x} {\rm
KS}_d(x)$ (see SI and \cite{clauset2009power} for details). Once we compute all
the parameters for all the functional forms analyzed we compare them with the
{\small\emph{likelihood ratio test}} $\mathcal{R}$ combined with the $p$-value
that gives the statistical significance of $\mathcal{R}$ (see SI for details).
The result of this procedure gives us the best candidate for the $F(a)$ for each
dataset. We find that a truncated power law is the best candidate for all the
APS datasets together with the MPN one. The only exception is the TMN that
displays a Log-Normal distribution of activity (see Fig.~\ref{fig:rhoa} and SI for details).
After we estimate the functional form and the parameters of the activity
distribution $F(a)$, Eq.~(\ref{rhok}) gives us the possibility to predict both
the functional form of the degree distribution $\rho(k)$ and the values of the
parameters of such a distribution (e.g. the $\alpha$ exponent in a power-law
with cutoff, see Table~\ref{tab:rhos} for details).
The degree distribution can then be fitted by optimizing over the non-scale-free
parameters for whose values we do not have an analytical or numerical prediction
(e.g. the cut-off $\tau$ in a power-law with cutoff).  Indeed, we are missing
the value of the constant in front of the $(at)^{1\over 1+\beta}$ term in the
growth of the average degree $\av{k(a,t)}$ in Eq.~(\ref{evol}).

\begin{center}
\begin{figure*}
    \centering
    \includegraphics[width=1.75\columnwidth,angle=0]{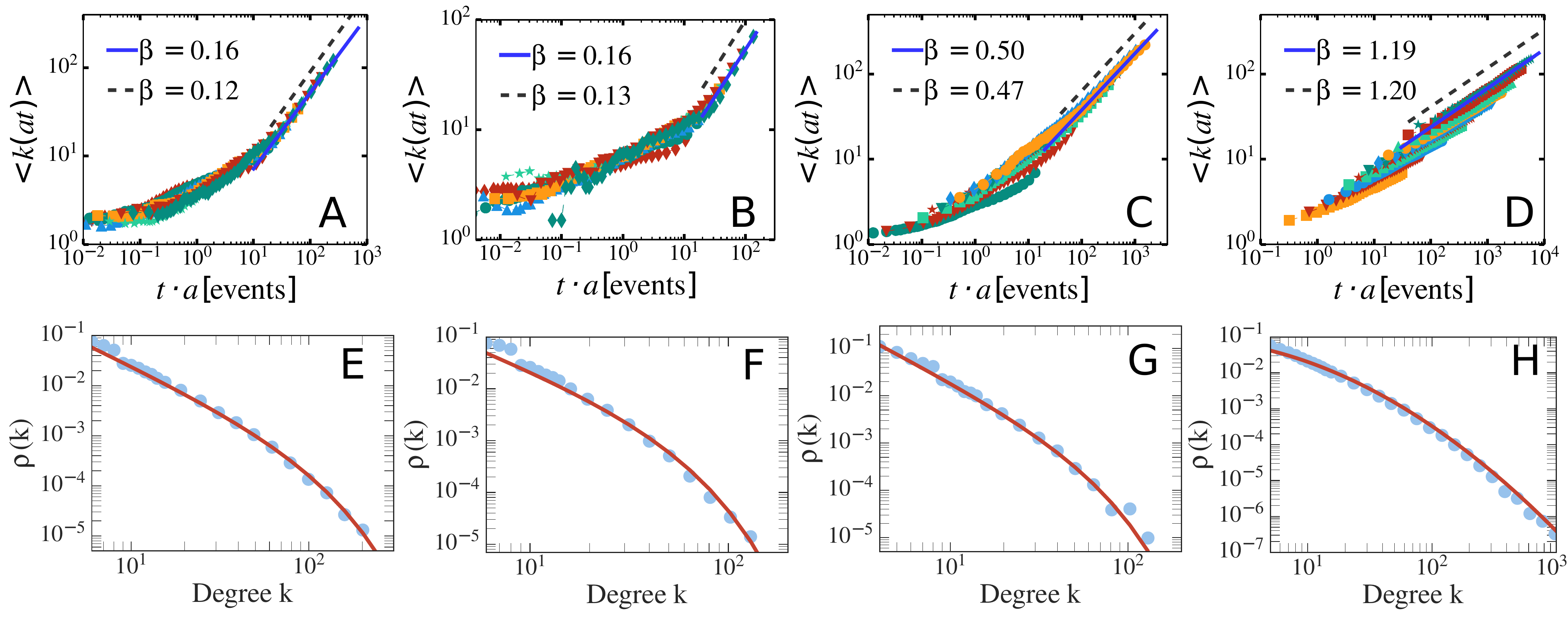}
    \caption{
        \label{fig:kat} ($A-D$) The rescaled $\av{k(at)}$ curves for selected
        nodes classes belonging to the PRB ($A$), PRL ($B$), TWT ($C$) and MPC
        ($D$) datasets.
        The time of the original data (symbols) is rescaled with the activity
        value $t\to at$. We also show the fitting curve $\av{k(t)} \propto t^{1
        \over 1 + \beta}$ (blue solid lines) and the expected asymptotic
        behavior (black dashed lines). In the MPC case $D$ we fit using $\beta =
        \beta_{\rm min} = 1.2$.
        ($E-H$) The degree distribution $\rho(k)$ for the PRB ($E$),
        PRL ($F$), PRA ($G$) and TMN ($H$) datasets. The predicted functional
        form of $\rho(k)$ found in Eq.~(\ref{rhok}) and Table~(\ref{tab:rhos})
        is shown for comparison (light blue solid lines).
        As in Fig. \ref{fig:rhoa} we show the data ranging from the lower bound
        of the degree distribution to the $99.9\%$ of the data range, thus
        excluding from the plot area the higher $0.1\%$ of the degree values.
    }
\end{figure*}
\end{center}

\begin{center}
\begin{figure*}
    \centering
    \def\arraystretch{1.05}
    \begin{tabular}{| l | c | c |}
        \hline
        PDF                         & $F(a)$                                &   $\rho(k)$                       \\
        \hline
        \small{Power Law}                   & \small{$a^{-\nu}$}     & \small{$k^{-[(1+\beta)\nu - \beta]}$}     \\
        \hline
        \small{Stret. Exp.}      & \small{$a^{\nu-1}\exp{\left[-\lambda a^{\nu}\right]}$} &  \small{$k^{[(1+\beta)(\nu-1) + \beta]} \exp{\left[-\tau k^{(1+\beta)\nu}\right]}$}     \\
        \hline
        \small{Trunc. PL}    & \small{$a^{-\nu}\exp{\left[-\lambda a\right]}$}  &  \small{$k^{-[(1+\beta)\nu - \beta]} \exp{\left[- \tau k^{(1+\beta)}\right]}$}     \\
        \hline
        \small{Log-Normal}                  & \small{${1\over a} \exp{\left[ -{(\ln(a) - \mu)^2
             \over 2\sigma_a^2} \right]}$}  &  \small{${1\over k} \exp{\left[ -{(\ln(k)-\gamma)^2
             \over 2\left( {\sigma_a\over 1+\beta}\right)^2} \right]}$}     \\
        \hline
    \end{tabular}
    \caption{
        \label{tab:rhos} The functional form of the activity PDF $F(a)$ and the
        predicted functional form of the $\rho(k)$ degree distribution as found
        in Eq.~(\ref{rhok}), i.e. by replacing $a\to k^{1+\beta}$.
        This substitution fixes the scale free parameters of the resulting
        distribution, i.e. the exponent of the power-law and of the $k$ term at
        the exponent in the first three cases, and the STD $\sigma_k =
        {\sigma_a\over 1+\beta}$ in the Log-Normal case.
        The free parameters over which we fit the degree distribution are: (i)
        the cut-off $\tau$ in the stretched exponential and power-law with
        cut-off and (ii) the $\gamma$ average value in the Log-Normal case.
        The selected PDF are, from top to bottom: power law, stretched exponential (Stret.
        Exp.), power law with cutoff (Trunc. PL) and the Log-Normal distribution.
    }
\end{figure*}
\end{center}

\bibliography{citation}{}
\bibliographystyle{unsrt}

\pagebreak
\onecolumn

\begin{center}
\textbf{\large Supplementary Information}
\end{center}
\setcounter{equation}{0}
\setcounter{figure}{0}
\setcounter{table}{0}
\setcounter{section}{0}
\setcounter{page}{1}
\renewcommand{\theequation}{S\arabic{equation}}
\renewcommand{\thefigure}{S\arabic{figure}}
\renewcommand{\thesection}{S\arabic{section}}

\section{Data-sets}
\label{sec:datasets}

\subsection{\textrm{A}merican \textrm{P}hysical \textrm{S}ociety}

The \textit{APS} dataset contains the five co-authorship networks of five
journals of the \textit{A}merican \textit{P}hysical \textit{S}ociety, i.e.,
\emph{Physical Review} \textit{A}, \textit{B}, \textit{D}, \textit{E} and
Letters (\textit{L}).

The various datasets contains the data referring to all the issues of the single
journals from their first issue up to a certain edition, specifically:
\begin{itemize}
    \item [-] \textit{PRA} from January $1970$ to December $2006$;
    \item [-] \textit{PRB} and \textit{PRD} from January $1970$ to December
        $2007$;
    \item [-] \textit{PRE} from January $1993$ to December $2006$;
    \item [-] \textit{PRL} from February $1960$ to December $2006$.
\end{itemize}

Each dataset is composed by several files (one per month). Each file has as many
lines as the number of papers published in that month. Finally, each line
contains the IDs of the authors of the specific paper.
For instance, the typical head of a file is:
\begin{verbatim}
        Author_000    Author_001    Author_002  #First Paper with 3 authors
        Author_003    Author_004                #Second Paper with 2 authors
          . . .         . . .         . . .            . . .
\end{verbatim}
The data are cleaned so as to not take into account the papers with a single author.

When analyzing this dataset we define the user's activity $a_i$ as the number of
engaged collaborations (e.g. an author $i$ that publish two papers, the first
with 3 co-authors and the second with a single co-author, has activity $a_i =
4$).

\subsection{\textit{T}witter \textit{M}ention \textit{N}etwork}

The dataset of \textit{Twitter} is composed by $273$ daily files covering the
period between January the $1^{\rm st}$ to September the $30^{\rm th} 2008$. The
dataset contains the so called \emph{fire-hose}, i.e., all the $16,329,466$
citations done by all the $536,210$ users in the given period. The nodes in the
network are connected via $2,620,764$ edges.

Each file contains the daily events with the structure:
\begin{verbatim}
    Citer_ID_00    Cited_ID_00    # Event 0
    Citer_ID_01    Cited_ID_01    # Event 1
    Citer_ID_02    Cited_ID_02    # Event 2
       . . .          . . .         . . .
\end{verbatim}

This dataset is not cleaned, as we have all the events that happened
on the platform in the selected period.

When analyzing this dataset we define the user's activity $a_i$ as the number of
citation made by $i$, i.e. the number of events actually engaged by the node $i$.

\subsection{\textit{M}obile \textit{P}hone \textit{N}etwork}

The dataset of the \textit{M}obile \textit{P}hone \textit{C}alls (\textit{MPC})
is composed by a single file containing the $1,949,624,446$ time ordered events
with $1$ second resolution covering the period between January and July of
$2008$ for $6,779,063$ users of a single operator with $20\%$ market share in an
undisclosed European country.

The dataset contains all the events from and toward users of the company (so
that even the calls from non-company users to company users and vice-versa are
taken into account).
As a result, we have $33,160,589$ nodes (of which $6,779,063$ are users of the
selected company) that are connected via $92,784,825$ edges.

We split the huge list of events in $98$ files (each of them containing more or
less the same number of events) for computing convenience.
Each file contains events with the structure:
\begin{verbatim}
    Caller_ID    Called_ID    Company_Caller   Company_Called   # Event 0
    Caller_ID    Called_ID    Company_Caller   Company_Called   # Event 1
    Caller_ID    Called_ID    Company_Caller   Company_Called   # Event 2
      . . .        . . .          . . .            . . .         . . .
\end{verbatim}
where \texttt{Company\_Caller} and \texttt{Company\_Called} are the value of the
provider company of the called and caller nodes, respectively (e.g. the value is
set to $1$ if the node is a customer of our company, $0$ otherwise).

When analyzing this dataset we define the user's activity $a_i$ as the number of
calls done by the node, i.e. the number of calls actually engaged by the node
$i$.


\section{Data analysis}
\label{sec:data_analysis}

\subsection{Activity distribution and the nodes binning}
\label{sec:act_bins}

For the datasets presented in Section \ref{sec:datasets} we first evaluate, for
each node $i$, the total number $u_i$ of events  engaged by the node itself.
For instance $u_i$ is the number of calls made by the node $i$ in the
\emph{MPC} dataset or the number of citations done by $i$ in the \emph{Twitter}
dataset.

\begin{figure}
    \centering
    \includegraphics[width=6.75in]{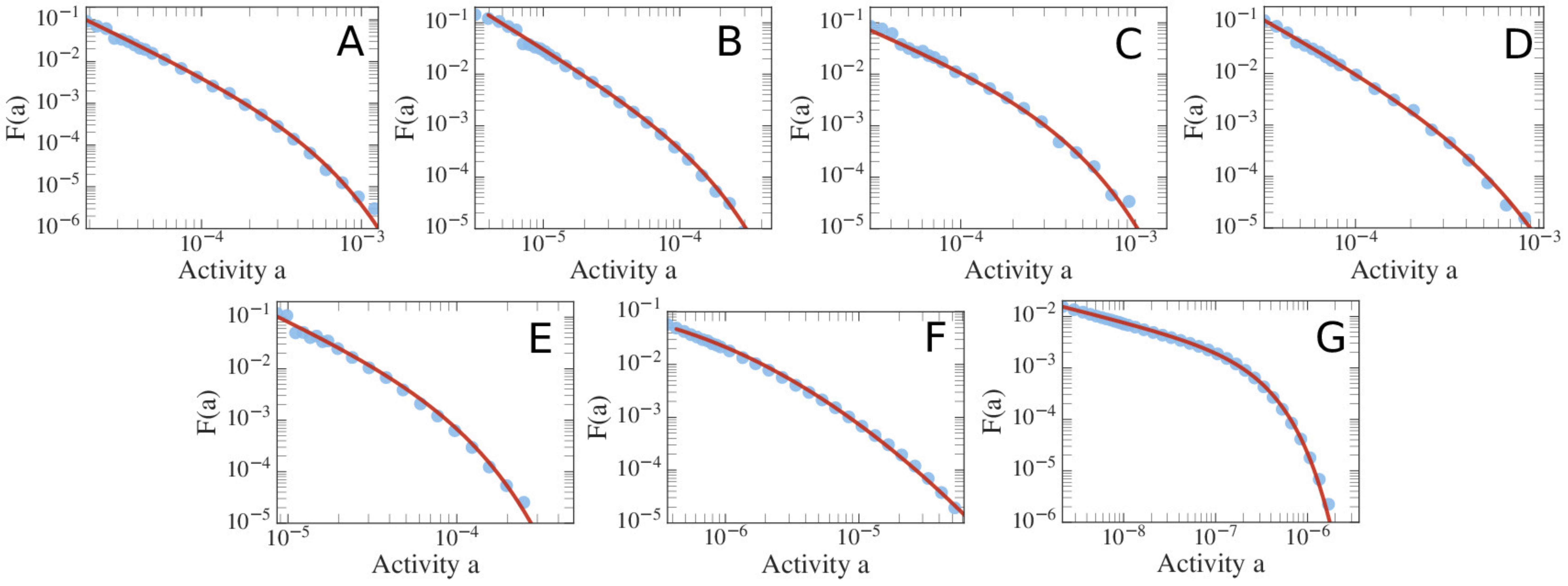}
    \caption{
        \label{fig:F_a} The experimental activity distribution $F(a)$ for (A) PRA, (B) PRB, (C) PRD, (D) PRE, (E) PRL, (F)
        TMN and (G) MPN (blue points).
        We also show the best candidate fit of the $F(a)$ distribution (blue solid lines)
        featuring the functional form and parameters found in Table (\ref{tab:Fa}).
        In all the plots we show the data and fit ranging from the lower bound
        $x_{min}$ to the $99.9\%$ of the measured data, thus excluding from the
        visible area the top $0.1\%$ of the activity values (see Table
        (\ref{tab:Fa}) for the lower bound details).
    }
\end{figure}

We then define the node activity  $a_i$ as the ratio between the $i$-th node's
number of events and the total number of events observed in the dataset, i.e. $a_i = u_i/u_{\rm
tot}$ where $u_{\rm tot} = \sum_j{u_j}$. Thus, $a_i$ falls in the range $a_i \in
[\epsilon, 1.0)$ with $\epsilon = {\rm min}_i(u_i)/u_{\rm tot}$.
We then introduce and compute the activity distribution $F(a)$.
In Fig. \ref{fig:F_a} we show the resulting activity distribution for each analyzed
dataset, while in Table (\ref{tab:Fa}) we show the best candidate functional form for the $F(a)$
distribution of each dataset. The latter is estimated using the methods found in
\cite{clauset2009power}.

\begin{table}
    \centering
    \def\arraystretch{1.15}
    \begin{tabular}{|l|c|c|c|c|c|}
        \hline
        Dataset & Distribution & Parameters  & $KS_d$    & $\%$ &  $\mathcal{L}$ \\
        \hline
        TMN     & Lognormal    & $a_{\rm min}=4.28e-7,\,\mu=-14.02,\,\sigma=1.71$       & $1.5e-2$  & $52$ & $-2.26e+3$  \\
        \hline
        PRA     & Truncated    & $a_{\rm min}=1.90e-5,\,\lambda=3.14e+3,\,\alpha=1.789$& $1.5e-2$  & $32$ & $-301$   \\
        \hline
        PRB     & Truncated    & $a_{\rm min}=6.31e-6,\,\lambda=7.96e+3,\,\alpha=1.638$& $1.4e-2$  & $41$ & $-744$  \\
        \hline
        PRD     & Truncated    & $a_{\rm min}=4.54e-5,\,\lambda=4.02e+3,\,\alpha=1.37$ & $1.6e-2$  & $27$ & $-286$  \\
        \hline
        PRE     & Truncated    & $a_{\rm min}=4.24e-5,\,\lambda=3.32e+3,\,\alpha=1.92$ & $1.5e-2$  & $23$ & $-264$  \\
        \hline
        PRL     & Truncated    & $a_{\rm min}=1.10e-5,\,\lambda=1.55e+4,\,\alpha=1.47$ & $1.7e-2$  & $31$ & $-577$  \\
        \hline
        MPN     & Truncated    & $a_{\rm min}=2.17e-9,\,\lambda=3.82e+6,\,\alpha=0.448$& $9.5e-3$  & $94$  & $-1.6e+4$   \\
        \hline
    \end{tabular}
    \caption{
        \label{tab:Fa} The candidate functional form of the activity
        distribution for each analyzed dataset, the evaluated parameters
        (see Table [$1$] in the main  for the analytical expressions), the
        Kolmgorov-Smirnov distance $KS_d$, the percent $\%$ of nodes in the
        dataset that have activity $a_i\ge a_{\rm min}$ and the normalized
        log-likelihood $\mathcal{L}$ . In the parameters we include $a_{\rm
        min}$ that is the value of the activity that minimizes the $KS$
        distance. This is the lower bound for the functional form behavior,
        i.e. the point at which data behave as the functional form.
    }
\end{table}

In particular, we compare the goodness of fit on the $F(a)$ distribution of the functional forms found in Table [1] of the main
paper, i.e. power-law, truncated power-law, stretched exponential and log-normal distribution.
The procedure for each dataset and each functional form reads as follows:
\begin{itemize}
    \item [-] we fit the $F(a)$ taking into account all the nodes featuring $a_i\ge
        x_{\rm min}$, where $x_{\rm min}$ is the lower bound of the distribution.
        The fit is performed using the maximum likelihood estimators (MLE) that return
        the optimal values of the parameters;
    \item [-] once the optimal parameters are found we compute the
        Kolmogorov-Smirnov distance ($KD_d(x_{\rm min})$) between the analytical and experimental complementary
        cumulative distribution function (CDF);
    \item [-] we then apply this procedure for different $x_{\rm min}$ and set the $a_{\rm min} =
        \min_{x_{\rm min}} KS_d(x_{\rm min})$ lower bound value as the one that minimizes the
        $KS_d$.
\end{itemize}
We then repeat this procedure for all the functional forms of the $F(a)$ and we
then compare them with the \emph{likelihood ratio test} $\mathcal{R}$ combined
with the $p$-value that gives the statistical significance of $\mathcal{R}$
\cite{clauset2009power,pypow.pone.0085777}.
The result of this procedure gives us the best candidate for the $F(a)$ for each
dataset as shown in Table (\ref{tab:Fa}).
We find that a truncated power law is the best candidate for all the
APS datasets together with the MPN one. On the other hand, in the TMN we find a log-normal
distribution as the best candidate for the dataset.

Our datasets provide evidence that nodes within the same activity class (i.e. node with similar
values of activity $a_i$) can feature very different memory behavior. In particular agents with
large activity may connect to very few different nodes (strong reinforcement) or establish new links at
almost every step (weak reinforcement).
For this reasons each node $i$ of the network is naturally classified according to her activity
$a_i$ and her final degree $k_i$, i.e. the total number of different agents that have been
connected to $i$ in the considered time window.

We then define a binning procedure that let us group together the similar nodes, i.e. nodes with
similar activity and final degree. We divide the nodes in $N_{\rm act}$ activity classes so that
within each activity class the most active node performs at most $1.5$ times the events of the least
active node.  Then, with the same procedure, we further group the nodes within each activity class
$a$ according to their final degree, thus defining $N_{\rm deg}(a)$ final degree classes.  The nodes
are therefore divided in $N_b = \sum_{a=1}^{N_{\rm act}} N_{\rm deg}(a)$ activity-degree classes. From
now on, unless differently stated, whenever we mention the nodes' class or bin $b$ we will be
referring to one of these $N_b$ classes.

\subsection{The reinforcement process}
\label{sec:memory}

To measure the reinforcement process of each system, we count all the communication events $e_b(k)$ engaged by every
node $i$ of the $b$-th class when it has degree $k_i = k$. In other words, $e_b(k)$ is the total number of events
engaged by the nodes of the $b$-th class at degree $k$.

Each time an event engaged by a node $i$ of the $b$-th class results in a degree increase $k_i = k\to
k_i = k+1$, we increment the counter $n_b(k)$ by $1$. In other words, $n_b(k)$ is the total number
of events that the nodes belonging to the $b$-th and featuring degree $k$ perform toward a new node.
Of course, if a node $i$ of the $b$-th class with degree $k_i = k$ increases its degree to $k_i = k +
1$ because it gets called by a new node, the $n_b(k)$ counter is not incremented.

The best estimate of the probability for a new node to get establish a new connection at degree $k$ then reads:
\begin{equation}
    f_b(k) = {n_b(k) \over e_b(k)},
    \label{eq:pk_data}
\end{equation}
where $n_b(k)$ and $e_b(k)$ are the event counters as defined above.
We can give an estimate of the uncertainty on $f_b(k)$, by assuming that at a
given degree $k$ the events are independent (i.e. there are no correlations between
users) and by checking that $1\ll n_b(k)\ll e_b(k)$ so that the STD
$\sigma(f_b(k))$ of $f_b(k)$ reads:
\begin{equation}
    \sigma(f_b(k)) = \sigma_b(k) = \sqrt{f_b(k)(1 - f_b(k)) \over e_b(k)}.
    \label{eq:sigma_data}
\end{equation}

We then fit $f_b(k)$ with the proposed reinforcement function $p_{b}(k, \beta)$:
\begin{equation}
    p_b(k, \beta) = \left( 1 + \frac{k}{c(b)} \right)^{-\beta},
    \label{eq:pn_chi2}
\end{equation}
where $c(b)$ is the social propensity of the $b$-th bin, $k$ is the cumulative degree and $\beta$ is
the reinforcement strength, that will be kept fixed for all the nodes in the system.
In particular, for each class $b$ and with a fixed $\beta$, we optimize the parameter $c(b)$, by minimizing
the function $\chi^2_b(\beta)$:
\begin{equation}
    \chi^2_b(\beta) = \sum_{k = 1}^{K_b}{\frac{\left[ f_b(k) - p_b(k,\beta)
    \right]^2}{\sigma_b(k)^2}},
    \label{eq:chi2_single_curve}
\end{equation}
where the index $k$ runs over the $K_b$ points of the $b$-th bin's
curve and $\sigma_b(k)$ is as defined in Eq. (\ref{eq:sigma_data}).
By repeating this procedure for each value of $\beta\in[0, 5.0]$ we find, for
each class $b$, a $\chi^2_b(\beta)$ curve.

\begin{figure}
    \centering
    \subfigure[]
    {\includegraphics[width=3.2in]{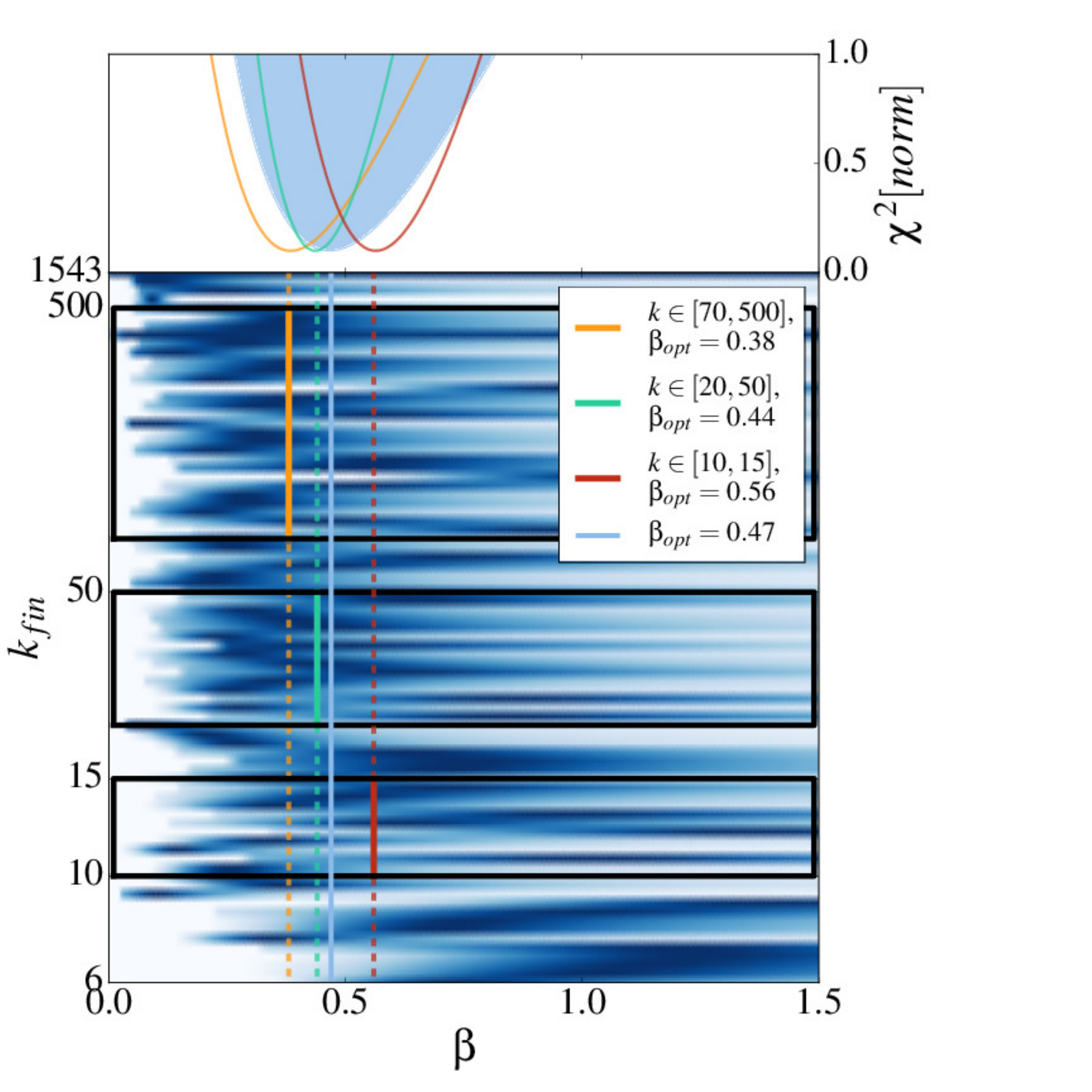}}
    \subfigure[]
    {\includegraphics[width=3.2in]{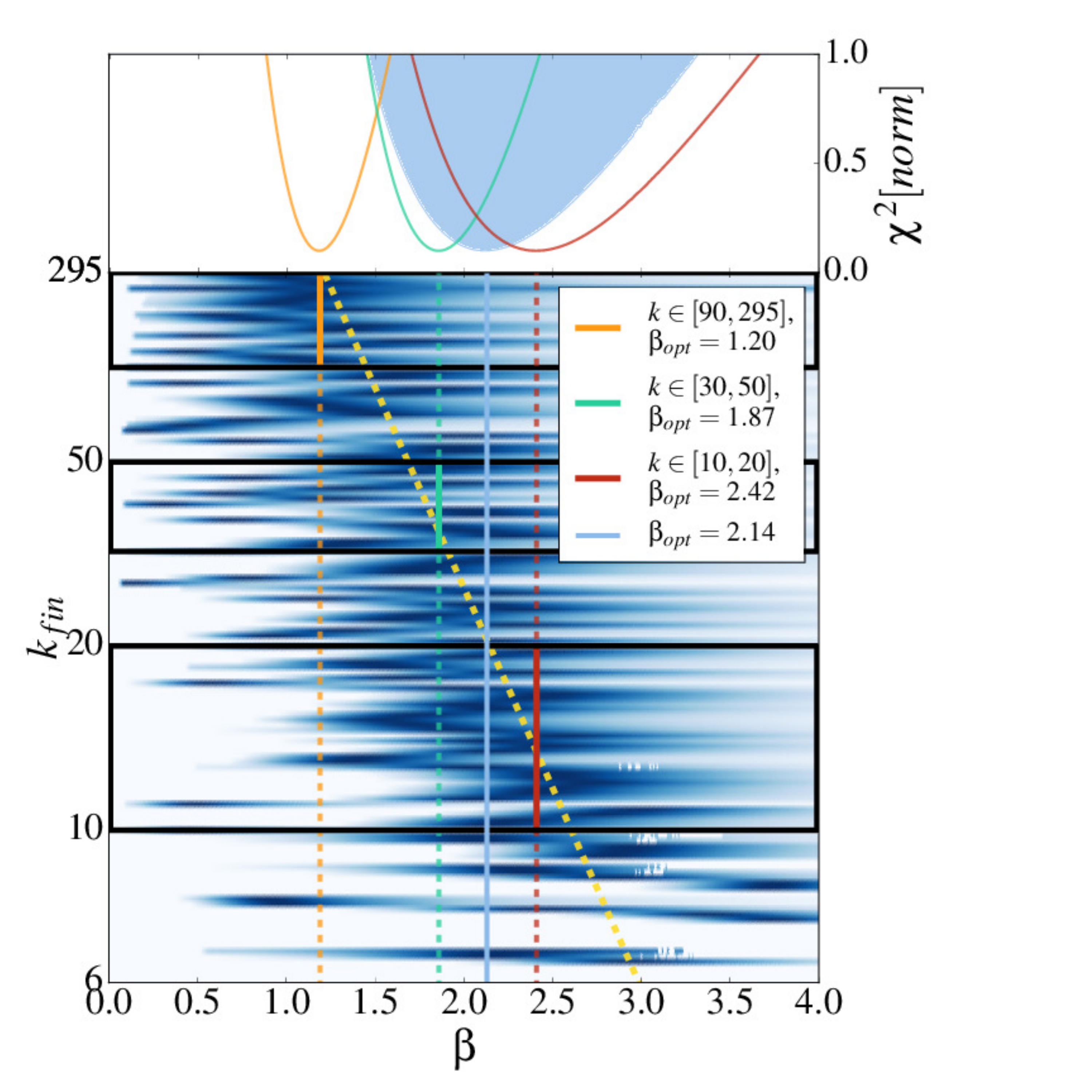}}
    \caption{
        \label{fig:HM}
        The heat-map-like value of $-\ln{[\chi^2_b(\beta)]}$  (bottom plots).
        We plot the exponent $\beta$ on the $x$-axes and the different bins $b$
        sorted by their final degree on the $y$-axes.
        The color-map is proportional to $-\ln{[\chi^2_b(\beta)]}$ representing
        the goodness of fit: the darker, the higher.
        The cyan vertical line is the value of $\beta_{opt}$ defined in Eq.
        (\ref{eq:beta_opt}), while the other vertical lines represent the same
        quantity evaluated in the three black boxes corresponding to different
        final degree intervals.
        (Top plots) The curve $\chi^2(\beta)$ as defined in Eq.
        (\ref{eq:chi2_beta}) (up-filled curve) and the same quantity for the
        three final degree intervals.
        For Twitter (a):
        a single value of $\beta_{opt} = 0.47$ fits most of the curves
        and only some bins $b$ deviate from the average behavior.
        (b) MPC:
        in this case we observe different behaviors depending on the final
        degree.  Thus, a single $\beta_{opt} = 2.14$ does not fit all the
        curves. We also show a ``guide-to-the-eye'' to highlight this feature
        (yellow dashed line).
    }
\end{figure}

In Fig. \ref{fig:HM} we show the behavior of $\chi^2_b(\beta)$.
For each class $b$ we find a minimum of $\chi^2_b(\beta)$ at a
certain $\beta_{opt}(b)$ (see the horizontal lines in the heat-map-like panels of
Fig. \ref{fig:HM}).

Moreover, Fig. \ref{fig:HM} shows that there are two different behaviors.
Specifically, in the TMN case (see Fig. \ref{fig:HM} (a)), one value of
$\beta_{opt} = 0.47$ fits most of the curves, exception made for some
outsiders: the value of $\beta_{\rm opt}(b)$ that maximizes the $1/\chi_b^2(\beta)$ is
practically the same for all the bins.
On the contrary, in the MPC case the maximum of the $1/\chi^2_b(\beta)$ function
follows a diagonal path ranging from a larger $\beta_{\rm opt}(b)$ for bins with lower
final degree to a smaller $\beta_{\rm opt}(b)$ for larger degree bins. In this case a
single $\beta_{\rm opt}$ cannot fit all the curves and we have to consider a
multi-$\beta$ model where each class $b$ features a different optimal value of $\beta$,
$\beta_{\rm opt}(b)$.

\begin{figure}
    \centering
    \subfigure[]
    {\includegraphics[width=2.1in]{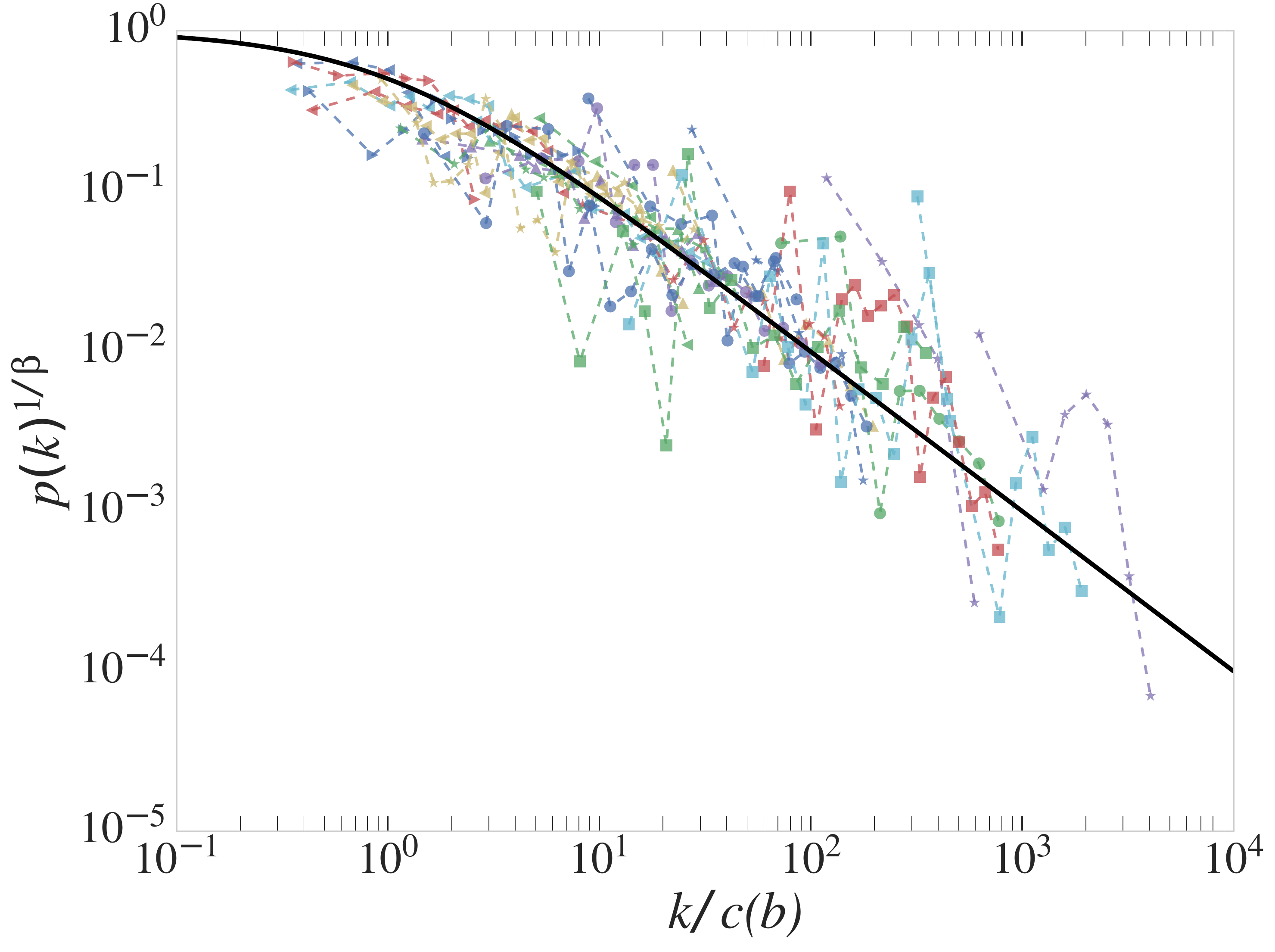}}
    \subfigure[]
    {\includegraphics[width=2.1in]{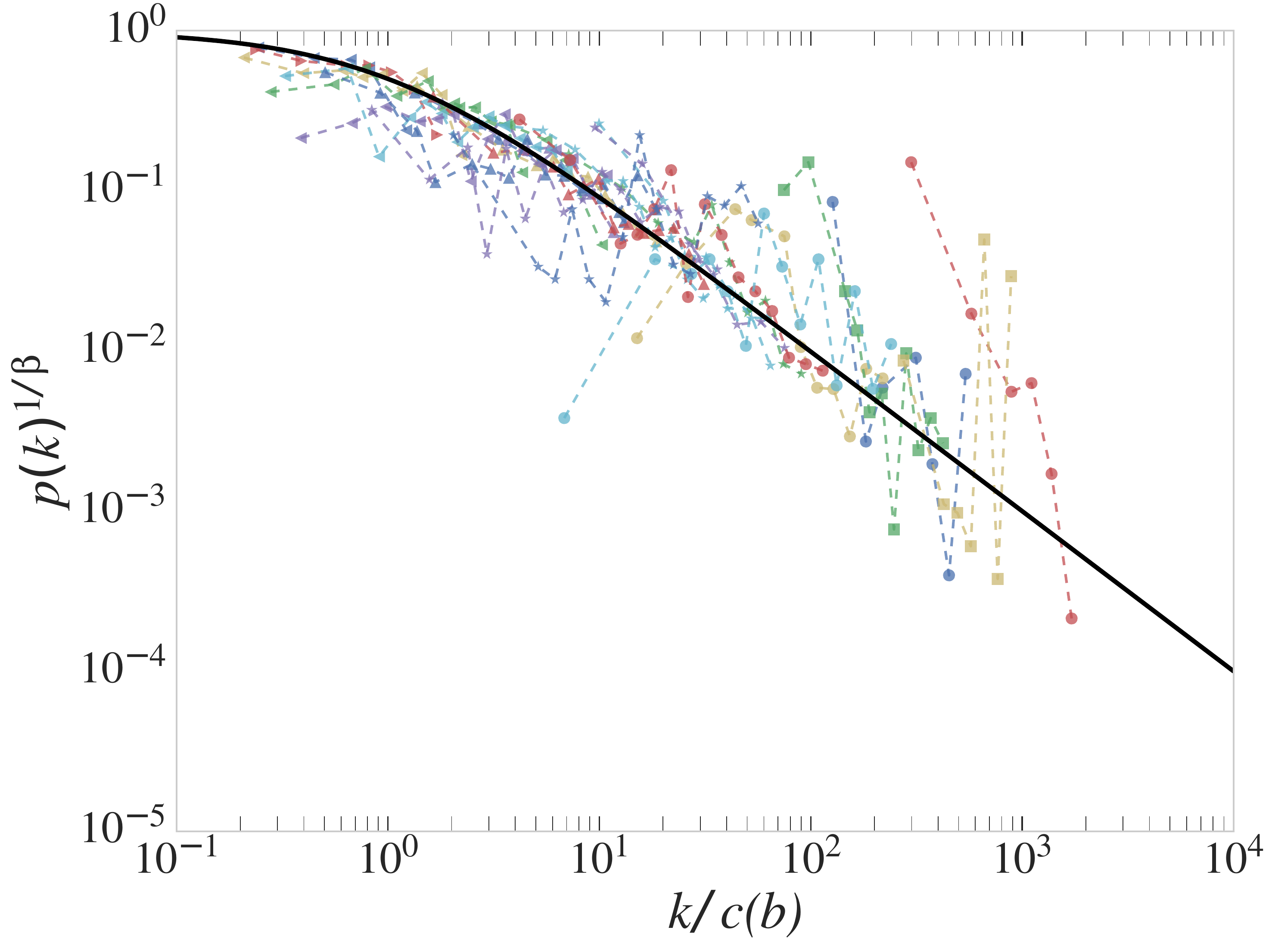}}
    \subfigure[]
    {\includegraphics[width=2.1in]{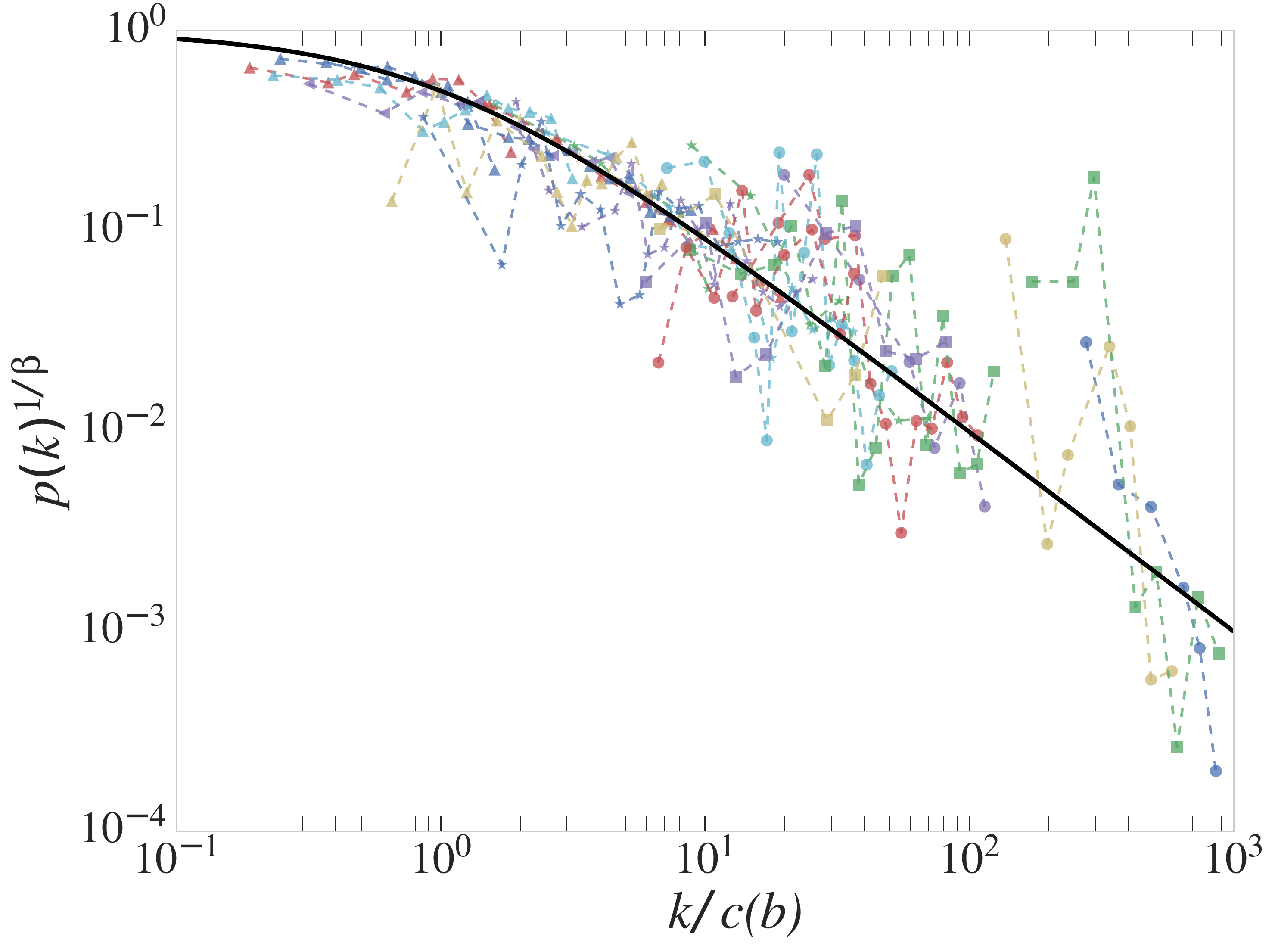}}
    \subfigure[]
    {\includegraphics[width=2.1in]{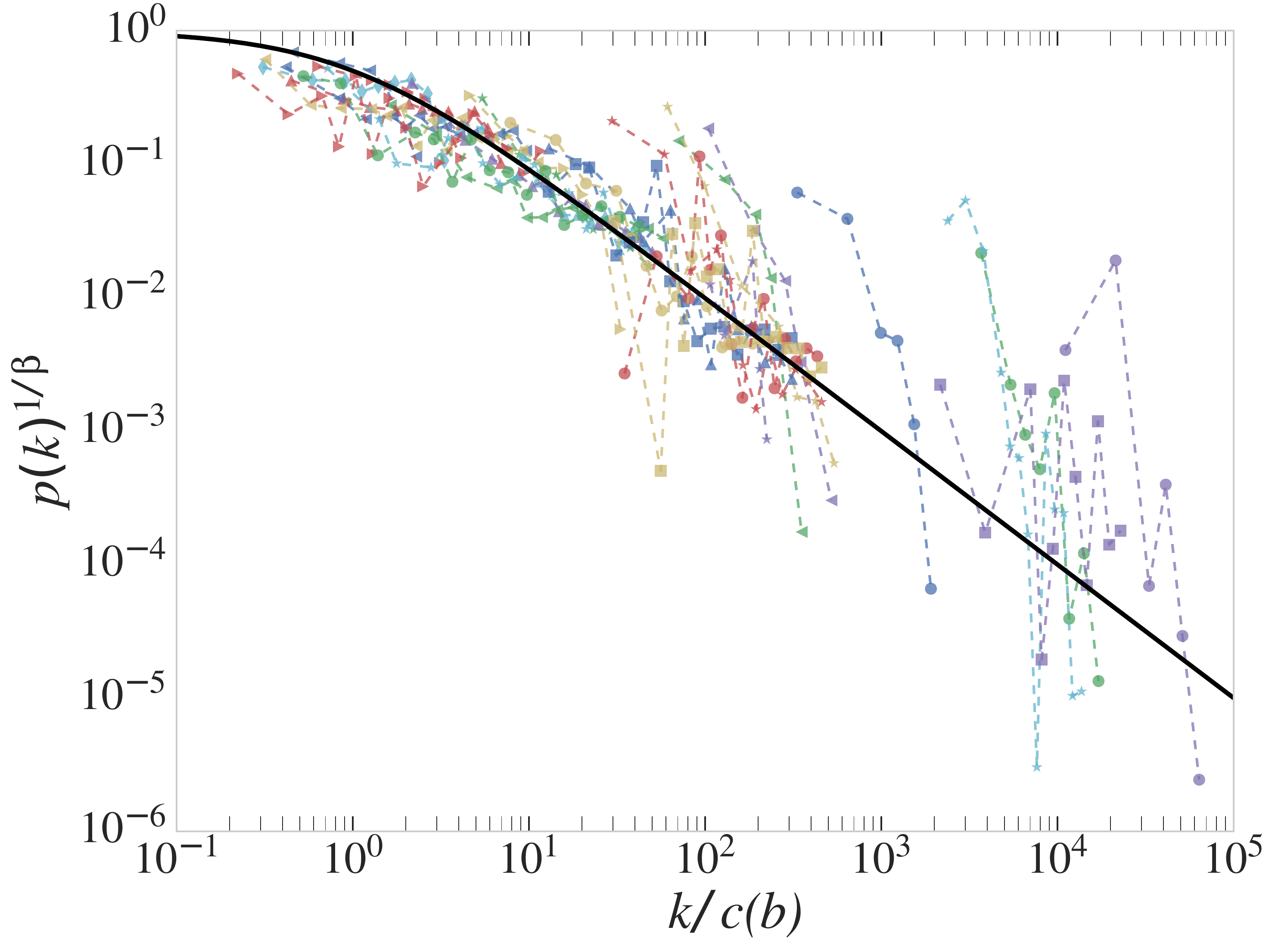}}
    \subfigure[]
    {\includegraphics[width=2.1in]{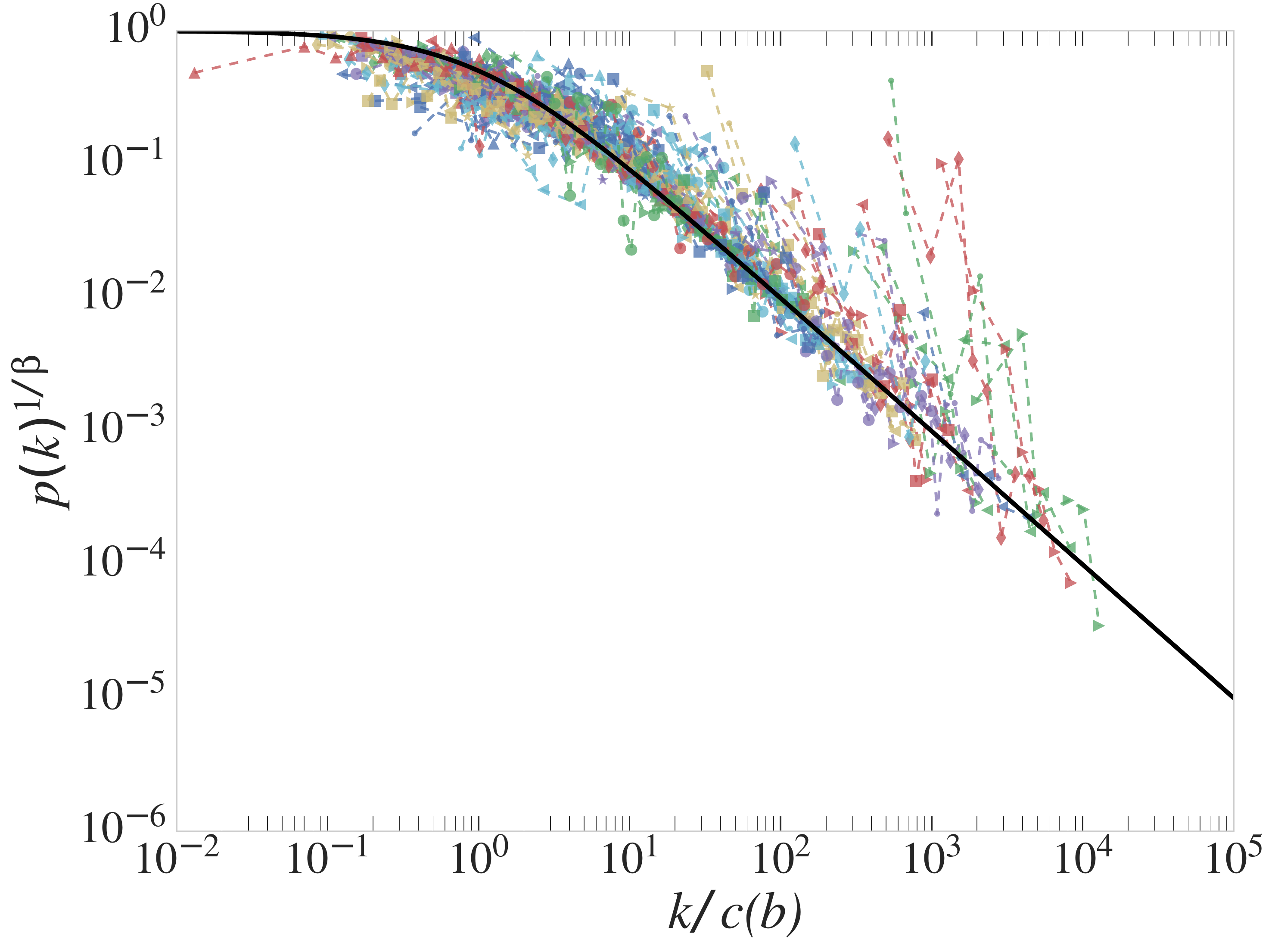}}
    \subfigure[]
    {\includegraphics[width=2.1in]{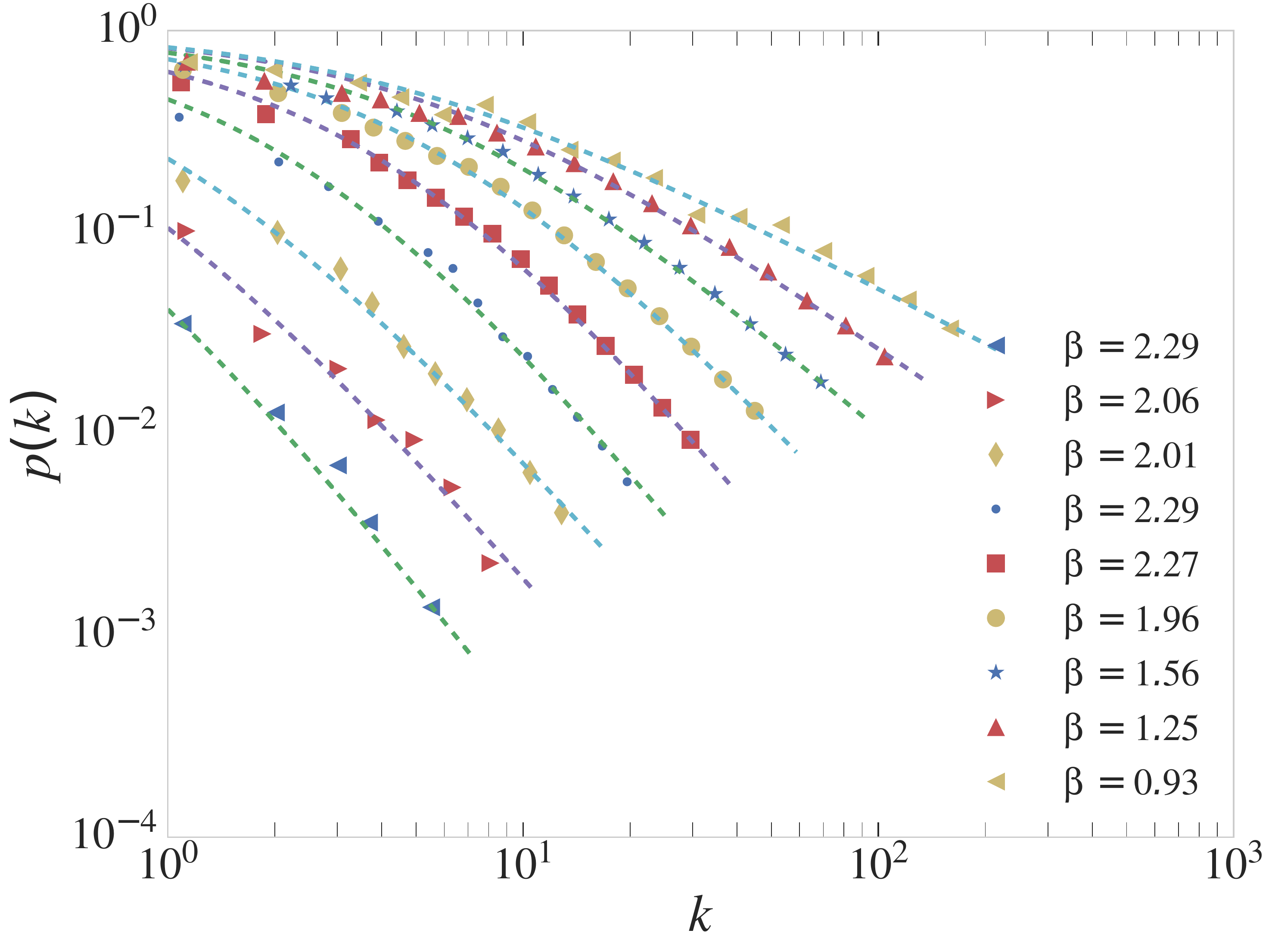}}
    \caption{
        \label{fig:pn}
        Plot of the experimental $p_b(k)$ curves for the (a) PRA, (b) PRD, (c) PRE,
        (d) PRL, (e) TMN and (f) MPC datasets.
        In the (a-e) cases the $k$ is rescaled as $k\to k/c(b)$, where $c(b)$ is the
        constant for the $b$-th class curve at $\beta = \beta_{\rm opt}$. The $p_b(k/c(b))$
        points are then rescaled sending $p_b(k/c(b))^{1/\beta_{\rm opt}}$.
        In the (f) panel for MPC we simply plot $p_b(k)$ as a function of $k$ with no
        rescaling, given that each curve features its own $\beta$ optimal value
        $\beta_{\rm opt}$ as shown in the legend.
    }
\end{figure}

In Fig. \ref{fig:pn} we present the rescaled $p_b(k)$ curves for the
PRA, PRD, PRE, PRL, TMN  and MPC datasets.
In the first five cases we show the rescaled curves obtained by substituting $k\to k/c_b$
and then plotting $p_b(k) \to p_b(k)^{1/\beta_{\rm opt}}$. As one can
see, the curves nicely collapse on the reference curve $(1 + k)^{-1}$.
In the MPC case we show instead the original curves, each one fitted with its own
$\beta_{\rm opt}(b)$. The latter parameter falls in the $1.2 \lesssim \beta_{\rm
opt}(b) \lesssim 3.0$ interval for most of the curves as we also show in Fig. \ref{fig:HM}.

To quantitatively define the $\beta_{\rm opt}$ parameter, let us  define the total mean
square deviation $\chi^2(\beta)$ as
\begin{equation}
    \label{eq:chi2_beta}
    \chi^2(\beta) = \sum_{b = 1}^{N_b}{\left[ \chi^2_b(\beta) \right]},
\end{equation}
where $N_b$ is the total number of curves, i.e. the number of activity-degree
bins $b$.
Then, for the single exponent case, the function $\chi^2(\beta)$ allows to
define $\beta_{\rm opt}$ as:
\begin{equation}
    \label{eq:beta_opt}
    \beta_{\rm opt} = \rm{min}_\beta(\chi^2(\beta)).
\end{equation}
In the multi-$\beta$ case instead, we compute the different values of the
exponent $\beta_{\rm opt}(b)$ found in the system by grouping the memory classes $b$
accordingly to their final degree as
shown in Fig. \ref{fig:HM}. The optimal value of $\beta_{\rm opt}(b)$ is found to be minimum
for the bins featuring a large final degree, i.e. $\beta_{{\rm min}}\equiv\beta_{\rm
opt}\sim 1.2$, which, as we will show in Section \ref{sub:multi_b}, is the exponent
driving the evolution of the network.

\begin{figure}
    \centering
    \subfigure[]{\includegraphics[width=.3\textwidth]{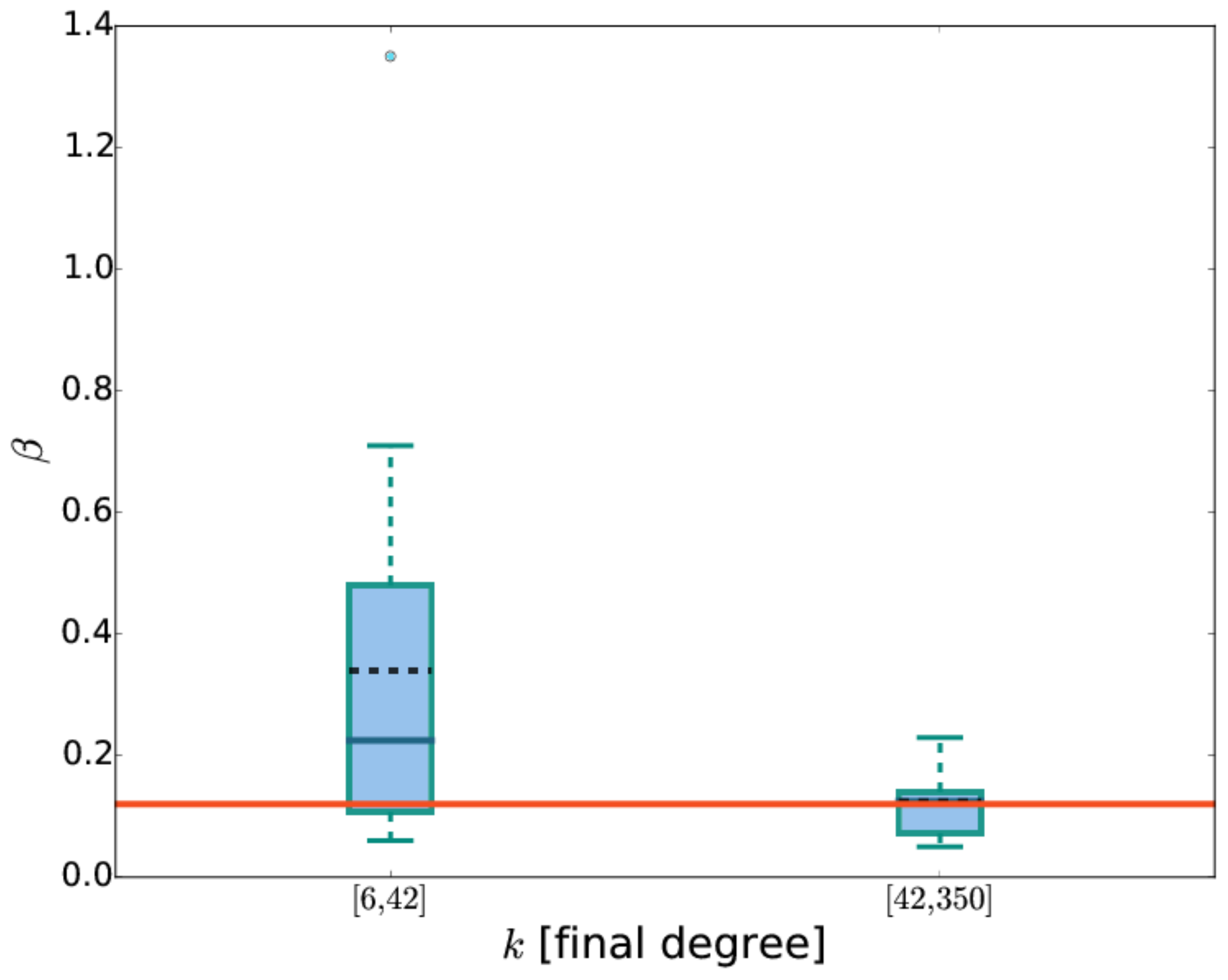}}
    \subfigure[]{\includegraphics[width=.3\textwidth]{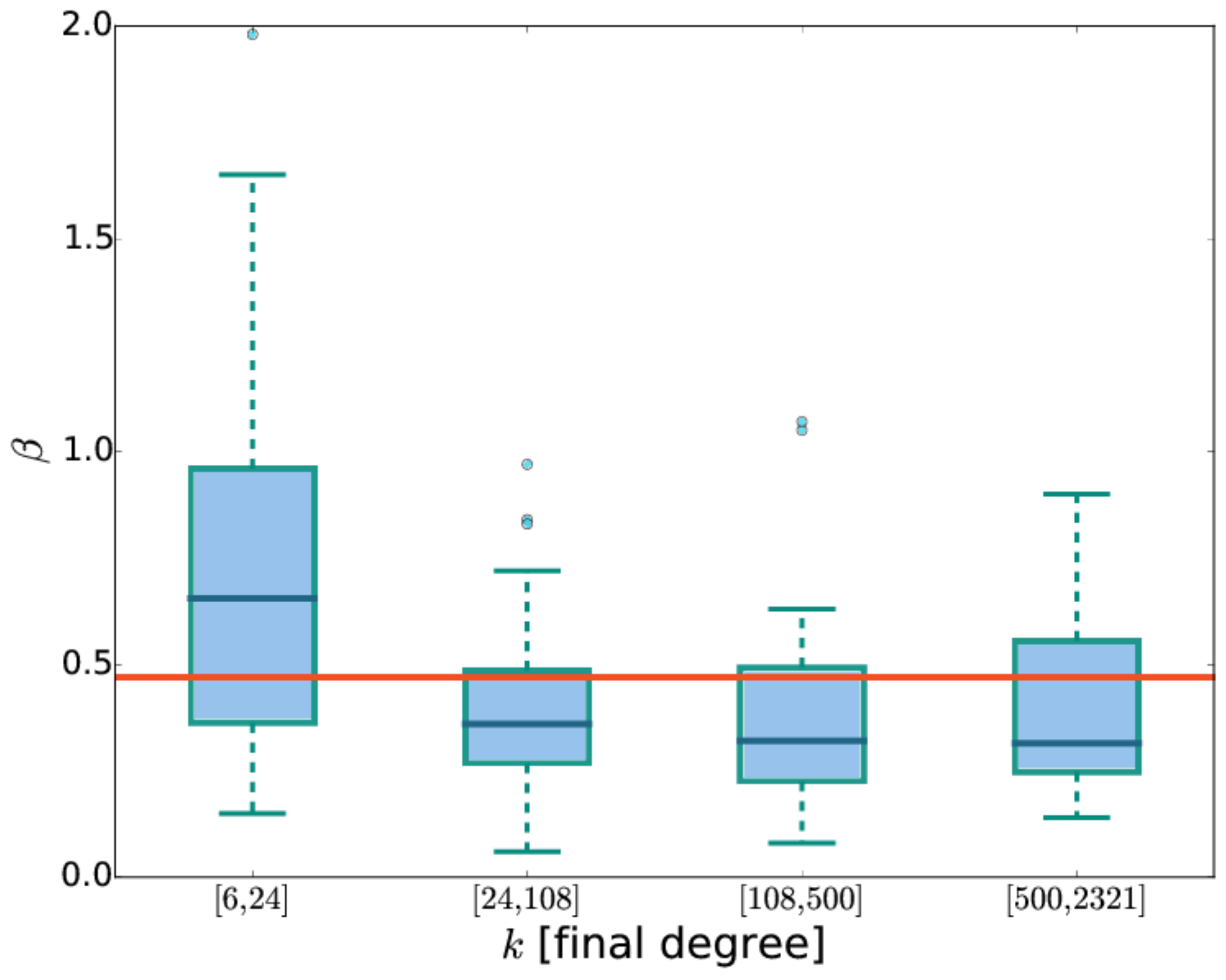}}
    \subfigure[]{\includegraphics[width=.3\textwidth]{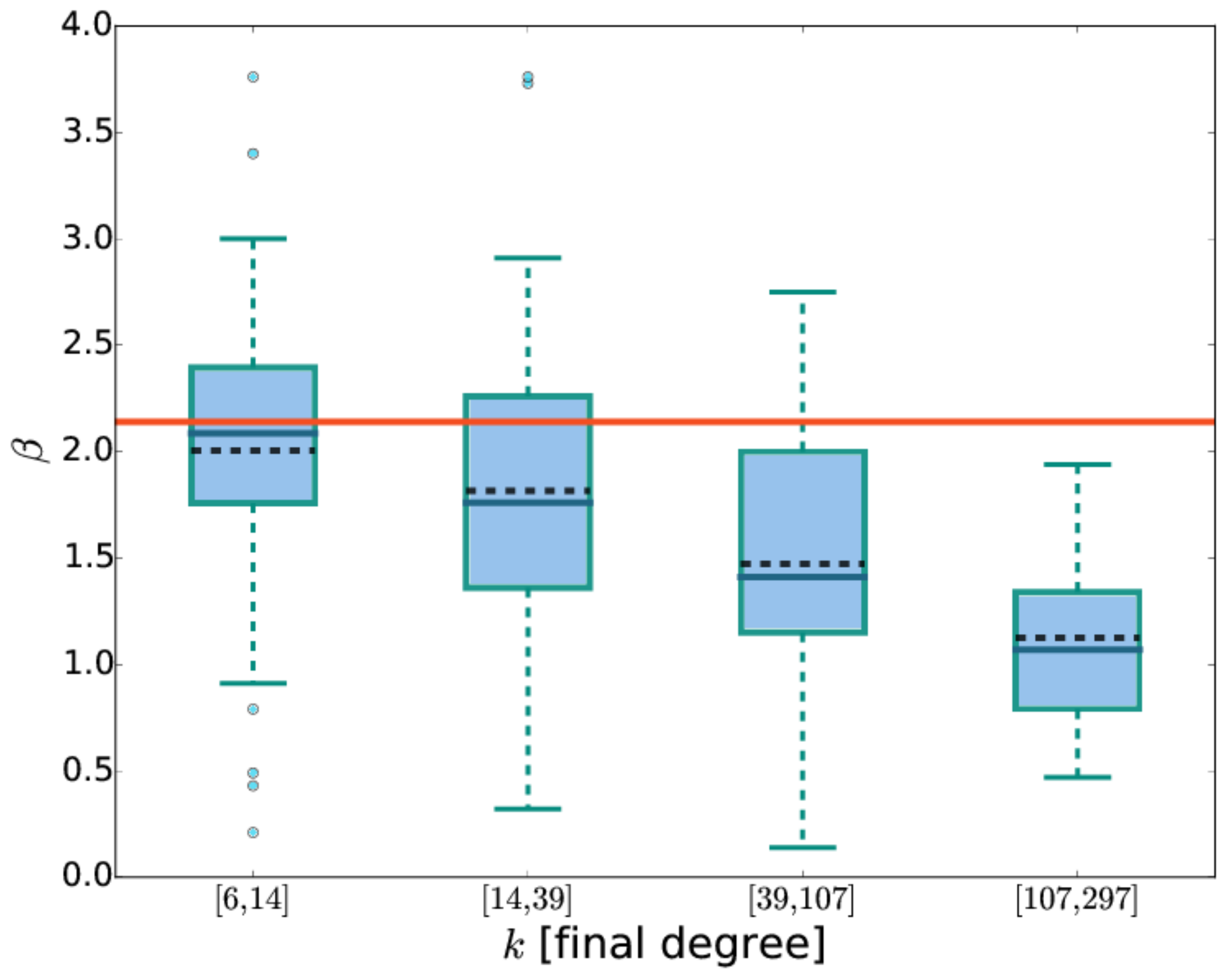}}
    \caption{
        \label{fig:chi2_curve}
        The box plot representing the distribution for different range of nodes classes
        $b$ of the $\beta_{\rm opt}(b)$ for (a) PRB, (b) TMN and (c) MPN. We also show the
        global optimal value $\beta_{\rm opt}$ (horizontal red line) as found in Eq.
        (\ref{eq:beta_opt}).
        The height of the box corresponds to the lower and upper quartile values of the
        distribution and the horizontal solid line corresponds to the distribution’s
        median, while the dashed lines indicates the average value for each range of final
        degree.  The whiskers extend from the box to values that are within $1.5x$ the
        quartile range.
        As one can see, in both the PRB and TMN
        datasets the optimal values $\beta_{\rm opt}$ is compatible with the distribution
        found in all the nodes class ranges (we find the same result for all the other APS
        datasets analyzed). On the other hand, in the MPN the
        distribution of $\beta_{\rm opt}(b)$ lowers as the final degree of the class
        increases. The last group of nodes classes is no more compatible with the overall
        optimal $\beta_{\rm opt}$, being the distribution centered around $\beta_{\rm
        opt}\sim 1.1$, in agreement with our estimation of $\beta_{\rm min} = 1.2$.
    }
\end{figure}

To corroborate the results just outlined, we show in Fig. \ref{fig:chi2_curve} the box
plot of the $\beta_{\rm opt}(b)$ distribution for different groups of nodes classes $b$
grouped by their final degree. We note that the APS and TWT datasets are well approximated
by a single $\beta_{\rm opt}$ as the distribution of $\beta_{\rm opt}(b)$ within each
sub-group of nodes is compatible with the global optimal value $\beta_{\rm opt}$.
On the other hand, in the MPN case we see that the large final-degree classes have their
$\beta_{\rm opt}(b)$ distribution centered around a smaller value of $\beta_{\rm
min}\sim 1.2$. As already anticipated, this value will lead the asymptotic growth of the
system as we will show in Section \ref{sub:multi_b}.

\begin{figure}
    \centering
    \subfigure[]
        {\includegraphics[width=2.8in]{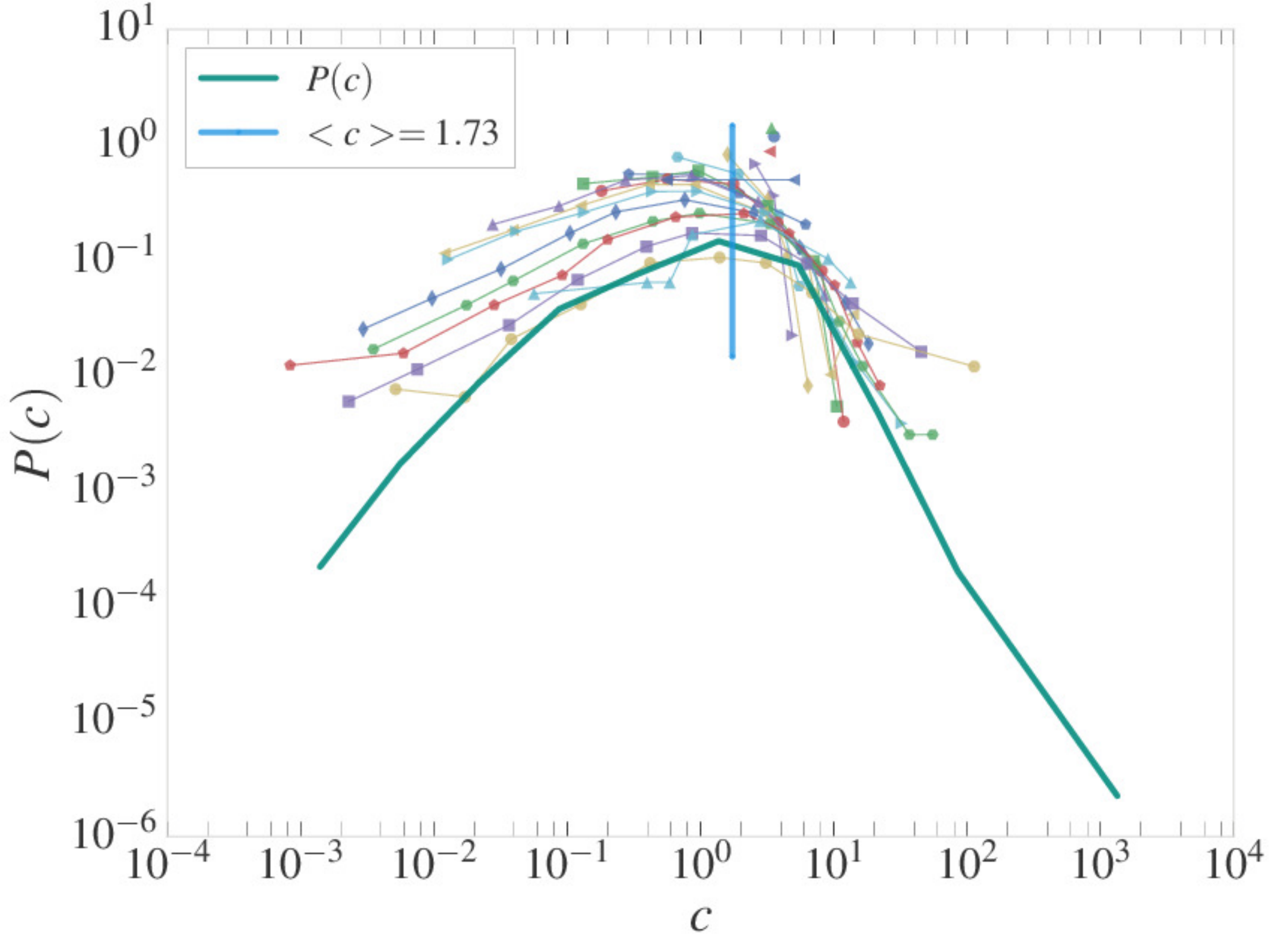}}
    \subfigure[]
        {\includegraphics[width=2.8in]{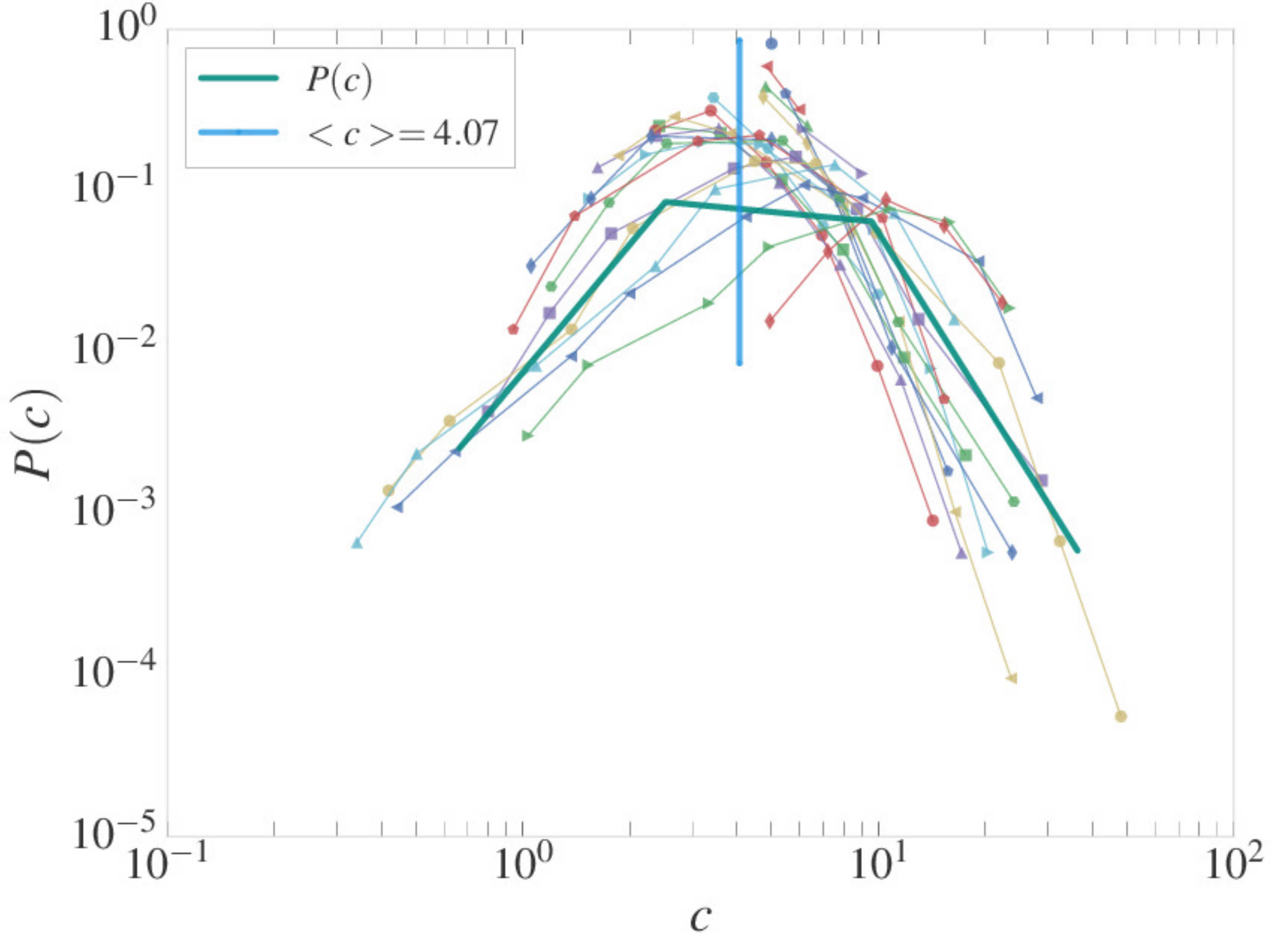}}
    \subfigure[]
        {\includegraphics[width=2.8in]{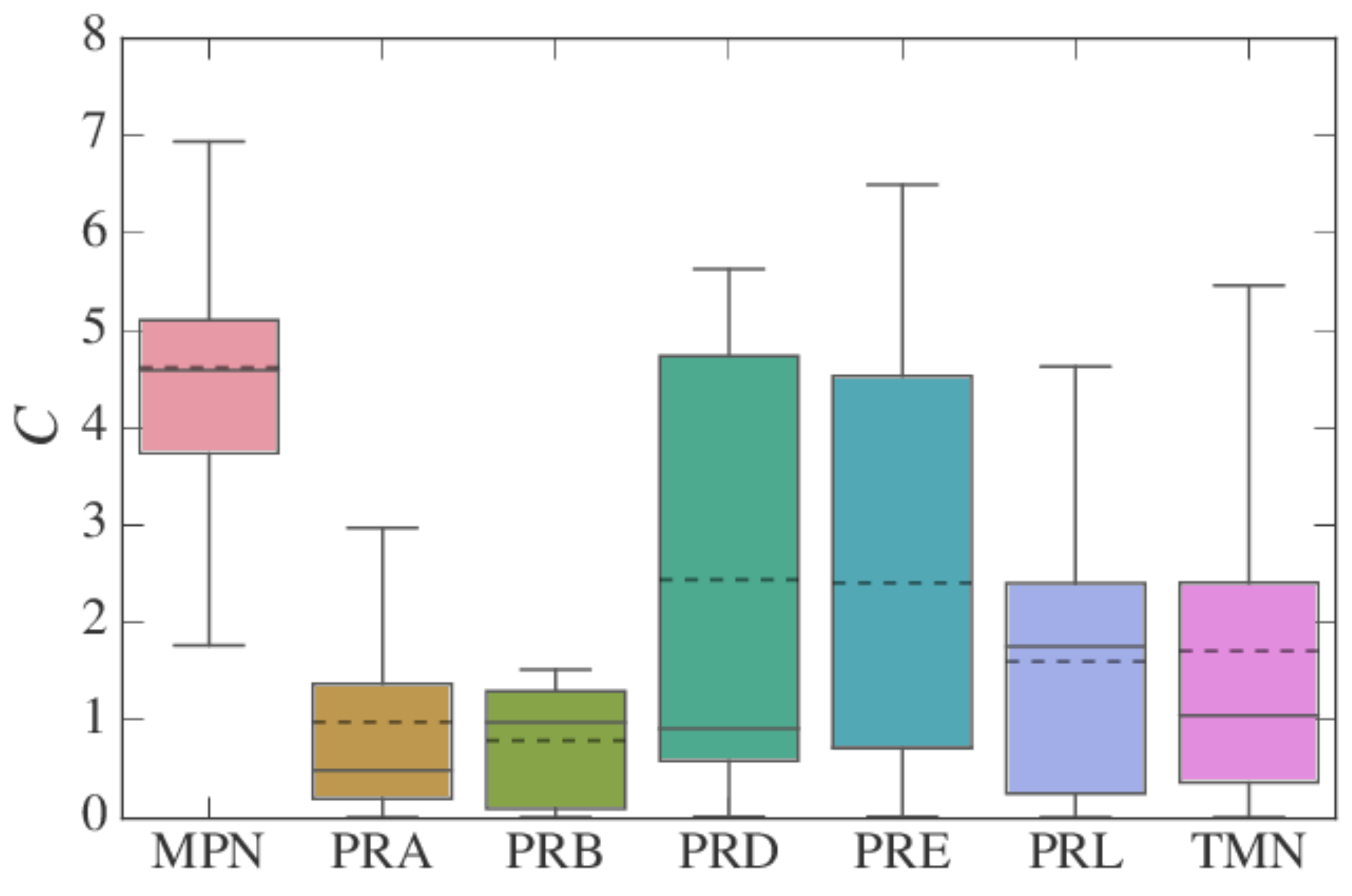}}
    \subfigure[]{
    \begin{minipage}{.45\columnwidth}
        \vspace{-1.8in}
        \centering
        \def\arraystretch{1.1}
        \begin{tabular}{|l|c|c|}
            \hline
            Dataset &   $\beta_{\rm opt}$   &   $\av{c(b)}$ \\
            \hline
            PRA     &   $0.20$              &   $1.5$       \\
            \hline
            PRB     &   $0.13$              &   $0.8$       \\
            \hline
            PRD     &   $0.28$              &   $2.3$       \\
            \hline
            PRE     &   $0.25$              &   $2.4$       \\
            \hline
            PRE     &   $0.15$              &   $1.6$       \\
            \hline
            TMN     &   $0.47$              &   $1.7$       \\
            \hline
            MPN     &   $1.20$              &   $4.6$       \\
            \hline
        \end{tabular}
    \end{minipage}
    }
    \caption{
        \label{fig:PConst} The $P(c)$ distribution of the constant $c(b)$ for the
        (a) TMN, and (b) MPN case (solid green line).
        We also compare the global $P(c)$ distribution with the distribution of the
        $c(b)$ values found within each activity class (solid lines and points):
        we find that the distribution of the $c(b)$ parameter is more or less activity
        independent as most of the distribution of the single activity classes follows the
        same functional form of the total distribution $P(c)$.
        We then report the average value $\av{c}$ of the $c(b)$ constant for each dataset
        (vertical cyan line). The latter reads $1.71$ for TMN and $4.62$ for
        MPN.
        Note that in the single $\beta$ case we evaluate $c(b)$ as the values of $c(b)$
        that best fits the $b$-th $p_b(k, \beta)$ curve fixing $\beta$ at its optimal value ($\beta =
        \beta_{opt}$). On the other hand, in the multi-$\beta$ case we evaluate $c(b)$ as
        the ones that best fits the $p_b(k, \beta(b))$ curve, where the exponent is now
        foxed to the $\beta(b)$ value for the memory class $b$, i.e. to the optimal value
        for the class $b$.
        (c) The box plot showing the global distribution of the constant $c(b)$ in
        all the datasets analyzed.
        The height of the box corresponds to the lower and upper quartile values of the
        distribution and the horizontal solid line corresponds to the distribution’s
        median, while the dashed lines indicates the average value for each range of final
        degree.  The whiskers extend from the box to values that are within $1.5x$ the
        quartile range.
        (d) In this table we report all the values of the reinforcement exponent
        $\beta_{\rm opt}$ and the average reinforcement constant $\av{c(b)}$. For the MPN
        case we report the $\beta_{\rm opt}=\beta_{\rm min}$ and the constant values are
        evaluated for each nodes class $b$ using its optimal value of $\beta$, $\beta_{\rm
        opt}(b)$.
    }
\end{figure}

As a last remark we present in Fig. \ref{fig:PConst} $(a,b)$ the measured distribution
of the constant $c(b)$ for the MPN and TMN datasets.
We show the distribution for all the nodes in the network and for each activity class $a$,
i.e. the group of nodes featuring similar activity.
The values of this constants are distributed but peaked around an average value.
Moreover, the distribution of the $c(b)$ parameter within each activity closely follows
the global one. The distribution of the social attitude $c(b)$ then appears to be a global,
activity independent feature of the nodes in the system.
Finally, in Fig. \ref{fig:PConst} (c) we show how the average value of the $c(b)$
constants, $\av{c}=\av{c(b)}_b$, differs from one dataset to the other varying from
$\av{c}=0.8$ in PRB to $\av{c}=1.7$ in TMN and $\av{c}=4.6$ in the MPN case, respectively.

\section{The model}

\subsection{Activity driven networks with no memory}
\label{sub:model_ML}

The activity driven networks are an effective framework to describe time varying networks.
The simplest memory-less model is defined as follows: the network consists of $N$ nodes
featuring an activity potential $a_i$, i.e. the probability for a node $i$ to get active in a
certain time interval $dt$ reads $a_i dt$.
The evolution rules are: \emph{(i)} at each time step we start with $N$ disconnected
nodes; \emph{(ii)} each node $i$ whether gets active with probability $a_i dt$ or does not
activate with probability $(1 - a_i) dt$. If a node gets active it calls a randomly
selected node $j$ in the network, thus creating an edge $e_{ij}$. \emph{(iii)} At the end
of the time step all the created connection are deleted and we start again from the
initial step \emph{(i)}.\\

These evolution rules define the Master Equation (ME) for $P_i(k,t)$, i.e. the
probability that a node $i$ of activity $a_i$ has degree $k$ at time $t$, where
the degree $k$ is the number of nodes that contacted $i$ up to time $t$. We
also set, without losing generality, $dt = 1$.
The discrete time equation for $P_i(k,t)$ then reads:
\begin{align}
    &P_i(k,t+1) =& \\ \nonumber
    &\qquad\qquad {a_i} \frac{N-k}{N} P_i(k-1,t) +  a_i \frac{k}{N} P_i(k,t)
    + P_i(k-1,t) {\sum_{j\nsim i}} a_j \sum_h  \frac{ P_j(h,t)}{N} + & \nonumber \\
    &\qquad\qquad P_i(k,t) {\sum_{j\nsim i}} a_j \sum_h  P_j(h,t) \frac{N-1}{N}+
    P_i(k,t) {\sum_{j\sim i}} a_j  + P_i(k,t)(1-\sum_j a_j).\label{memoryless}& 
\end{align}
The equation is obtained in the approximation where $a_i\ll 1$, so that between
two consecutive times $t_i = t$ and $t_{i+1} = t + 1$ only one site
can be active.
We will assume that the activity $a_i$ of a node $i$ is small, i.e.
$0<a_i\ll1$, and we wil also consider the approximation $1\ll k \ll N$ i.e. the
integrated number of neighbors of a site is much larger than $1$ but much smaller
than the total number of agents $N$. The first term of the sum represents the
probability that the site $i$ is active and a new link is added to the system.
The second term is the probability that the site $i$ is active but this site
connects to a site that has been already linked. In the third and fourth terms,
the symbol $\sum_{j\nsim i}$ denotes the sum over the sites that are not yet connected
to $i$. In particular, the third term represents the probability that one of
these sites is active and that it connects to $i$. The fourth term is the
probability that one of these sites is active but no link between $j$ and $i$ is
established. The fifth term is the probability that one of the sites already
connected to $i$ is active (being $\sum_{j\sim i}$ the sum over the nodes
already connected to $i$); in this case no new link is added to $i$.
Finally, the last term represents the probability that at time $t$ all the sites
are not active. For $k \ll N$, the second term can be neglected. After some
algebra we obtain the equation:
\begin{equation*}
    P_i(k,t+1)-P_i(k,t) = -\left( P_i(k,t)-P_i(k-1,t) \right)
    \left( a_i +\frac{1}{N} {\sum_{j\nsim i}} a_j \right)
\end{equation*}
given that $\sum_h P_j(h,t)=1$.
For $k \ll N$, we assume that $\frac{1}{N} {\sum_{j\nsim i}} a_j = \langle a \rangle$
i.e.  the average value of the activity. In the limit of large time and large
$k$ we can write a continuous equation in $t$ and $k$ obtaining:
\begin{equation}
    \frac{\partial P_i(k,t)}{\partial t} = ({a_i}+\langle a \rangle) \left(-
    \frac{\partial P_i(k,t)}{\partial k} +  \frac{\partial^2 P_i(k,t)}{\partial k^2} \right).
    \label{eq:partial_ml}
\end{equation}
The solution of Eq. (\ref{eq:partial_ml}) is straightforward:
\begin{equation}
    P_i(k,t)=(2 \pi (a_i+ \langle a\rangle) t)^{-\frac 1 2} \exp
    (- \frac {(k-(a_i+ \langle a\rangle)t)^2}{2t  (a_i+ \langle a\rangle)}).
    \label{eq:Pakt_ml}
\end{equation}
In the large time limit this solution reduces to a delta function: $P(a,k,t)=
\delta (k-(a+\langle a \rangle)t)$ 
Therefore, the average degree $\av{k(a,t)}$ of the nodes of activity $a$ grows as:
\begin{equation}
    \av{k(a,t)} \propto (a+\av{a})t.
    \label{eq:avgk_ML}
\end{equation}
as already found in \cite{Perra:2012uq,PhysRevE.87.062807}.
Moreover the asymptotic degree distribution
$\rho(k)$ of a network with activity distribution $F(a)\propto a^{-\nu}$ is:
\begin{equation}
    \rho(k) \propto k^{-\nu}.
    \label{eq:rhok_ML}
\end{equation}

\subsection{Plugging in the reinforcement process}

The model presented in Sec. \ref{sub:model_ML} is a basic model as it contains no
correlations on an agent's story at all.
In particular, the probability for a node $i$ to re-call an already contacted node is
independent of the node degree.  While simple to describe and solve analytically, this
model is not realistic, as there are no correlations in the each agent's history.
Moreover, the probability to call an already contacted node is always small as $k/N \ll 1$
(and thus the probability to call a new node remains $\sim 1$ even at large degree $k$).
However, as shown in Sec. \ref{sec:memory}, real-world systems features a strong reinforcement
process, as the probability $p_i(k)$ to call a new node at degree $k$ decreases as
the degree $k$ increases.\\

For this reason we introduce an extended version of the model described in
\cite{Karsai:2014aa} et al.  which includes a reinforcement function $p_i(k)$ that measures the
probability for an active node $i$, that has already contacted $k$ different nodes in the
network, to call a new node instead of an already contacted one.

\subsubsection{The single $\beta$ case}
\label{sub:single_b}

As already shown in Sec. \ref{sec:memory} the functional form for the reinforcement
process $p_i(k)$, i.e. the probability of adding a new link for the node $i$ of degree
$k$, reads:
\begin{equation}
    p_i(k) = (1 + k/c_i)^{-\beta}.
    \label{eq:pni}
\end{equation}

By plugging Eq. (\ref{eq:pni}) into Eq. \eqref{memoryless} for node $i$, we get:
\begin{align}
    \label{eq:ME_single_b}
    P_i(k,t+1) = P_i(k-1,t) \bigg[ a_i p_i(k-1) + \sum_{j\nsim
    i}{a_j} \sum_{h}{p_j(h) \over (N - h)} P_j(h,t) \bigg]+ \\ \nonumber
    P_i(k,t) \bigg[ a_i[1-p_i(k)] + \sum_{j\nsim i}{a_j \sum_{h}
    \bigg( 1 - {p_j(h) \over {N - h}} P_j(h, t)\bigg)}\bigg]
    +\\\nonumber
    P_i(k,t) \bigg[ 1 - \sum_{j}{a_j}\bigg],
\end{align}
where $N$ is the number of nodes in the network, $\sum_{i\nsim j}$ is the sum
over the nodes not yet connected to $i$ and $\sum_{j}$ is the sum over all the
$N$ nodes of the network.
Each term of Eq. (\ref{eq:ME_single_b}) corresponds to a particular event that
may take place in the system, as already presented in the paper.
For instance, the first term of the l.h.s. of Eq. (\ref{eq:ME_single_b}) takes
into account the increment of the node $i$'s degree from $k - 1$ to $k$.
This may happen whether because node $i$ gets active and contacts a new node in
the system with probability $a_i p_i(k - 1)$ or because a node $j$ never contacted
before gets active and calls exactly node $i$ with probability $a_j
p_j(h)/(N - h)$, being $h$ the degree of $j$.
In the same way, the second line takes into account that node $i$ does not
change degree $k$ whether because it calls an already contacted node or
because the non contacted nodes call other nodes in the network.
The last line of Eq. (\ref{eq:ME_single_b}) considers the possibility that no
node in the network gets active.

If we now substitute Eq. (\ref{eq:pni}) in Eq. (\ref{eq:ME_single_b}), after
some algebra we get:
\begin{align}
    P_i(k,t+1) - P_i(k,t) = \frac{a_i c_i^\beta}{(k-1+c_i)^\beta}
    P_i(k-1,t) - \frac{a_i c_i^\beta }{(k+c_i)^\beta} P_i(k,t)
    \nonumber \\
    -\left( P_i(k,t) - P_i(k-1,t) \right) {\sum_{j\nsim i}} a_j
    \sum_{h} \frac{P_j(h,t) c_j^\beta}{(N-h)(h+c_j)^\beta}.
\end{align}
Then, by applying the same approximations of large degree $k$ and time $t$
we obtain the continuous equation:
\begin{align}
    \label{eq:diff_single_b}
    \frac{\partial P_i(k,t)}{\partial t}  = & - a
    \frac{c_i^\beta}{k^\beta} \frac{\partial P_i(k,t)}{\partial k} +
    \frac{a_i c_i^\beta}{2 k^\beta} \frac{\partial^2 P_i(k,t)}{\partial
    k^2} + \frac{a_i \beta c_i^\beta}{k^{\beta+1}} P_i(k,t)+ &
    \\\nonumber
    & \bigg( \frac{1}{2} \frac{\partial^2 P_i(k,t)}{\partial k^2}-
    \frac{\partial P_i(k,t)}{\partial k} \bigg) \int da_j F(a_j) a_j
    \int dc_j \rho(c_j | a_j) \int dh \frac{c_j^\beta
    }{h^\beta}P_j(h,t),&
\end{align}
where $\rho(c_j|a_j)$ is the probability for a node $j$ of activity $a_j$ to
have reinforcement constant $c_j$.\\

The long time asymptotic solution of Eq. (\ref{eq:diff_single_b}) is of the
form:
\begin{equation}
    P_i(k,t) \propto \exp{\Bigg[ - A{(k - C(a_i,c_i)t^{1 \over 1+\beta})^2 \over
    t^{1/(1 + \beta)}} \Bigg]},
    \label{eq:sol_single_b}
\end{equation}
Moreover, $C(a,c)$ is a constant depending on the activity $a$ and the reinforcement
constant $c$ that follows the:
\begin{equation}
    {C(a,c) \over 1 + \beta} = {ac^\beta \over C(a,c)^\beta} + \int{da'
        F(a')\int{dc' \rho(c',a') {a'c'^\beta \over C(a',c')^\beta}}}.
    \label{eq:CA}
\end{equation}
We do not have an exact solution for $C(a,c)$, however $C(a,c) \simeq (a
c^\beta)^{1/(1+\beta)}$ for large $a$.

Let us note that Eq. (\ref{eq:sol_single_b}) can be obtained setting the variable $x = k-C(a)
t^{\frac{1}{1+\beta}}$ and substituting it in Eq. (\ref{eq:diff_single_b}) and
imposing that $|x| \ll t^{\frac 1 {1+\beta}}$ from Eq. (\ref
{eq:diff_single_b}):
\begin{align}
    \frac{\partial P_i(x,t)}{\partial t} = & \frac{a_i \beta
        c_i^{\beta}}{C(a_i,c_i)^{1+\beta} t} \left( x \frac{\partial P_i(x,t)}{\partial x}+
        P_i(x,t)\right) + \frac{C(a_i,c_i)}{2 (1+\beta) t^{\frac\beta{1+\beta}}}
        \frac{\partial^2 P_i(x,t)}{\partial x^2}& \\\nonumber
        ~& - \frac{\partial P_i(x,t)}{\partial x} \int
        da_j F(a_j) \int dc_j \rho(c_j|a_j)
        \int dy \frac{ a_j \beta c_j^{\beta}}{C(a_j,c_j)^{1+\beta} t} P_j(y,t) y. &
\end{align}
The solution of the latter equation is of the form
\begin{eqnarray}
P_i(x,t)\approx {t^{- \frac1{2(1+\beta)}}}\exp\left( -\frac {Ax^2}{t^{1/(1+\beta)}} \right)
\end{eqnarray}
thus confirming that $x$ can be considered much smaller than $t^{1 \over
1+\beta}$.\\

An important consequence of equations (\ref{eq:sol_single_b}) and (\ref{eq:CA}) is that,
for a system featuring a reinforcement strength $\beta$, the average degree of the nodes 
belonging to a class $b$ of activity $a$ and constant $c$ grows as:
\begin{equation}
    \av{k(a,c,t)} \propto C(a,c)\cdot t^{\frac 1 {1+\beta}}.
    \label{eq:avgk_single_b}
\end{equation}
In particular, $\av{k(a,c,t)}\propto (at)^{1\over 1+\beta}$ for large values of the activity
$a$.

As expected, the average degree grows slower than in the memoryless case ($\beta
= 0$) where the average degree grows linearly in time, as found in Eq.
(\ref{eq:avgk_ML}).
Moreover, the presence of a reinforcement process also affects the asymptotic behavior of
$\rho(k)$.
Indeed, as already shown in the main paper, Eq. (\ref{eq:avgk_single_b}) gives us the
relation between the degree $k$ and the activity $a$ at a given time $t$, as $k \propto
a^{1\over 1+\beta}$.
Thus, given an activity distribution $F(a)$, we can infer the functional form of the
degree distribution $\rho(k)$ by substituting $a\to k^{1\over 1+\beta}$, finding:
\begin{align}
    \rho(k) dk \propto   F(k^{(1+\beta)}) k^{\beta} dk.
    \label{rhok}
\end{align}
Specifically, by supposing a power-law activity distribution $F(a)\propto a^{-\nu}$
and considering that the degree distribution for a class $b$ is described by Eq.
(\ref{eq:sol_single_b}), we obtain
\begin{equation}
    \rho(k) \propto k^{-[(1 + \beta)\nu - \beta]}.
    \label{eq:rhok_single_b}
\end{equation}
where we integrated over time $t$ and reinforcement constant $c(b)$ and we considered the
asymptotic regime of large time and activity.

\subsubsection{Numerical results}
\label{sub:num_fixb}

We performed numerical simulations to check the result of Section \ref{sub:single_b}.
We fix the following parameters:
\begin{itemize}
    \item [-] $N=10^6$ nodes;
    \item [-] activity $a\in[\epsilon, 1.0]$ with $\epsilon=10^{-3}$, power -law
        distributed so that $F(a)\propto a^{-\nu}$ with $\nu=2.1$;
    \item [-] single value of the reinforcement exponent $\beta = \left\{0.5, 1.0,
        1.5, 2.0\right\}$ and a fixed $c=1$ for all the nodes;
    \item [-] $T=10^5$ evolution steps.
\end{itemize}

We start with no edge in the system and we draw for each node
the activity $a_i$ from the distribution $F(a)$.
At each step a randomly chosen node gets active with probability $a_i$.
An active node then connects with probability $p_i(k)$ with a randomly
chosen node which have not been yet connected to $i$ or, with probability $1-p_i(k)$, the
node calls an already contacted node and no new connection is added to the system. An
evolution step corresponds to $N$ of these elementary steps, i.e. for each evolution step
we give, on average, the possibility to make a call to every node in the network.

\begin{figure}
   \centering
   \subfigure[]
       {\includegraphics[width=3.2in]{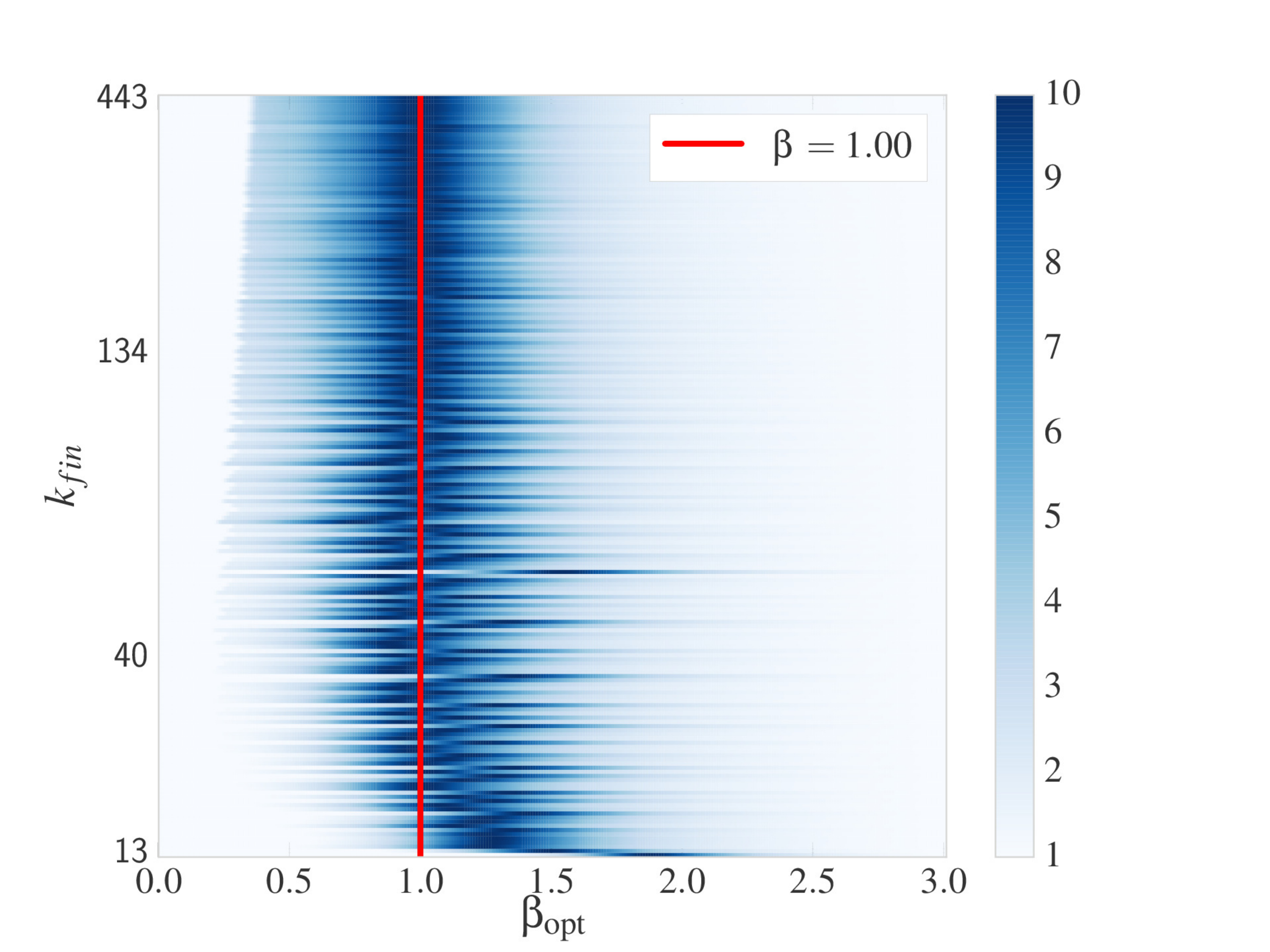}}
   \subfigure[]
       {\includegraphics[width=3.2in]{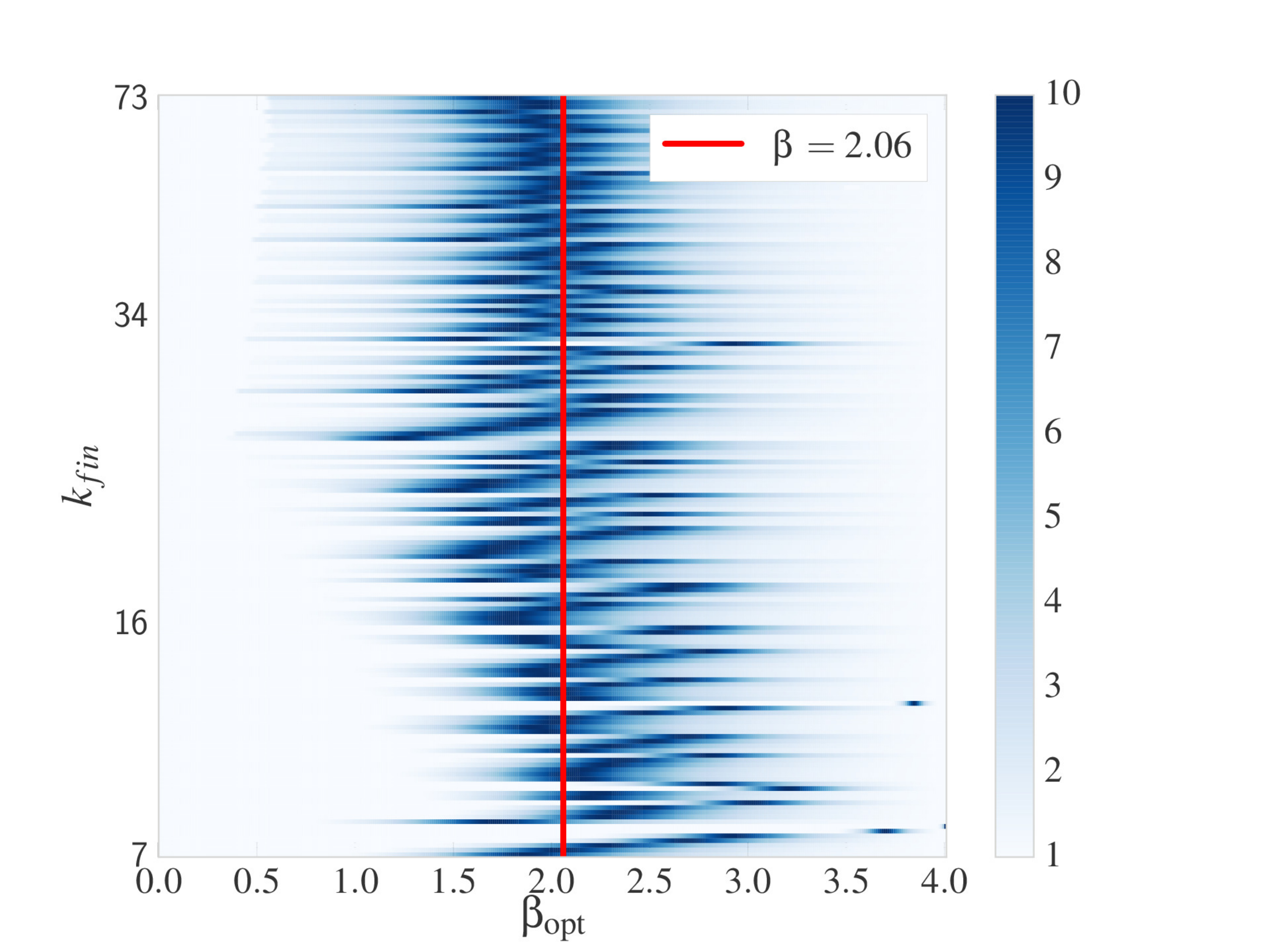}}
   \caption{
       \label{fig:recover_beta} The heat-map of $-\ln(\chi^2_b(\beta))$
       obtained from the simulation in the same way as in Figure \ref{fig:HM}
       from real data.  As one can see the recovered $\beta_{\rm opt}$ is in
       excellent agreement with the value used in the simulation:
       $1.0$ in the (a) panel and $2.06$ in the in the (b) panel.
   }
\end{figure}

The results are in excellent agreement with the analytical predictions.
First, in Fig. \ref{fig:recover_beta} we show that the analysis presented in
Section \ref{sec:act_bins} correctly recovers the reinforcement exponent $\beta_{\rm opt}$.
Indeed the minimum of the $\chi^2_b(\beta)$ are vertically aligned with the value of
$\beta$ fixed in the simulations.

\begin{figure}
    \centering
    \subfigure[]
    {\includegraphics[width=2.99in]{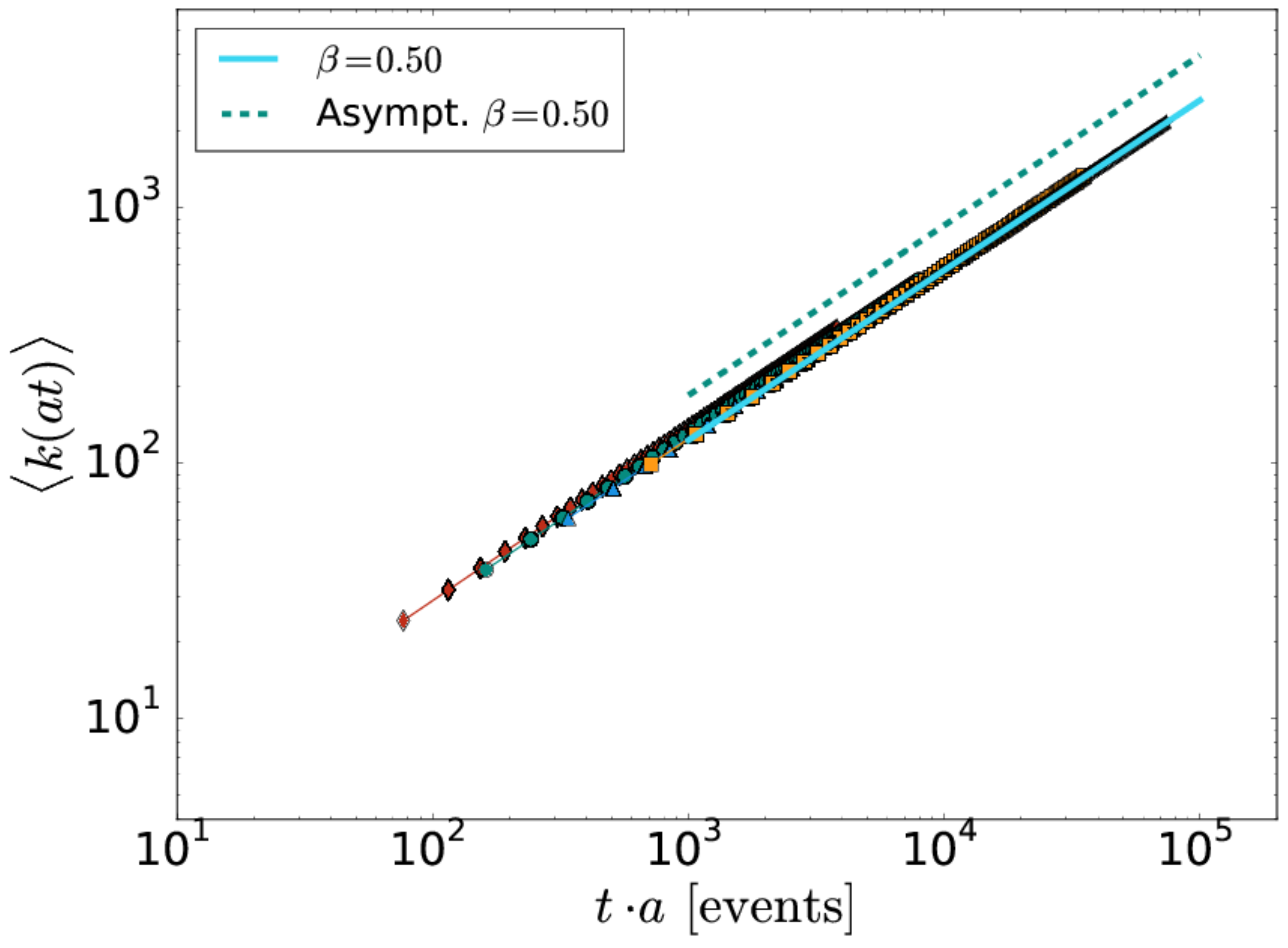}}
    \subfigure[]
    {\includegraphics[width=2.99in]{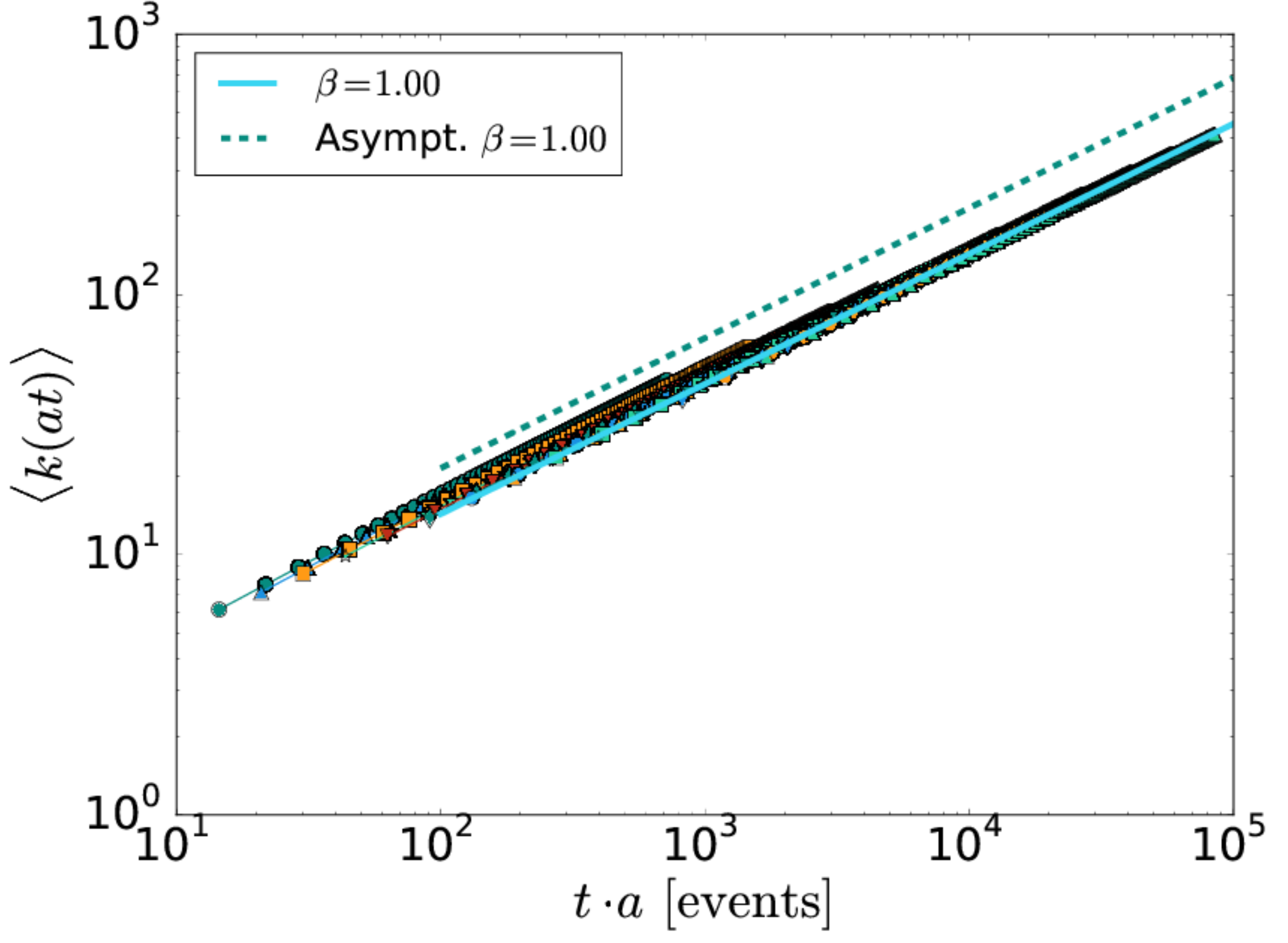}}
    \subfigure[]
    {\includegraphics[width=2.99in]{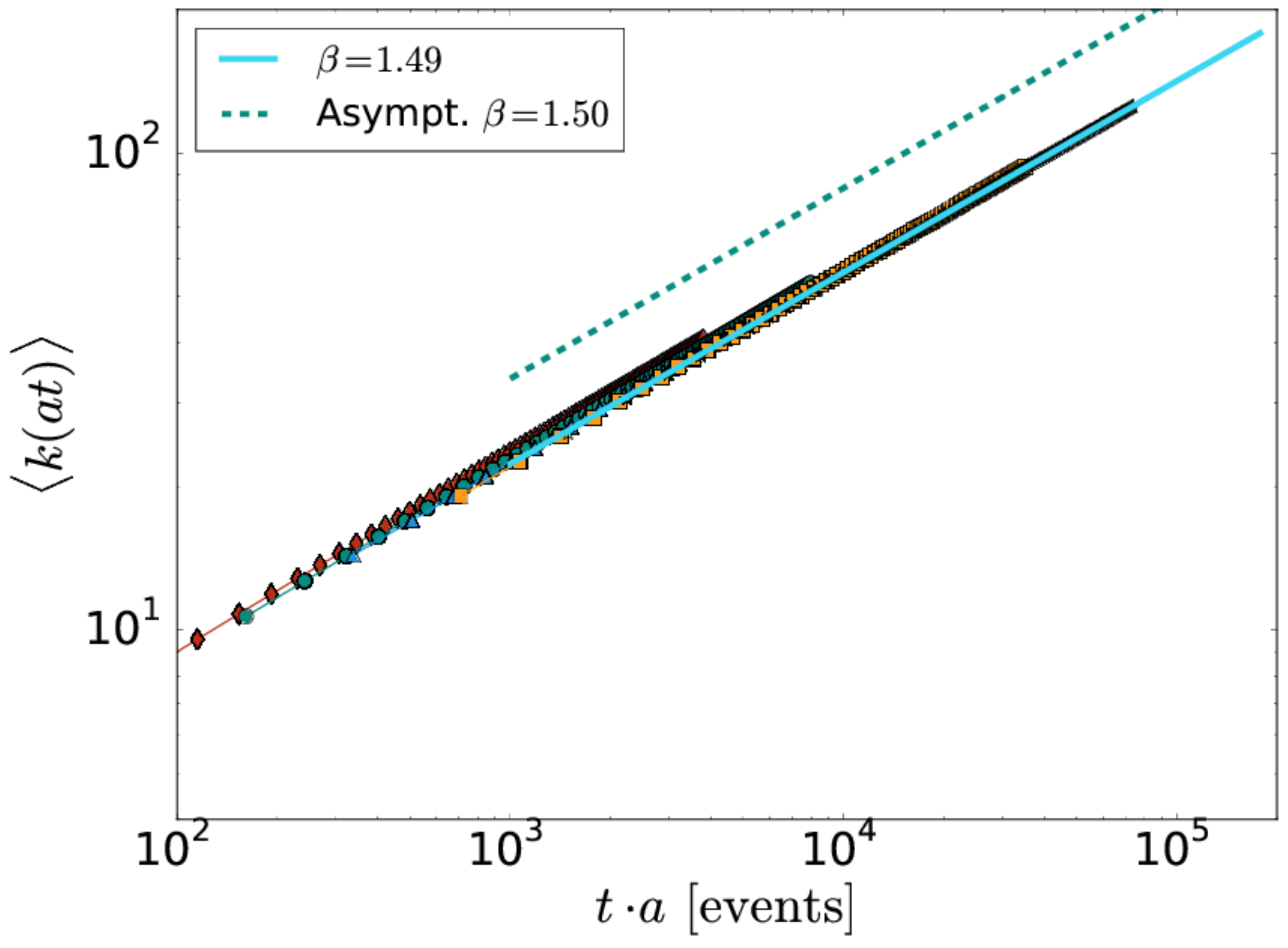}}
    \subfigure[]
    {\includegraphics[width=2.99in]{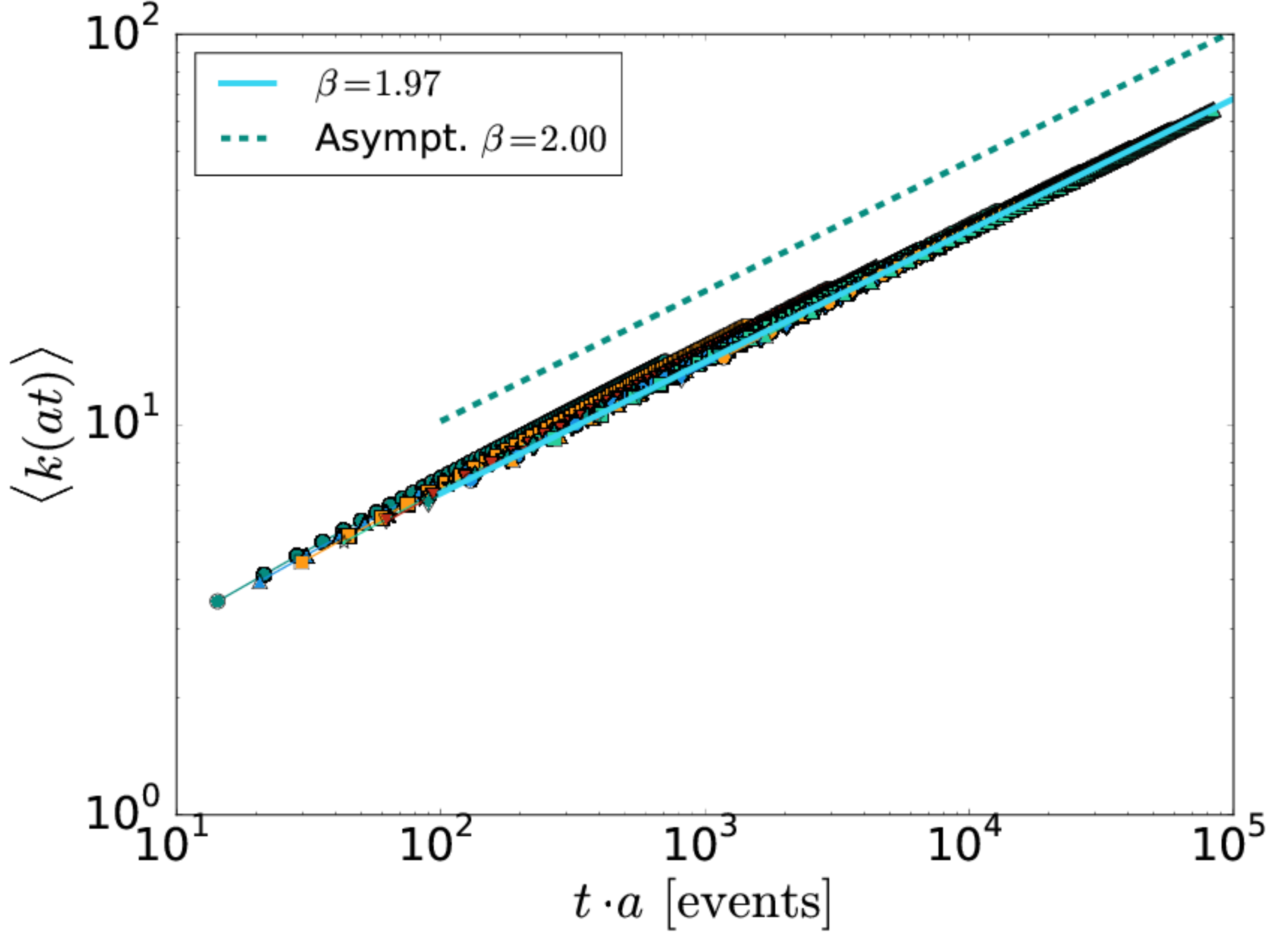}}
    \caption{
        \label{fig:kat_fixb} The average degree $\av{k(at)}$ for different
        activity classes in the $\beta=0.5$ (a), $\beta=1.0$ (b), $\beta=1.5$ (c) and
        $\beta=2.0$ (d) case.
        The time is rescaled with activity $t\to at$, so that all the curves
        collapse on a single behavior. We also fit $\av{k(at)} \propto
        (t/A)^{1\over 1+\beta^*}$ (cyan solid line) and compare the simulation
        with the analytical result $\av{k(at)} = A\cdot t^{1\over 1+\beta}$
        (blue dashed line).
    }
\end{figure}

\begin{figure}
    \centering
    \subfigure[]
        {\includegraphics[width=2.8in]{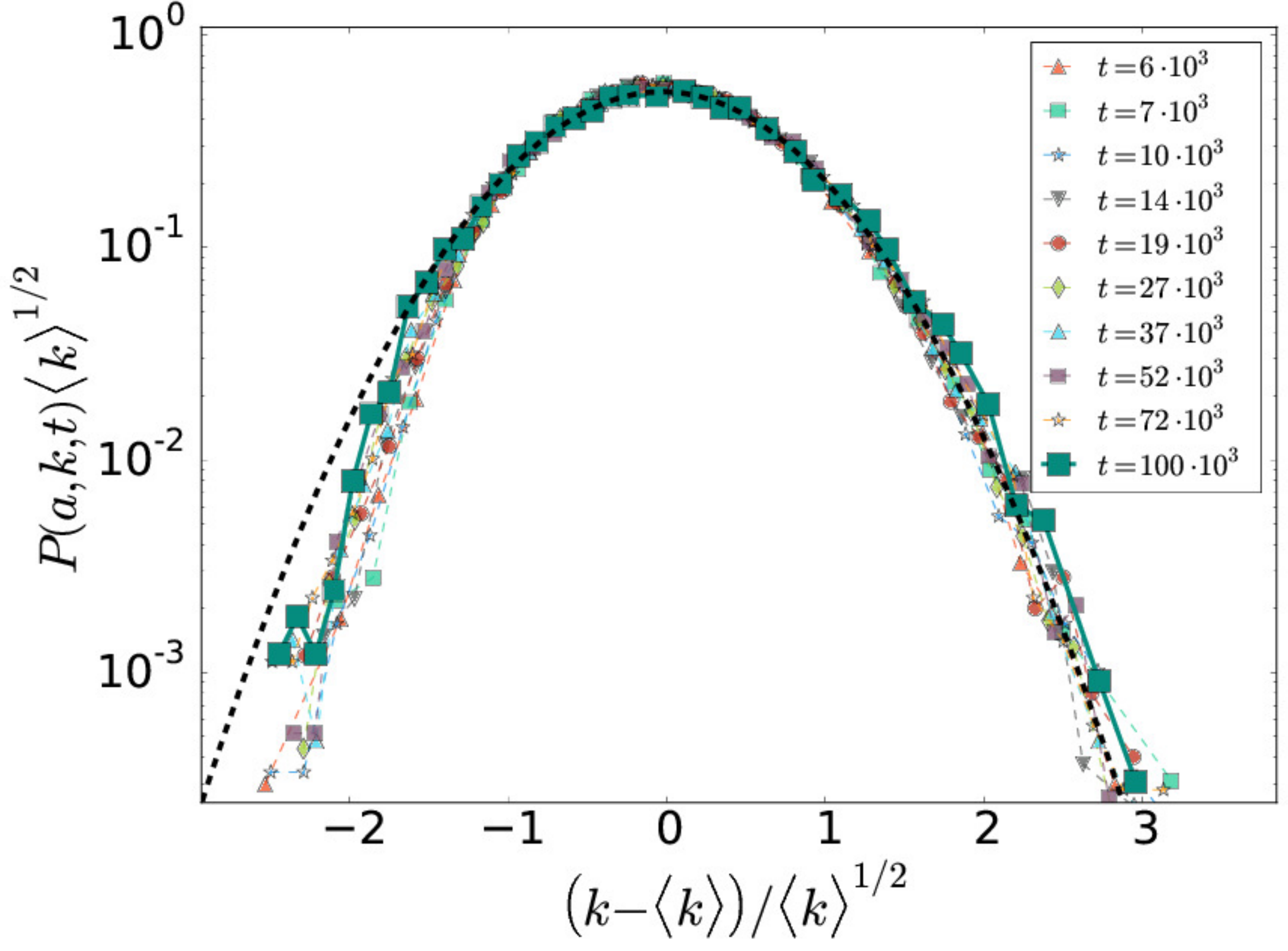}}
    \subfigure[]
        {\includegraphics[width=2.8in]{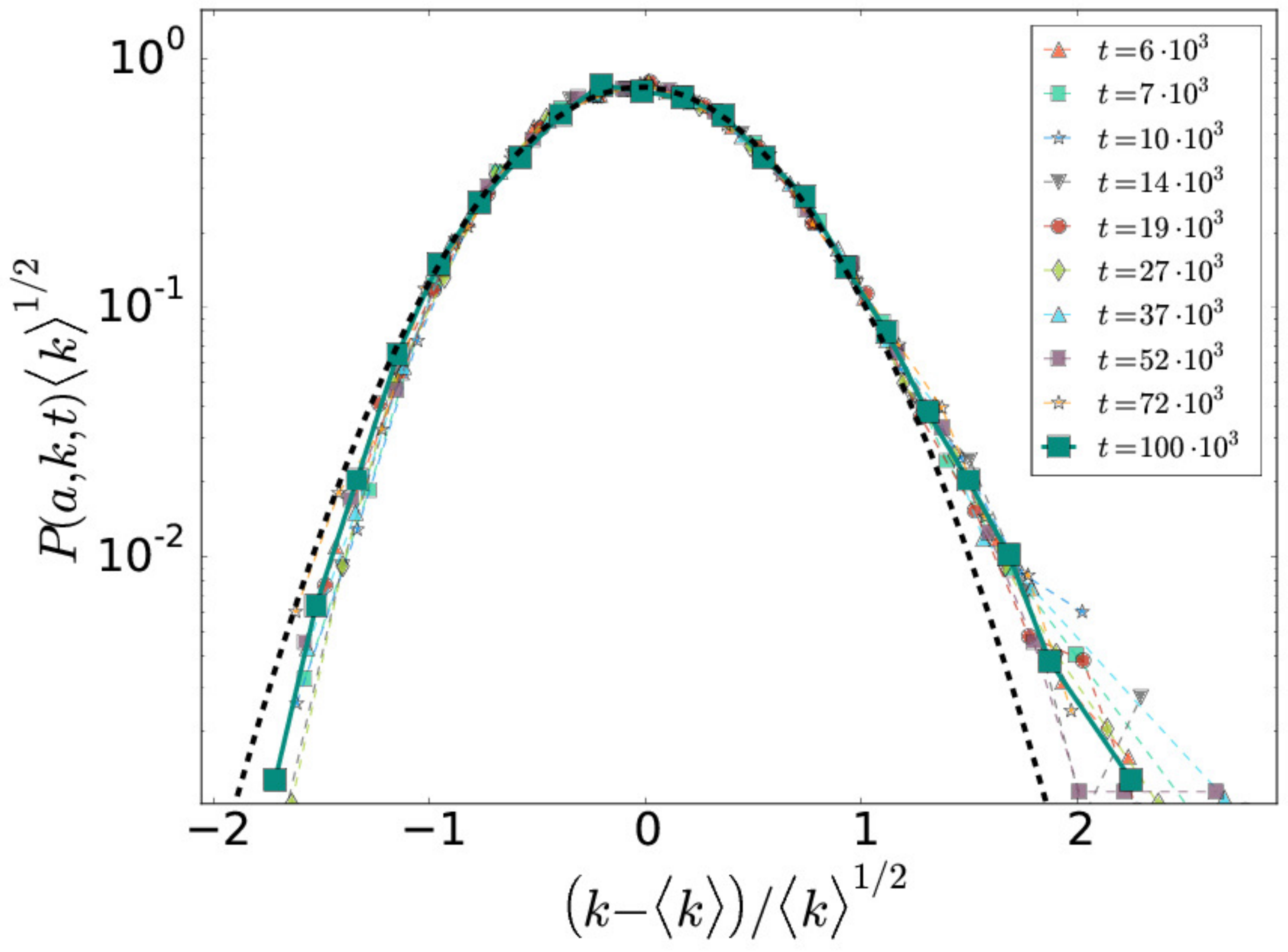}}
    \caption{
        \label{fig:pakt_fixb} The probability distribution $P_a(k,t)$ for a selected activity
        class $a$ in the simulations with exponent $\beta=1.0$ (a) and $\beta=2.0$
        (b). We compare different evolution times (see legend) by rescaling the degree
        $k\to\tilde k = (k-\av{k(a,t)})/\av{k(a,t)}^{1/2}$ on the $x$-axis and the distribution
        itself $P_a(k,t)\to\av{k(a,t)}^{1/2}P(a,\tilde k, t)$ on the $y$-axis, where
        $\av{k(a,t)}$ is the average degree at time $t$ for the nodes belonging to the
        activity class $a$. We also show the fit of the large time $P(a,k,t)$ with a Gaussian 
        curve (black dashed line) as predicted in Eq. (\ref{eq:sol_single_b}).
    }
\end{figure}

Then, in Fig. \ref{fig:kat_fixb} we present the asymptotic growth of the average
degree for an activity class (i.e. a collection of nodes bins $b$ featuring similar
activity values) and we compare it with the analytical prediction 
 $\av{k(a,t)} \propto (at)^{1\over 1+\beta}$.
In Fig. \ref{fig:pakt_fixb} we show that the shape and the evolution of
the $P_i(k,t)$ distribution follows the predicted form of Eq. (\ref{eq:sol_single_b}).

\begin{figure}
    \centering
    \subfigure[]
        {\includegraphics[width=2.in]{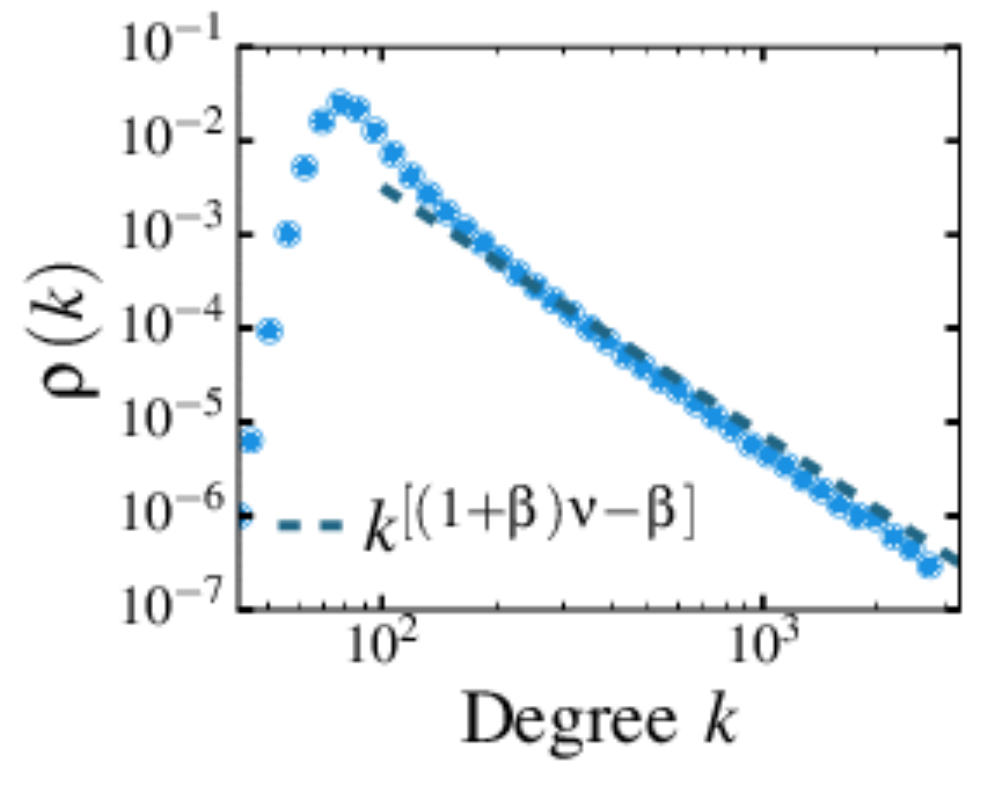}}
    \subfigure[]
        {\includegraphics[width=2.in]{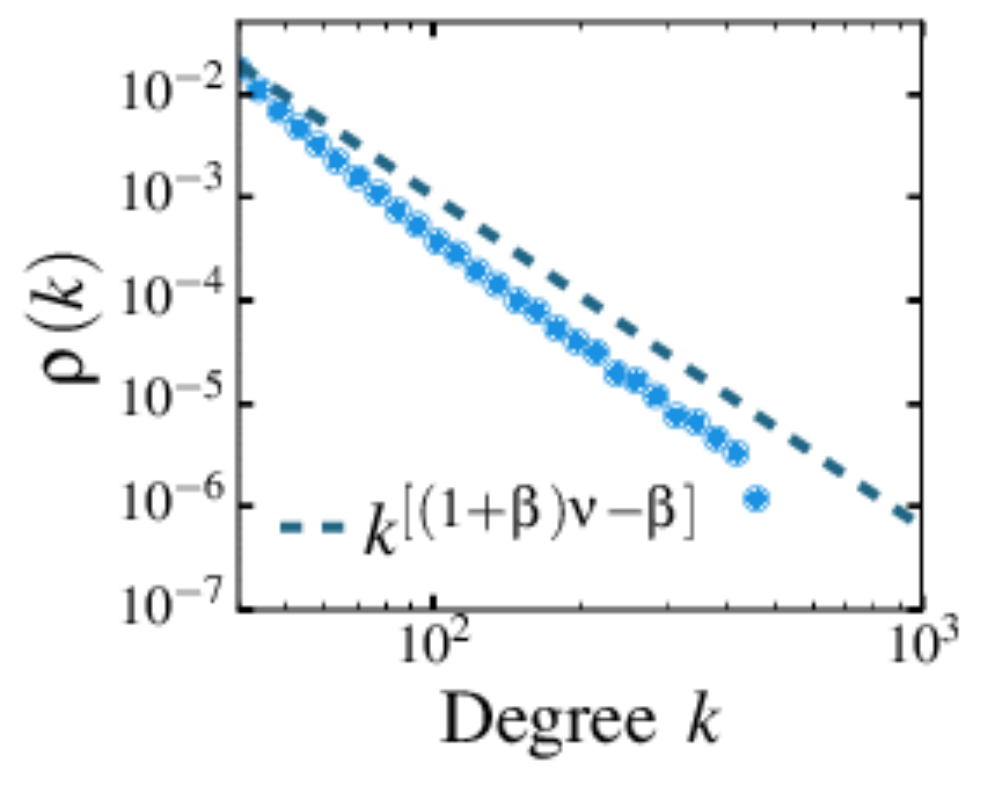}}
    \hspace{3in}
    \subfigure[]
        {\includegraphics[width=2.in]{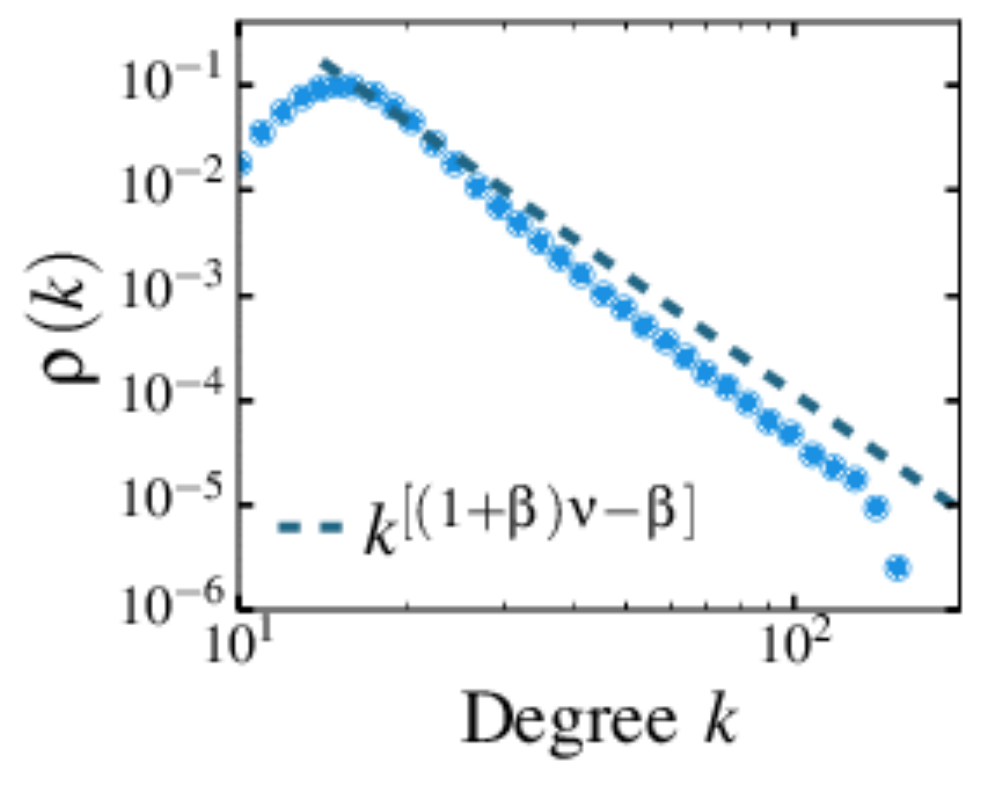}}
    \subfigure[]
        {\includegraphics[width=2.in]{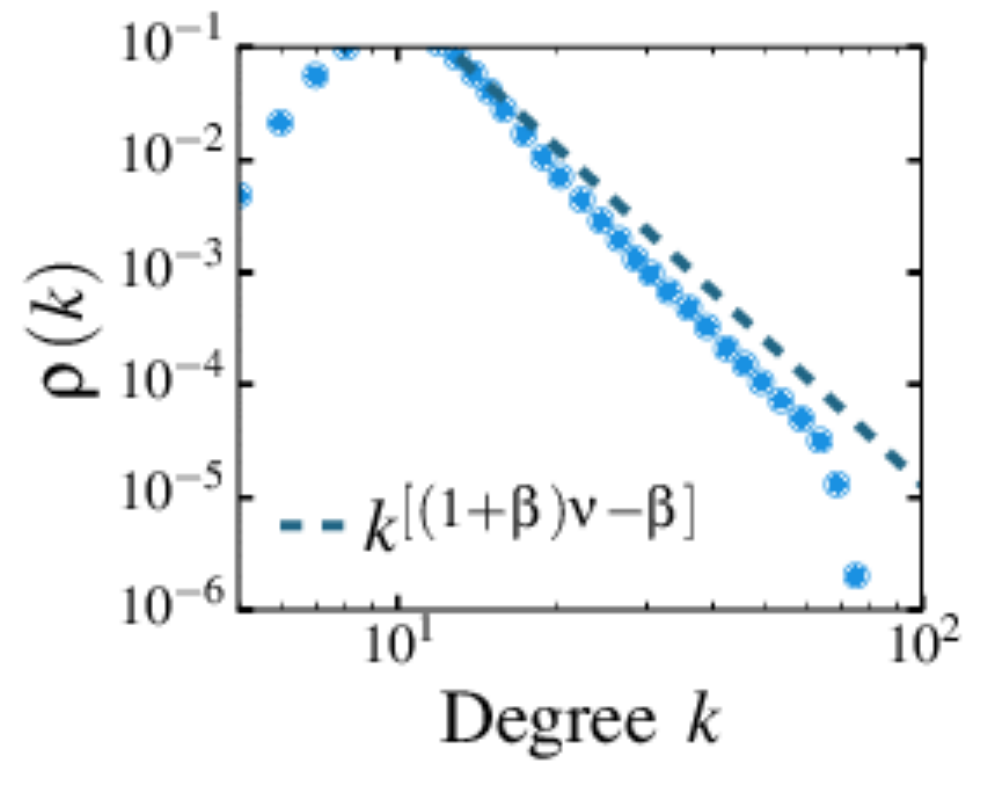}}
    \caption{
        \label{fig:rhos_fixb} The resulting degree distribution of simulations featuring
        $\beta=0.5$ (a), $\beta=1.0$ (b), $\beta=1.5$ (c) and $\beta=1.0$ (d).
        The analytical predictions (given $F(a)\propto a^{-\nu}$, with $\nu=2.1$) for the
        scaling exponent are sown (blue dashed lines).
    }
\end{figure}

The last check regards the overall degree distribution $\rho(k)$ that should
follow Eq. (\ref{eq:rhok_single_b}). In Fig. \ref{fig:rhos_fixb} we compare the
activity distribution $F(a)\propto a^{-\nu}$ and the degree distribution
$\rho(k)$. The exponent $\mu$ leading the $\rho(k)\propto k^{-\mu}$ is in good
agreement with the analytically predicted value $\mu = [(1+\beta)\nu-\beta]$.

\subsubsection{The multi-$\beta$ case}
\label{sub:multi_b}

As shown in Section \ref{sec:data_analysis} a single value of the reinforcement exponent $\beta$
is found to fit most of the $p_b(k)$ curves in both the \emph{APS} and \emph{TMN}
datasets, while for $MPN$ a single value of $\beta$ cannot fit all the $p_b(k)$ curves for
each activity-degree class $b$ at once.
For this reason we further develop the model, letting each node $i$ to feature three
parameters: the activity $a_i$ and the reinforcement constant $c_i$ together with the
exponent $\beta_i$ of the underlying reinforcement process.

Since the model with three parameters per node is difficult to handle, we apply some
approximations in order to get analytical insight.
In particular, we work in simplified single-agent framework, where we focus on a single agent that can only
connect to other nodes and never get called. Within this approximation the master equation
for the node $i$ reads:
\begin{align}
    \label{eq:ME_multi_b}
    P_i(k,t+1) = a_i p(k-1) P_i(k-1,t) + P_i(k,t)\left[ a_i
        (1-p(k)) + (1-a_i) \right].
\end{align}
The continuum limit for large degree $k$ and time $t$ of Eq.
(\ref{eq:ME_multi_b}) is:
\begin{align}
    \label{eq:ME_cont_multi_b}
    {\partial P \over \partial t}= -a\left({c\over k}\right)^{\beta}\left[
        {\partial P\over\partial k} - {1\over2}{\partial^2 P\over \partial
        k^2}\right].
\end{align}
The solution for $P_i(k,t)$ is:
\begin{equation}
    P_i(k, t) \propto \exp{\left[-A\frac{\left(k-C_i t^{{1\over
    1+\beta_i}}\right)^2}{t^{1/(1+\beta_i)}}\right]},
    \label{eq:sol_multi_b}
\end{equation}
where the $C_i$ now reads:
\begin{equation}
    C_i = [(1+\beta_i)c_i^\beta a_i]^{1\over 1+\beta_i}.
    \label{eq:ca_multi_b}
\end{equation}

Again, the average degree $\av{k_i(t)}$ grows as:
\begin{equation}
    \av{k_i(t)} \propto C_i t^{1\over 1+\beta_i}.
    \label{eq:kat_multi_b}
\end{equation}

The result found in Eq. (\ref{eq:kat_multi_b}) holds for a single class of nodes
with a given set of activity $a_i$ and reinforcement constant $c_i$ and strength $\beta_i$.

The average degree $\av{k(a,t)}$ for the activity class
$a$ can be computed by integrating over the different values of $\beta_i$ and $c_i$:
\begin{equation}
    \av{k(a,t)} = \int dc'  \int d\beta' \rho(\beta',c'|a) C(a,c',\beta')(t)^{1 \over
    1+\beta' }
    \label{eq:avg_kact_multib}
\end{equation}
where $\rho(\beta, c|a)$ is the probability for a node of activity $a$ to have a
reinforcement exponent and constant equal to $\beta$ and $c$.
By assuming that the distribution of the exponent $\beta$ is independent from $a$ and $c$
we can factor out the time-depend term obtaining for the activity class $a$:
\begin{equation}
    \av{k(a,t)} \propto \int{d\beta' \rho(\beta') t^{1\over 1+\beta'}},
    \label{eq:kat_multib}
\end{equation}
where $\rho(\beta)$ is the probability distribution of the $\beta$ parameter.

Let us assume that $\rho(\beta)$ can be written as a sum of Kroenecker 
$\delta$-functions, i.e.:
\begin{equation}
    \rho(\beta) = {1\over \sum_i{C_i}}\sum_{i=1}^{N_{\beta}}{C_i
    \delta(\beta - \beta_i)}.
    \label{eq:rhob}
\end{equation}

By plugging Eq. (\ref{eq:rhob}) in Eq. (\ref{eq:kat_multib}) we find that:
\begin{equation}
    \av{k(a,t)} \propto \sum_{i=1}^{N_{\beta}}{C_i t^{1\over 1+\beta_{i}}}
    \;\xrightarrow{t\to\infty}\; t^{1\over1+\beta_{\rm min}},
    \label{eq:avg_kat_multib}
\end{equation}
so that the minimum value of $\beta_i$, $\beta_{\rm min}$ leads the asymptotic
behavior of the $\av{k(a,t)}$ function.


\subsubsection{Numerical results}

To investigate the multi-$\beta$ case we performed further numerical
simulations considering networks with the following parameters:
\begin{itemize}
    \item [-] $N=10^6$ nodes;
    \item [-] activity $a\in[\epsilon, 1.0]$ with $\epsilon=10^{-3}$, power -law
        distributed so that $F(a)\propto a^{-\nu}$ with $\nu=2.1$;
    \item [-] (a) reinforcement exponent $\beta = [0.5, 1.5, 2.5]$ with
            probability $[1/6, 1/3, 1/2]$ (i.e. one sixth of the nodes has
            $\beta=0.5$, one third $\beta=1.5$ and a half of them $\beta=2.5$
            regardless of their activity) and (b) $\beta=[1.0,1.5,2.0]$ with equal
            probability $1/3$.
    \item [-] fixed $c=1$ for all the nodes;
    \item [-] $T=2\cdot 10^5$ evolution steps.
\end{itemize}

The numerical procedure is similar to the one described in Section \ref{sub:num_fixb},
the difference being that we compute the attachment probability $p_i(k)$
taking into account the reinforcement exponent $\beta_i$ of the node itself.

\begin{figure}
    \centering
    \subfigure[]
        {\includegraphics[width=3.2in]{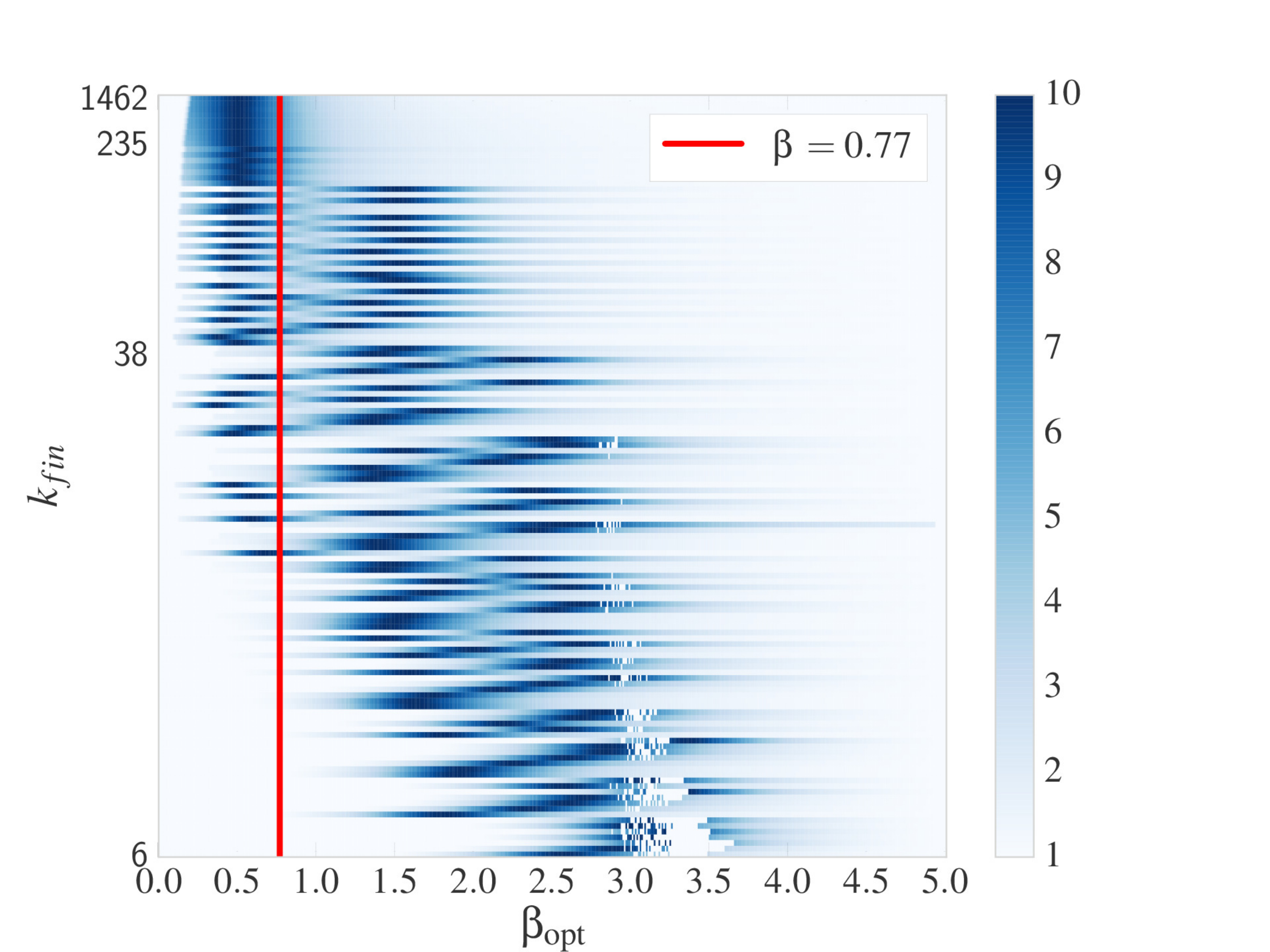}}
    \subfigure[]
        {\includegraphics[width=3.2in]{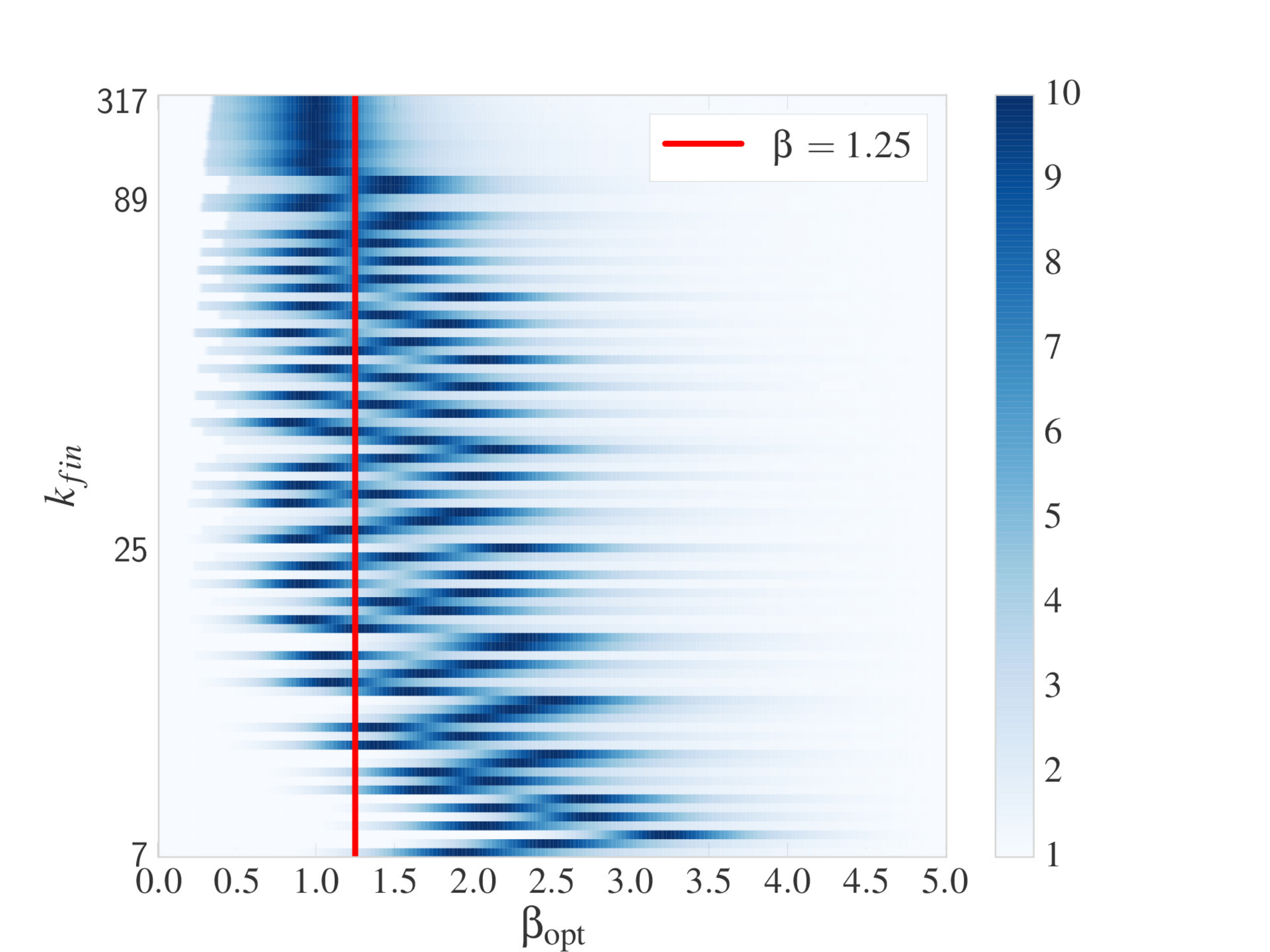}}
    \caption{
        \label{fig:recover_multib} The heat-map like matrix of
        $-\ln(\chi^2(\beta))$ for the simulation with $\beta\in[0.5, 1.5, 2.5]$
        (a) and $\beta\in[1.0,1.5,2.0]$ (b); analogous to Figure \ref{fig:HM}
        for real data.
    }
\end{figure}

In Fig. \ref{fig:recover_multib} we show that, in both the cases, we can
recover the behavior described in Section \ref{sec:act_bins} for real data.
In particular Fig. \ref{fig:recover_multib}(a) (related to the $\beta\in[1,2]$
case) we observe a clear diagonal pattern of the optimal values of the exponent
$\beta(b)$ for the $b$ bins that minimize the $\chi^2_b(\beta)$. In particular $\beta(b)$ varies
from $\beta\sim2.0$ values for lower degree nodes bins up to $\beta\sim1.0$ values for the larger
final degree nodes bins.
The figure recalls the situation of the MPC dataset presented in Fig. \ref{fig:HM} (b) and in
the main paper.

\begin{figure}
    \centering
    \subfigure[]
        {\includegraphics[width=2.7in]{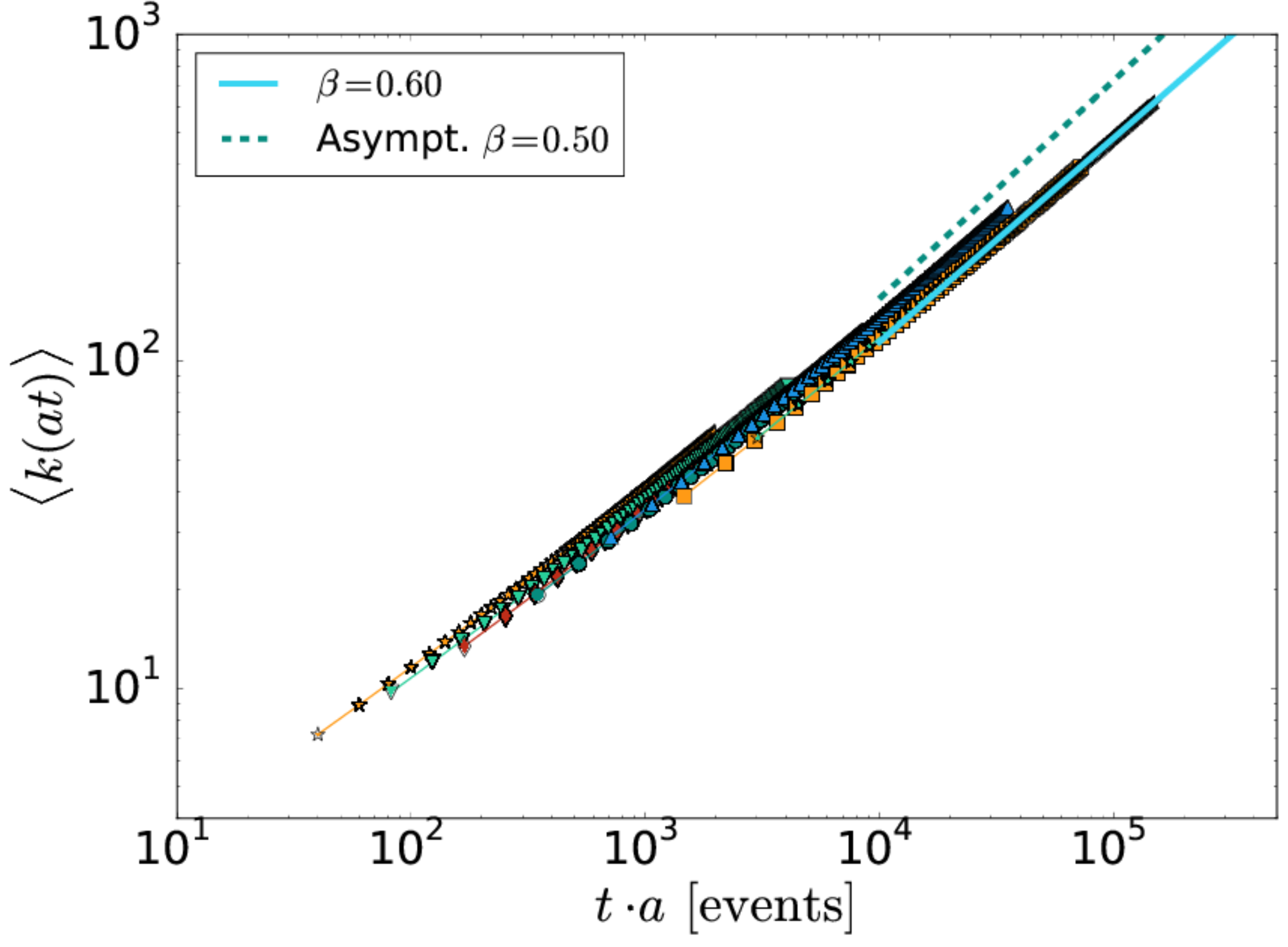}}
    \subfigure[]
        {\includegraphics[width=2.7in]{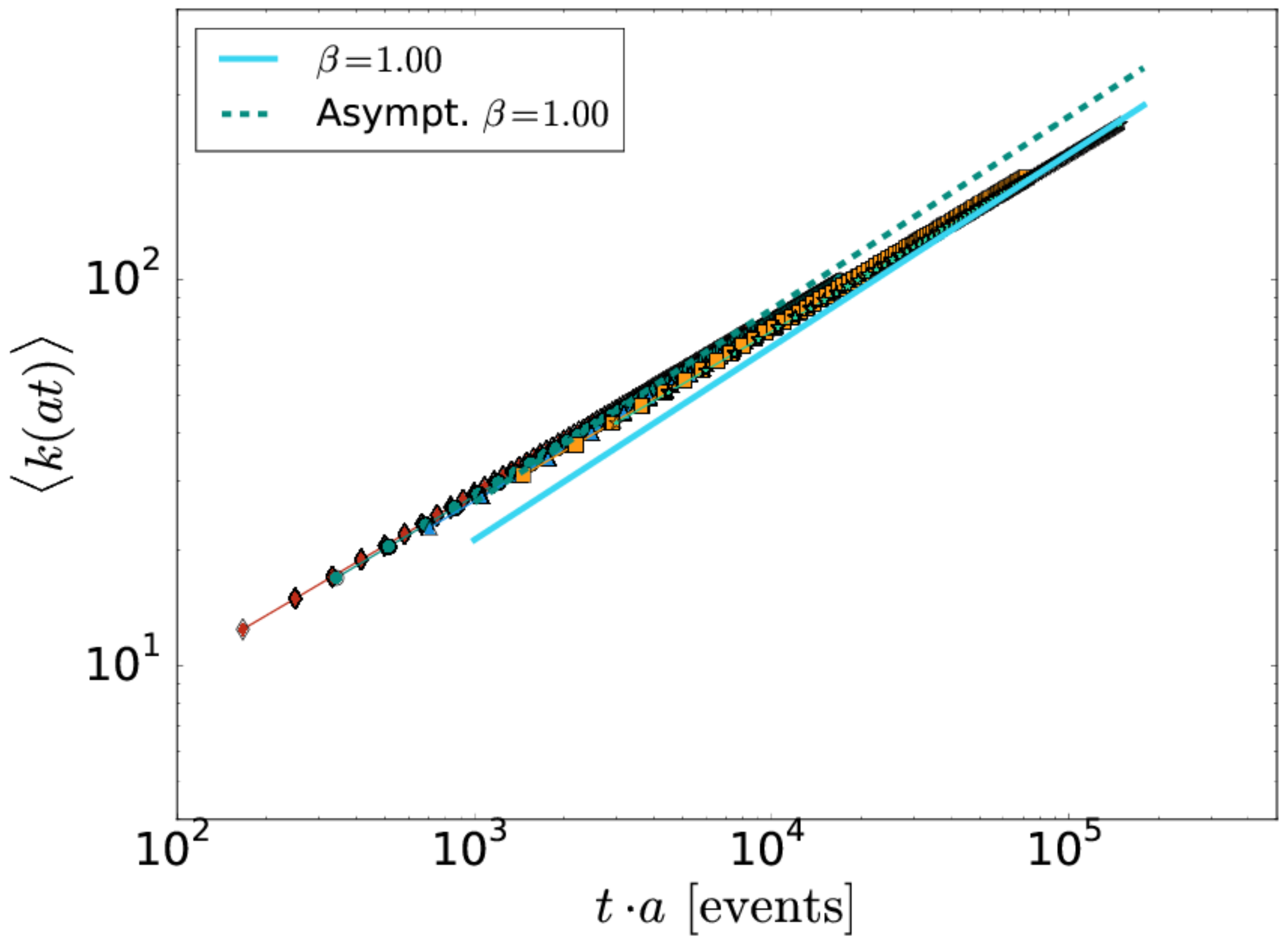}}
    \caption{
        \label{fig:kat_distb} The average degree $\av{k(at)}$ for different
        activity classes in the (a) $\beta\in[0.5,1.5.2.5]$ and (b)
        $\beta\in[1.0, 1.5, 2.0]$ case. The time is rescaled with activity
        $t\to at$, so that all the curves collapse. We also plot the fit
        $\av{k(at)} \propto (t/A)^{1\over 1+\beta^*}$ in the long time limit
        (cyan solid line) and the predicted asymptotic growth $\av{k(at)} =
        A\cdot t^{1\over 1+\beta_{\rm min}}$ (dashed line).
    }
\end{figure}

In Fig. \ref{fig:kat_distb} we show the asymptotic growth of the average degree
$\av{k(a,t)}$ together with the predicted asymptotic behavior proportional to
$t^{1\over 1+\beta_{\rm min}}$. As one can see, numerical results and the
suggested analytical solution are in very good agreement in both the cases.

\subsection{Comparison with real data}

In Fig. \ref{fig:kat} we present the comparison between prediction and
real data for the PRA, PRB, PRD, PRE, TMN and MPN datasets.
The $\av{k(a,t)}$ curve of each activity class is shown with the time rescaled
with the activity of each activity class, i.e. $t\to at$.
In the MPC case we use as $\beta_{\rm opt} = \beta_{\rm min} = 1.2$ (the
$\beta$ value found in the largest degree bins of Fig. \ref{fig:HM} (b)). In
all the other cases, as the $\beta_{\rm opt}$ fits correctly most of the curves,
we use the $\beta_{\rm opt}$ value returned by our analysis.

\begin{figure}
    \centering
    \subfigure[]
    {\includegraphics[width=1.75in]{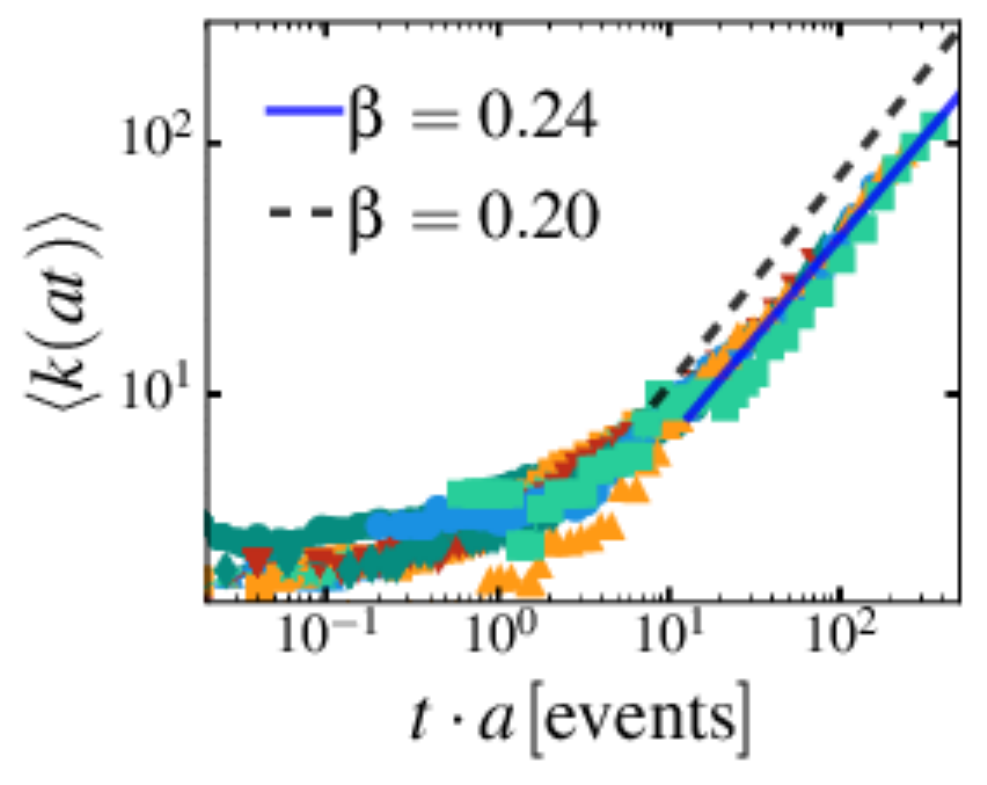}}
    \subfigure[]
    {\includegraphics[width=1.75in]{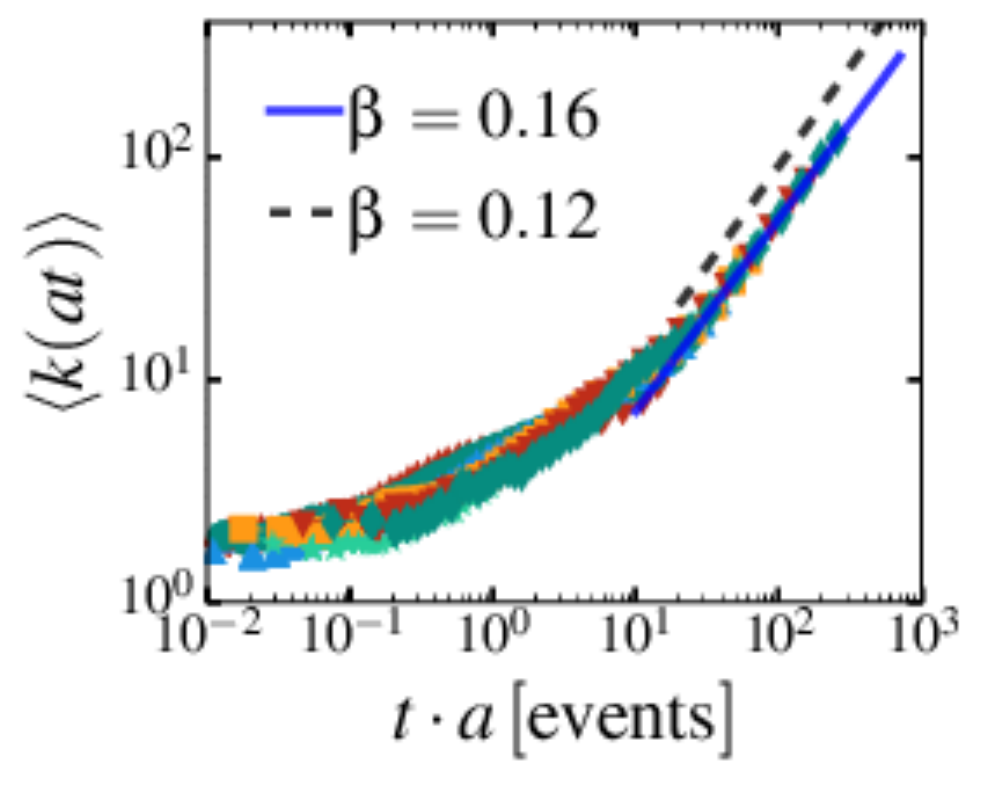}}
    \subfigure[]
    {\includegraphics[width=1.75in]{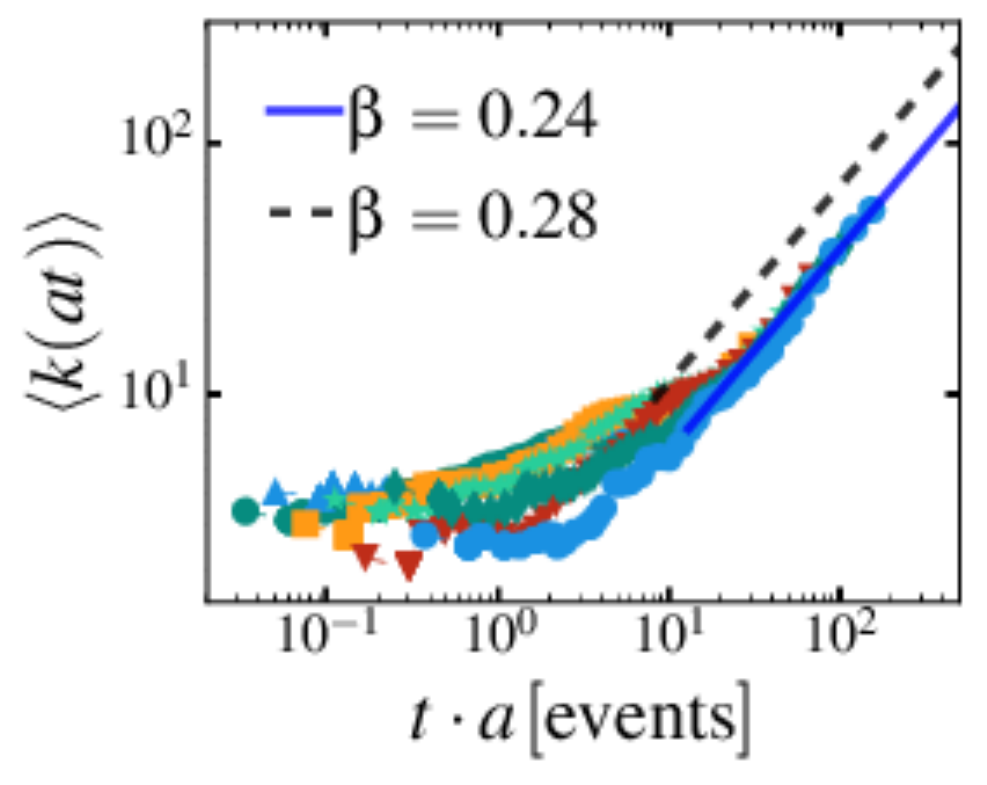}}
    \subfigure[]
    {\includegraphics[width=1.75in]{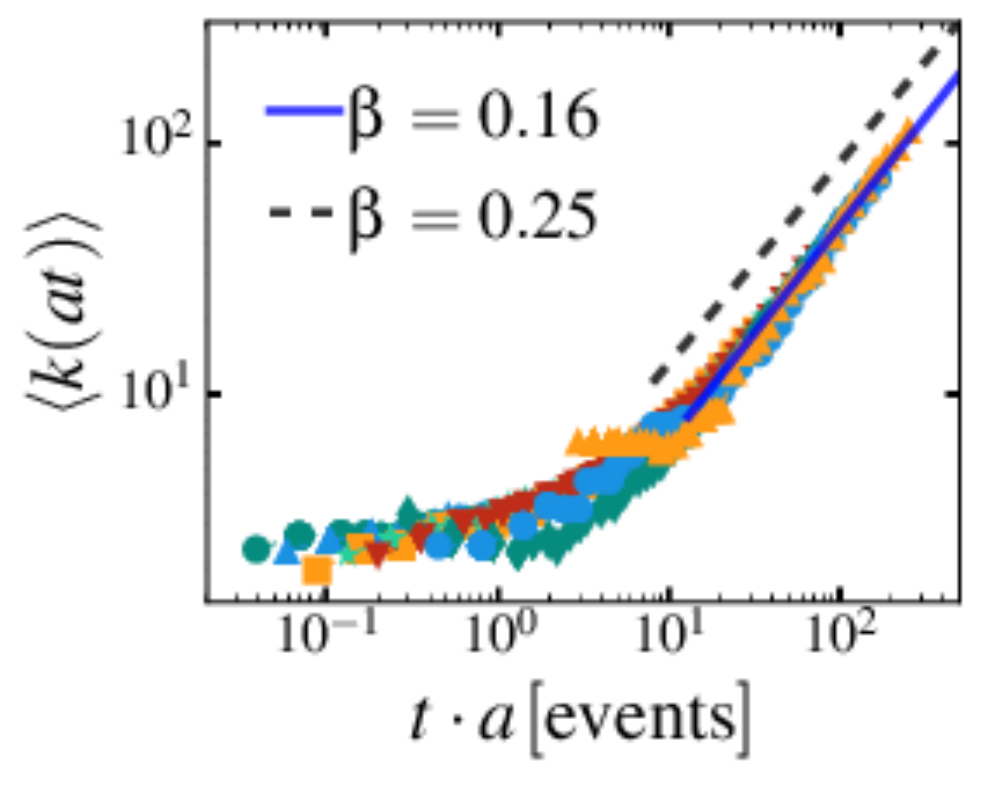}}
    \subfigure[]
    {\includegraphics[width=1.75in]{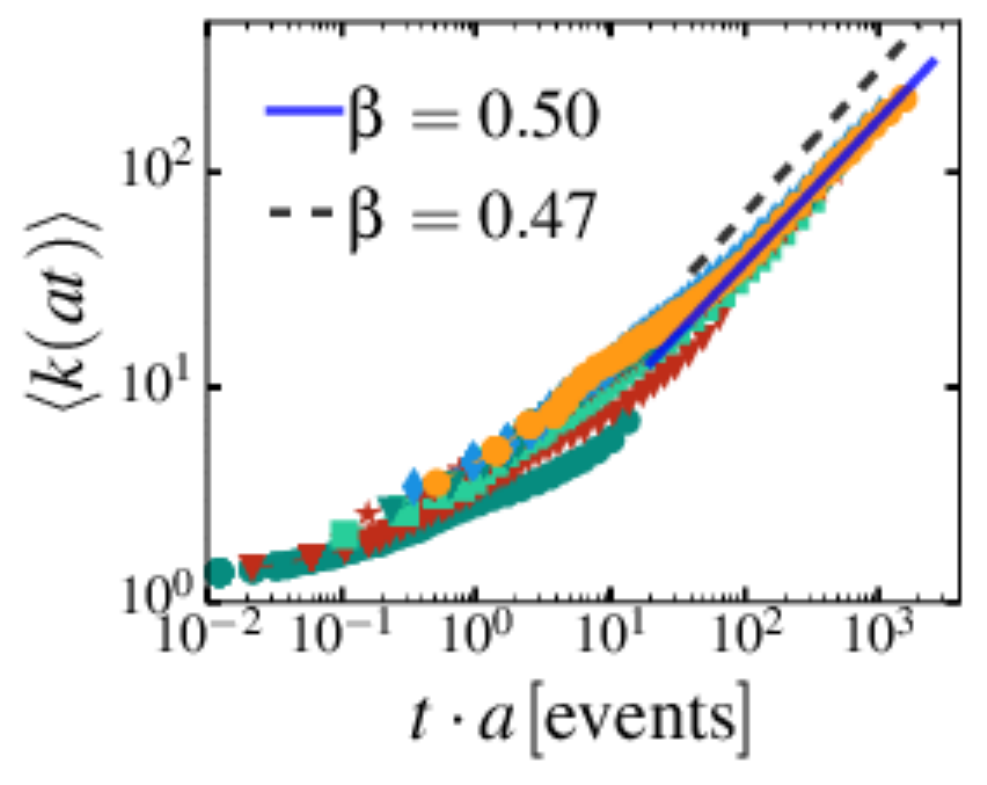}}
    \subfigure[]
    {\includegraphics[width=1.75in]{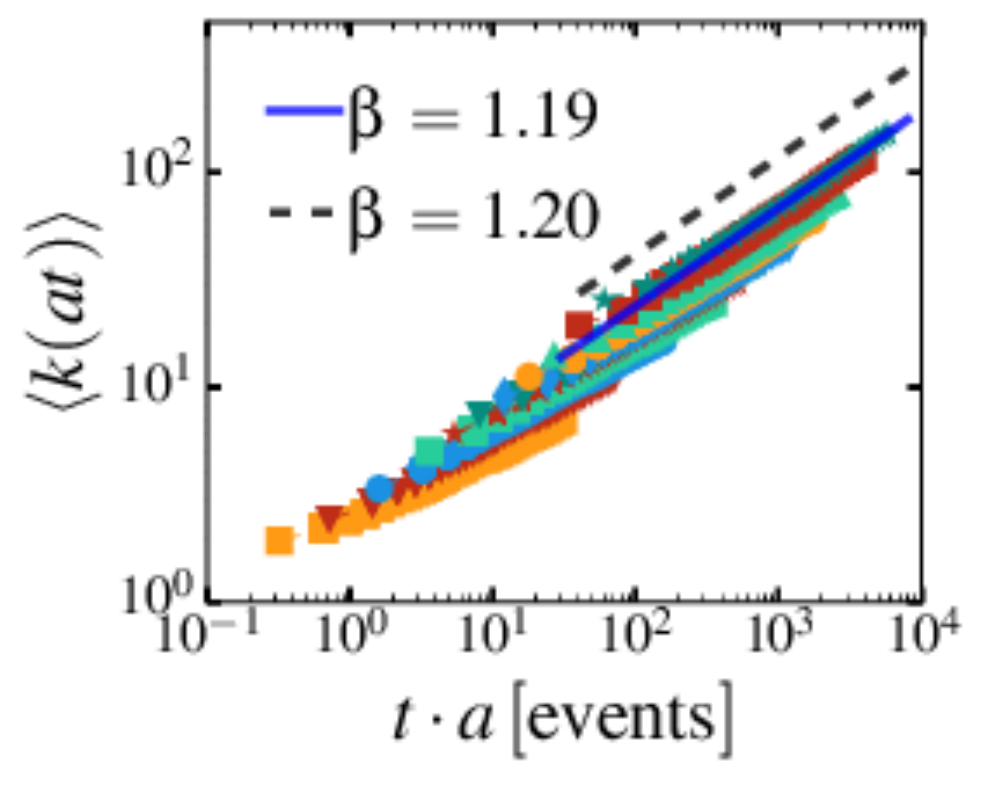}}
    \caption{
        \label{fig:kat}
        The average degree $\av{k(at)}$ (each data series corresponds to a
        different activity class) for: (a) PRA, (b) PRB, (c) PRD, (d) PRE, (e) TWT and (f)
        MPC. We compare the data for $\av{k(a,t)}$ with the expected behavior (dashed
        lines) $(at)^{1/(1+\beta_{\rm opt})}$: in panels (a-e)
        $\beta_{\rm opt}$ has been evaluated according Eq. (\ref{eq:beta_opt}),
        while in the (f) case we use $\beta_{\rm opt} = \beta_{\rm min} = 1.2$.
        We also plot the power-law fit $\av{k(a,t)} \propto (at)^{1/(1+\beta^*)}$
        (solid lines) for comparison.
    }
\end{figure}

\begin{figure}
    \centering
    \subfigure[]
    {\includegraphics[width=1.75in]{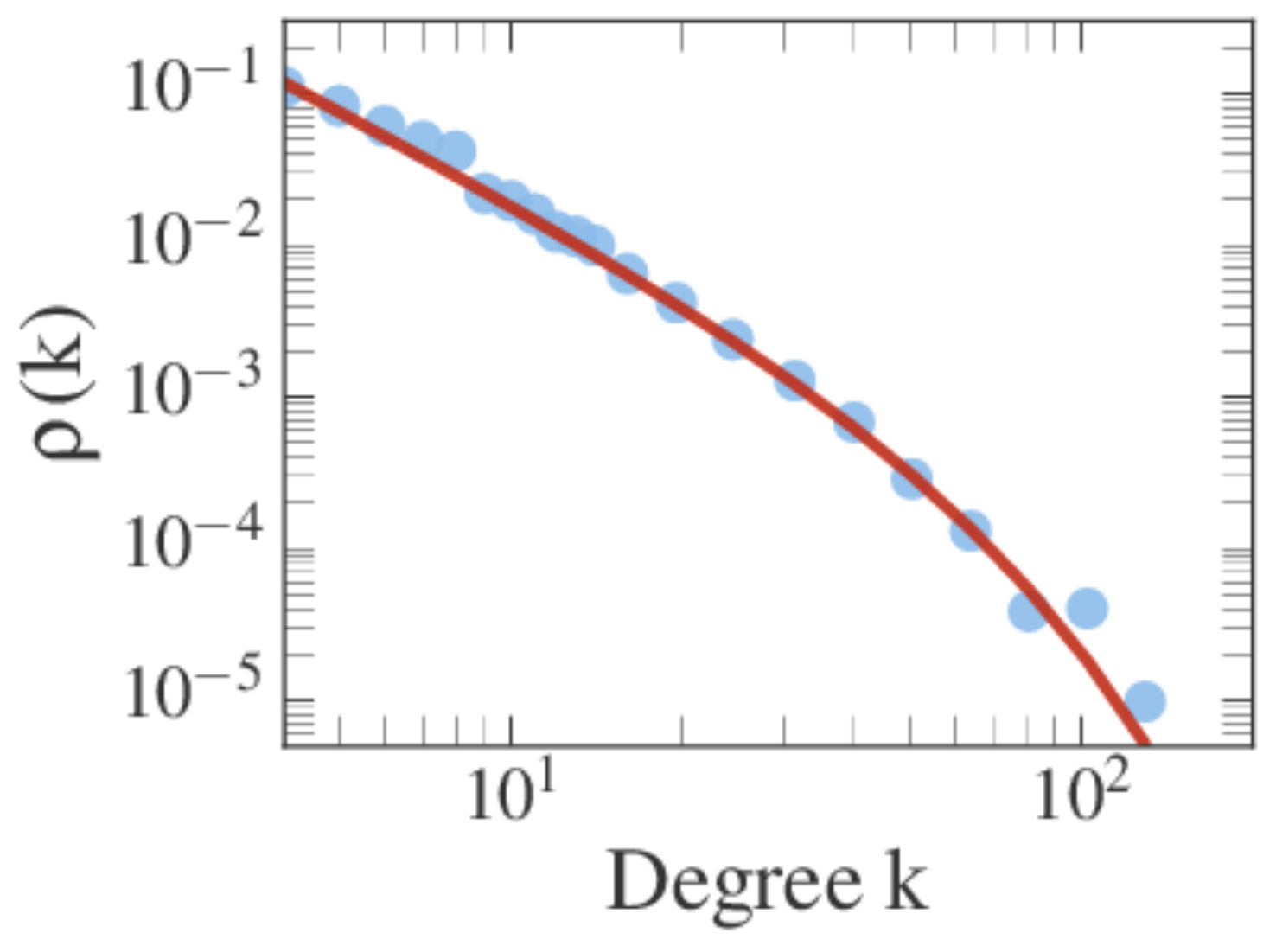}}
    \subfigure[]
    {\includegraphics[width=1.75in]{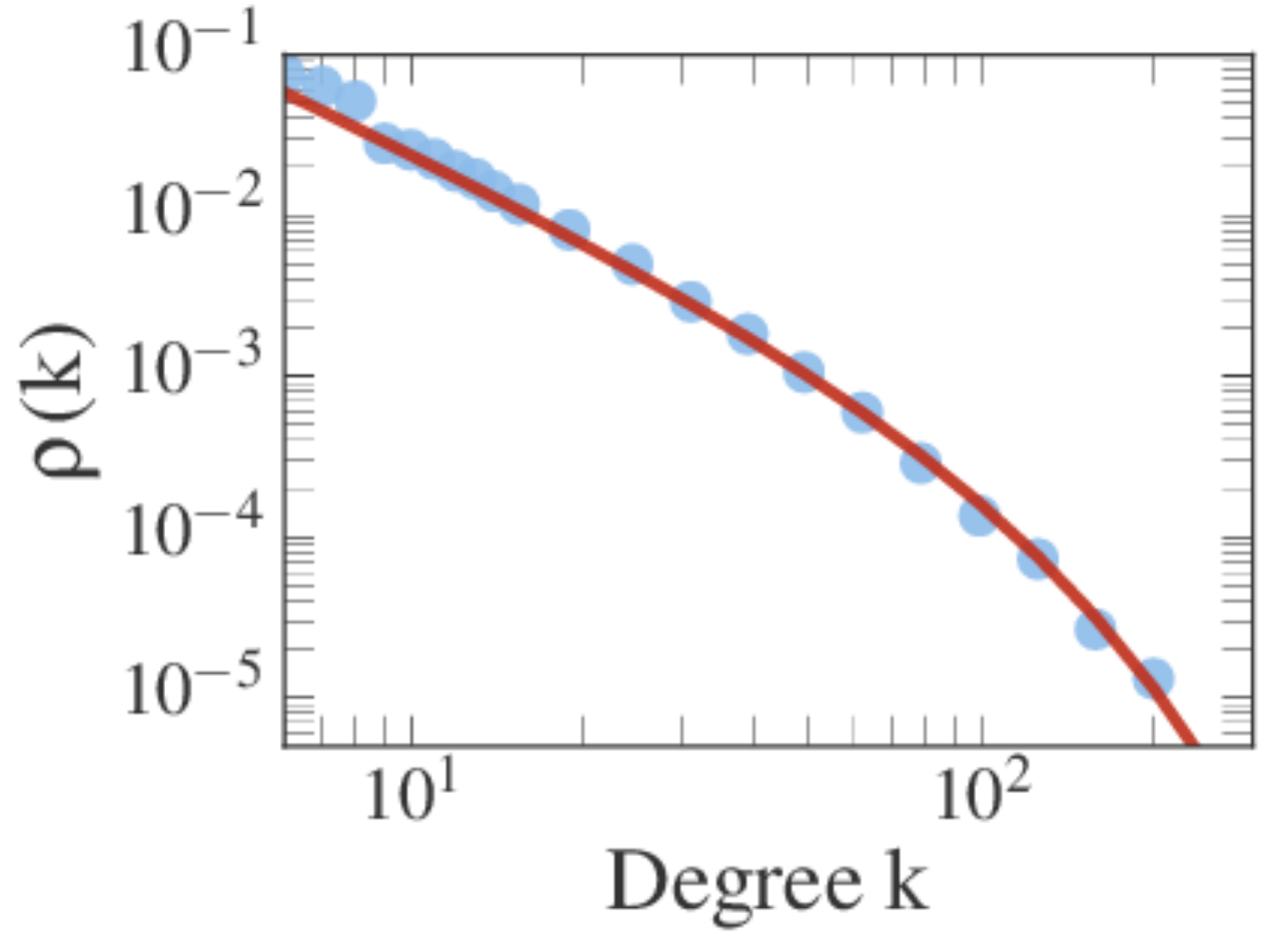}}
    \subfigure[]
    {\includegraphics[width=1.75in]{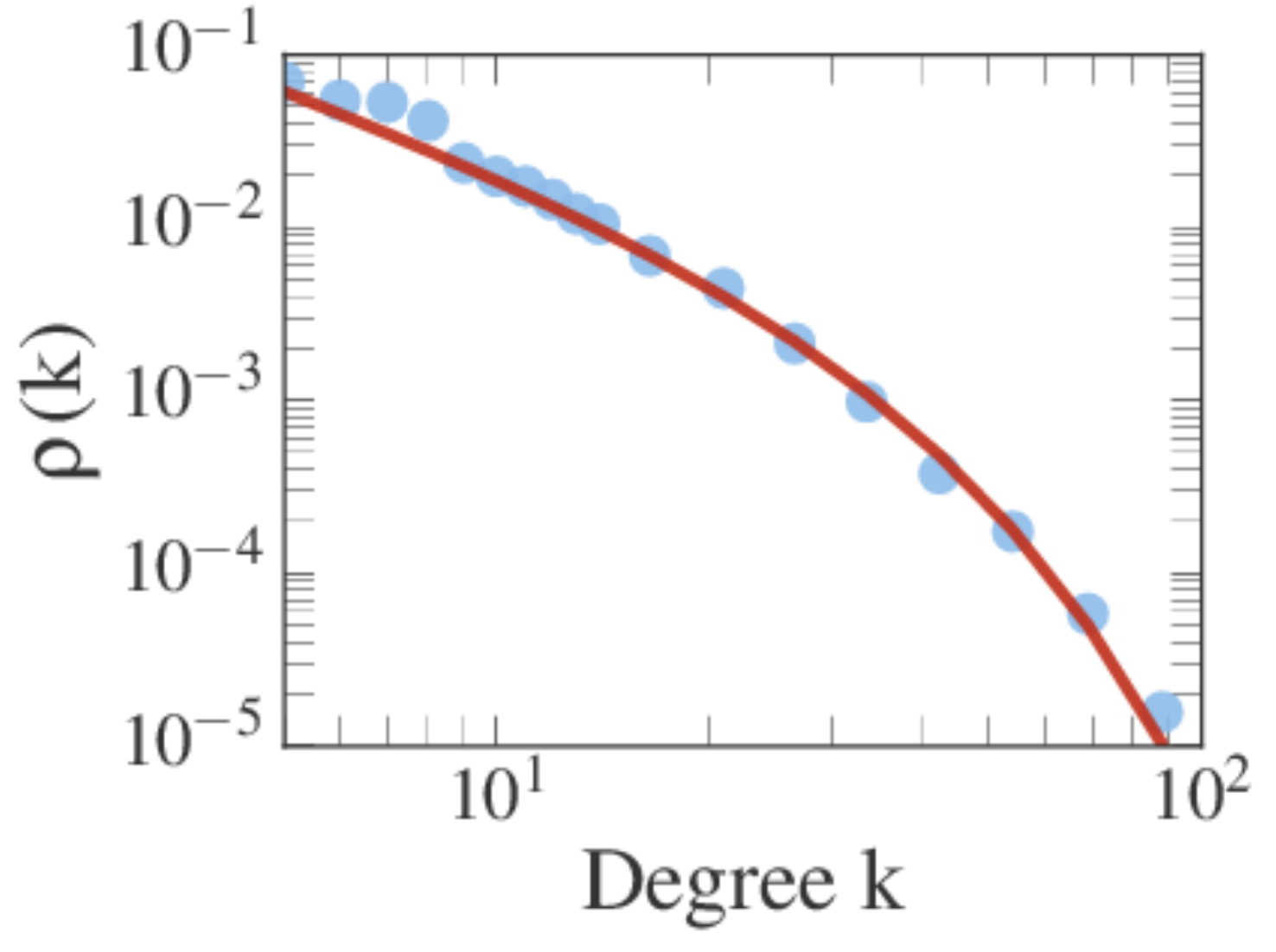}}
    \subfigure[]
    {\includegraphics[width=1.75in]{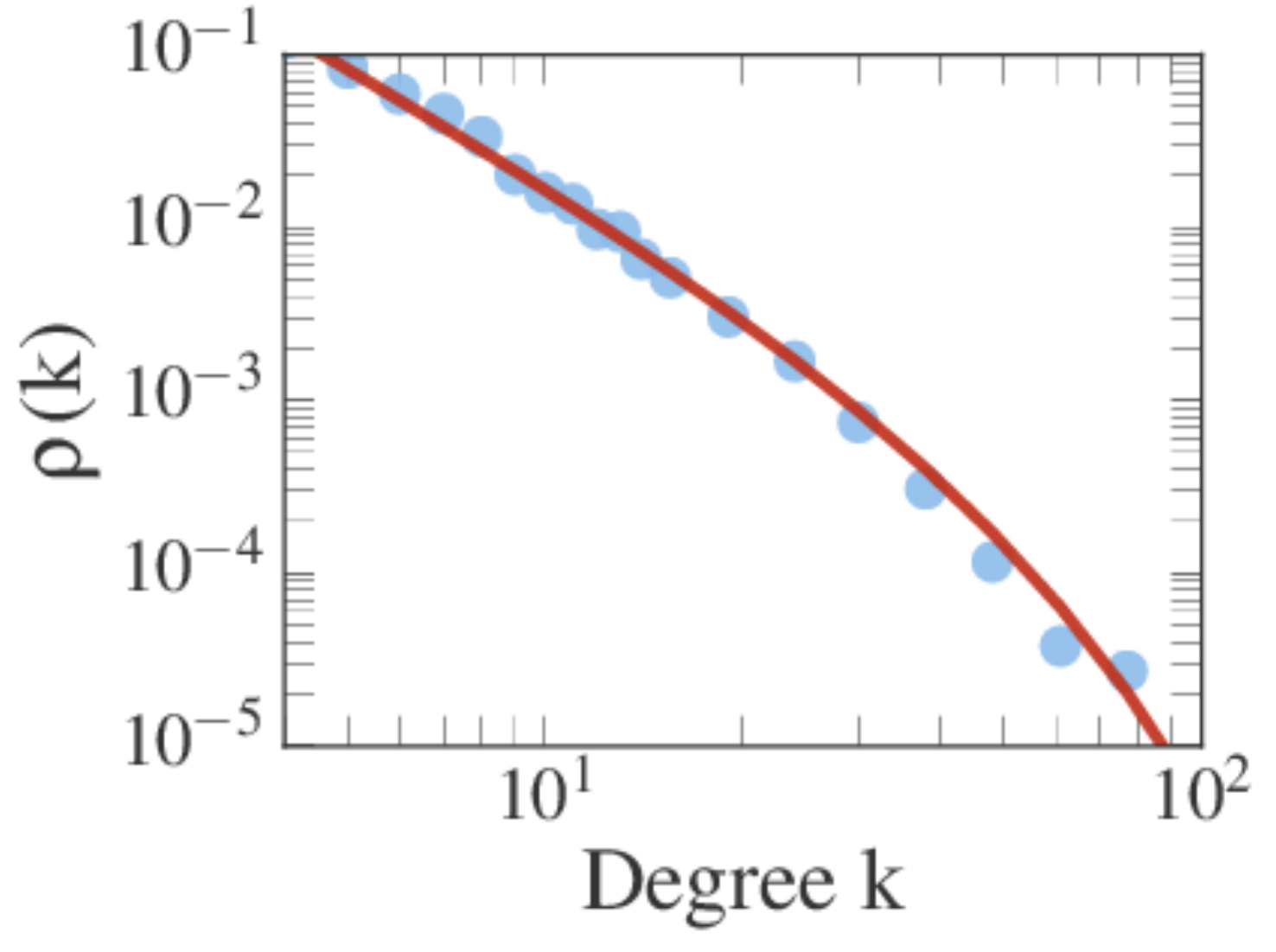}}
    \subfigure[]
    {\includegraphics[width=1.75in]{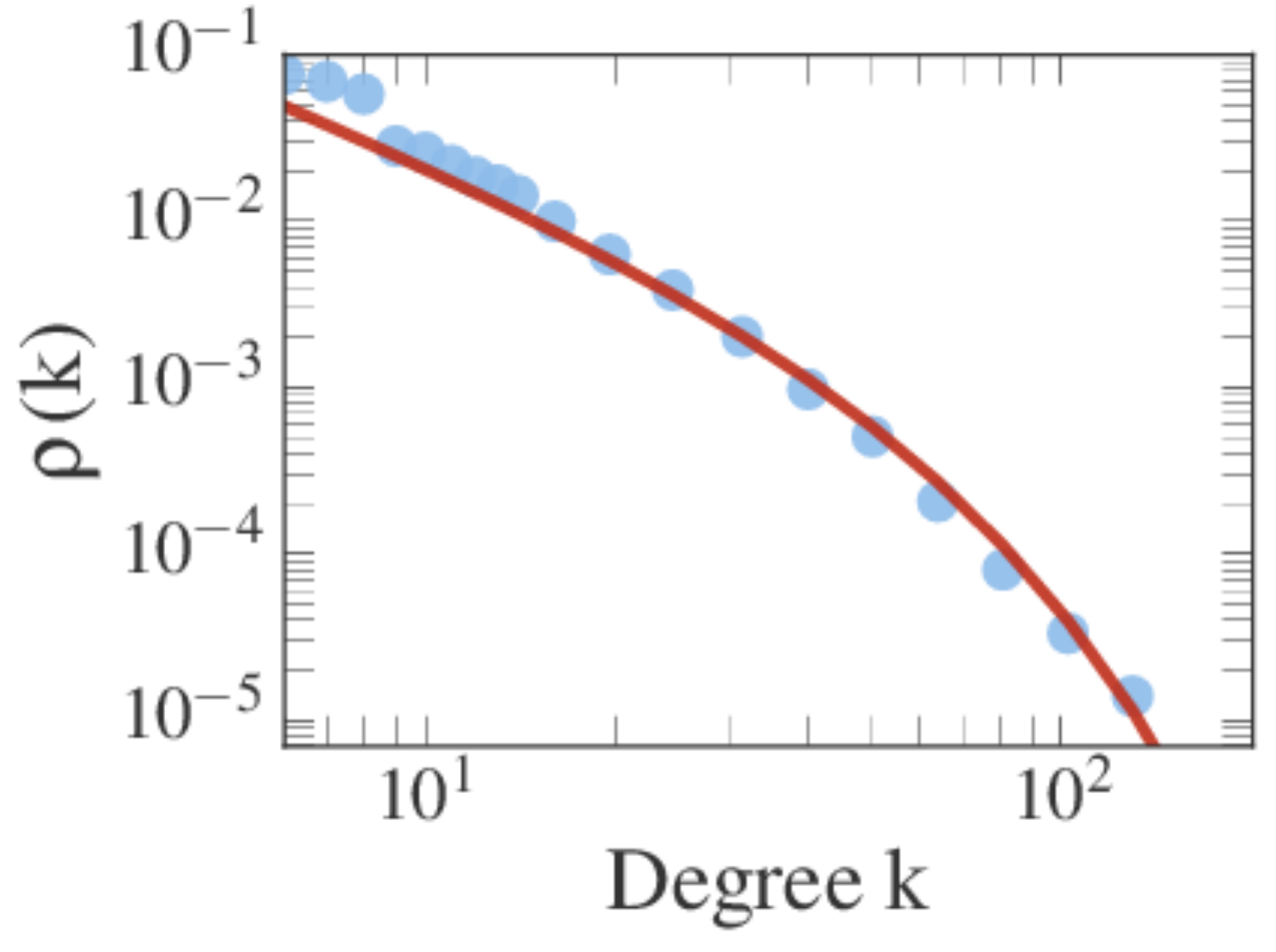}}
    \subfigure[]
    {\includegraphics[width=1.75in]{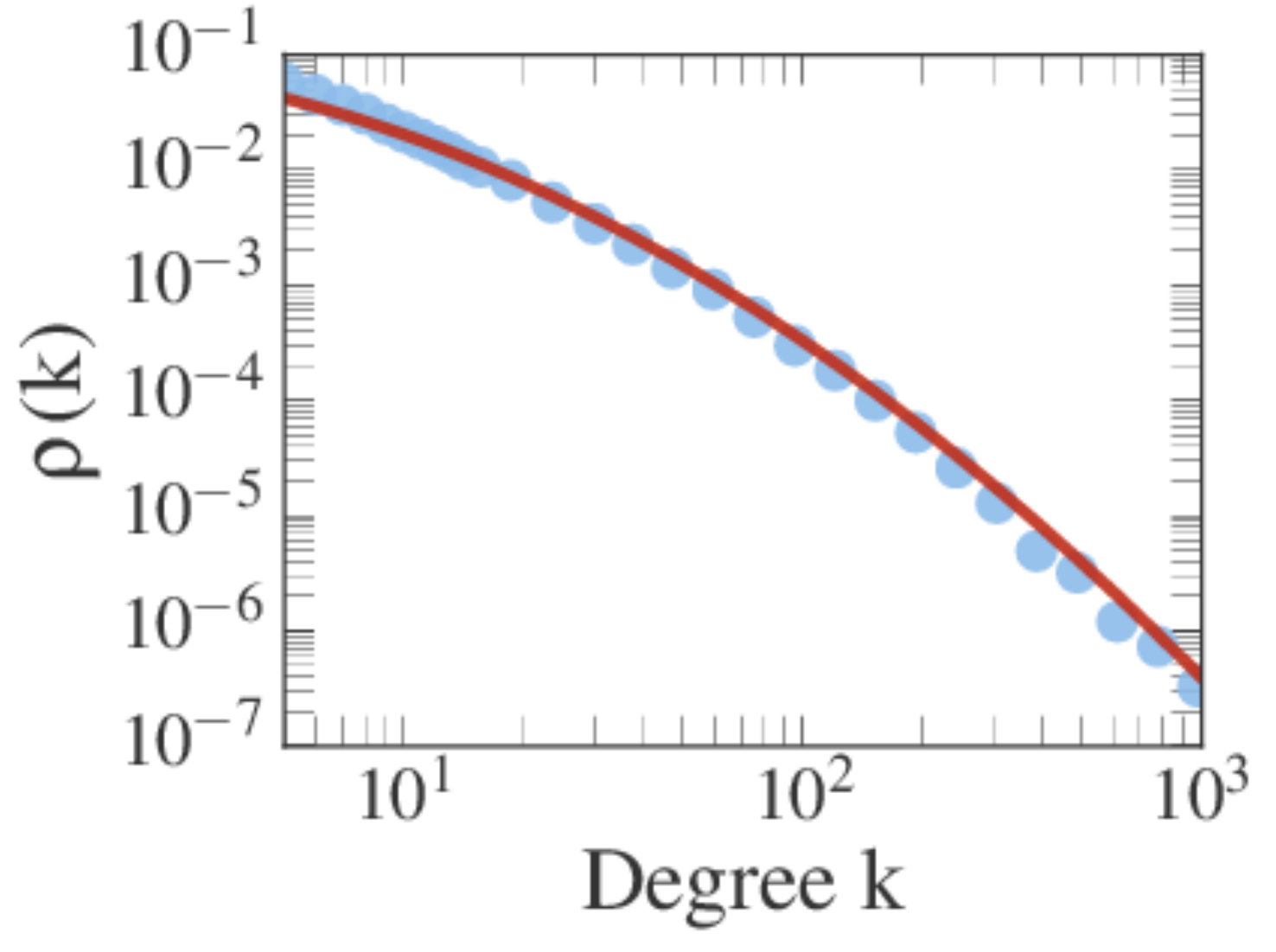}}
    \caption{
        \label{fig:rhos}
        The degree distribution $\rho(k)$ for: (a) PRA, (b) PRB, (c) PRD, (d) PRE, (e) PRL
        and (f) TMN (blue circles).
        We compare the results with the predicted behavior of Table (1) of main paper
        given the parameters of Table (\ref{tab:Fa}) (red solid lines).
        We use the single value of $\beta_{\rm opt}$ defined by Eq.
        (\ref{eq:beta_opt}) in all the cases.
        As in Fig. \ref{fig:F_a} we show the data and fit ranging from the lower
        bound to the $99.9\%$ of the measured data, thus excluding
        from the visible area the top $0.1\%$ of the degrees values.
    }
\end{figure}

Finally, in Fig. \ref{fig:rhos}  we present the degree distributions, together with the
predicted functional form of degree distribution as found in Table (1) in the main paper.

\end{document}